\documentclass[iicol, pdflatex,sn-mathphys-num]{sn-jnl}

\usepackage{graphicx}%
\usepackage{multirow}%
\usepackage{amsmath,amssymb,amsfonts}%
\usepackage{amsthm}%
\usepackage{mathrsfs}%
\usepackage[title]{appendix}%
\usepackage{xcolor}%
\usepackage{textcomp}%
\usepackage{manyfoot}%
\usepackage{booktabs}%
\usepackage{algorithm}%
\usepackage{algorithmicx}%
\usepackage{algpseudocode}%
\usepackage{listings}%
\usepackage{ulem}

\theoremstyle{thmstyleone}%
\theoremstyle{thmstyletwo}%

\theoremstyle{thmstylethree}%

\begin{document}

\title[Article Title]{ A unified framework for identifying influential nodes in hypergraphs}

\author[1,4]{\fnm{Yajing} \sur{Hao}}

\author*[2,4,5,6,7]{\fnm{Longzhao} \sur{Liu}}\email{longzhao@buaa.edu.cn}

\author*[2,4,5,6,7]{\fnm{Xin} \sur{Wang}}\email{wangxin\_1993@buaa.edu.cn}

\author[2,4]{\fnm{Zhihao} \sur{Han}}

\author[1,4]{\fnm{Ming} \sur{Wei}}

\author[2,3,4,5,6,7,8,9]{\fnm{Zhiming} \sur{Zheng}}

\author*[2,3,4,5,6,7,8,9]{\fnm{Shaoting} \sur{Tang}}\email{tangshaoting@buaa.edu.cn}

\affil[1]{\orgdiv{School of Mathematical Sciences}, \orgname{Beihang University}, \orgaddress{\city{Beijing}, \postcode{100191}, \country{China}}}

\affil[2]{\orgdiv{School of Artificial Intelligence}, \orgname{Beihang University}, \orgaddress{\city{Beijing}, \postcode{100191}, \country{China}}}

\affil[3]{\orgdiv{Hangzhou International Innovation Institute}, \orgname{Beihang University}, \orgaddress{\city{Hangzhou}, \postcode{311115}, \country{China}}}

\affil[4]{\orgdiv{Key laboratory of Mathematics, Informatics and Behavioral Semantics}, \orgname{Beihang University}, \orgaddress{\city{Beijing}, \postcode{100191}, \country{China}}}

\affil[5]{\orgdiv{Beijing Advanced Innovation Center for Future Blockchain and Privacy Computing}, \orgname{Beihang University}, \orgaddress{\city{Beijing}, \postcode{100191}, \country{China}}}

\affil[6]{\orgdiv{Zhongguancun Laboratory}, \orgaddress{\city{Beijing}, \postcode{100094}, \country{China}}}

\affil[7]{\orgdiv{State Key Laboratory of Complex \& Critical Software Environment}, \orgname{Beihang University}, \orgaddress{\city{Beijing}, \postcode{100191}, \country{China}}}

\affil[8]{\orgdiv{Institute of Trustworthy Artificial Intelligence}, \orgname{Zhejiang Normal University}, \orgaddress{\city{Hangzhou}, \postcode{310012}, \country{China}}}

\affil[9]{\orgdiv{Beijing Academy of Blockchain and Edge Computing}, \orgaddress{\city{Beijing}, \postcode{100085}, \country{China}}}


\abstract{
Identifying influential nodes plays a pivotal role in understanding, controlling, and optimizing the behavior of complex systems, ranging from social to biological and technological domains. Yet most centrality-based approaches rely on pairwise topology and are purely structural, neglecting the higher-order interactions and the coupling between structure and dynamics. Consequently, the practical effectiveness of existing approaches remains uncertain when applied to complex spreading processes. 
To bridge this gap, we propose a unified framework, Initial Propagation Score (IPS), to directly embed propagation dynamics into influence assessment on higher-order networks. 
We analytically derive mechanism-aware influence measures by relating the early-stage dynamics and local topological characteristics to long-term outbreak sizes, and such explicit physical context endows IPS with robustness, transferability, and interpretability.
Extensive experiments across multiple dynamics and more than 20 real-world hypergraphs show that IPS consistently outperforms other leading baseline centralities.
Furthermore, IPS estimates node influence with only local neighborhood information, yielding computational efficiency and scalability to large-scale networks.
This work underscores the necessity of considering dynamics for reliable identification of influential nodes and provides a concise principled basis for optimizing interventions in epidemiology, information diffusion, and collective intelligence.

}

\maketitle

\section{Introduction}\label{sec1}
Complex systems in society, biology, communication, and beyond emerge from the interrelations among their constituent units and are naturally represented as complex networks, where entities are modeled as nodes and their pairwise relations are modeled as edges~\cite{csreview, brain, prx, microbiomeNet, unifying}. 
In such systems, a small fraction of “critical” nodes exerts a significant influence on global function. For instance, analyses of Twitter have suggested that roughly 1\% of accounts drive about 80\% of rumor diffusion~\cite{baribi2024supersharers}; in functional brain networks, removing only 7\% of nodes with high Collective-Influence scores can nearly abolish network integration~\cite{brainIC}; and in Large Language Models, a small number of instruction examples corresponding to high-influence nodes in a learned instruction graph can substantially improve instruction-tuning performance~\cite{llmic}. 
To a large extent, the importance of critical nodes lies in their ability to propagate impact across the system, and they are also called influential nodes. Identifying such influential nodes has become a fundamental problem that has attracted sustained attention~\cite{app}, and a broad spectrum of approaches has been developed to score node influence. Among them, the most prominent are centrality-based methods, which quantify a node’s influence based on network structure, such as degree, betweenness, etc.~\cite{ degree, kcore, Pei, IC, diffusioncapacity, domirank}. 

In recent years, traditional pairwise graphs have been proven insufficient to capture group interactions that transcend dyads. For example, under social reinforcement, consensus within a group can lead to a higher adoption rate than simple pairwise propagation, and such higher-order mechanisms profoundly alter macroscopic dynamics~\cite{unexplained, 19nc,cpseeding, Multistability}.  
This motivates the study of higher-order networks (e.g., hypergraphs), in which a single “hyperedge” can join an arbitrary number of nodes~\cite{higherReview}. 
Consequently, identifying influential nodes under higher-order interactions becomes a research focus. Recent work has extended classical centralities to higher-order settings: hyper-degree naturally generalizes degree to count incident hyperedges~\cite{higherReview}; hyper-betweenness and hyper-closeness incorporate higher-order distance notions~\cite{hyperbet,swalk}; and hyper-eigenvector centrality models influence transfer through coupled nodes and hyperedges~\cite{hypereigen-l,hypereigen-s,hypereigen-ne}. Notably, hyper-coreness has shown effectiveness in ranking influential nodes under higher-order dynamics~\cite{hypercore}. 
Beyond such extensions, several measures analyze higher-order structures directly: The gravity-based centrality couples node degree with higher-order distance to score node influence~\cite{hypergravity}; The von Neumann entropy method quantifies a node’s importance via the entropy change~\cite{hyperfuzzy}; And, recently, a method extends message passing to hypergraphs and jointly tracks the activation probabilities of nodes and hyperedges to maximize the influence of node sets~\cite{hypertIC}.

Despite such progress, their identification performance in higher-order settings varies markedly across domains~\cite{hypereigen-l}. 
The coreness-type measures perform well in social spreading~\cite{hypercore}, whereas path-based ranking may perform better in transportation analysis~\cite{trans}.
Even within contexts of social contagion, changes in the underlying propagation mechanism would also affect measure performance~\cite{hip}. 
As a result, current methods, largely rooted in structural characteristics or a certain dynamics, struggle to stay effective across real-world scenarios, calling for a unified framework that is robust and applicable to various higher-order interactions.

In this work, we unify structural and dynamical information and propose a general centrality framework called Initial Propagation Score (IPS), which makes use of the early-stage propagation range. Surprisingly, we find that such a simple measure can effectively reflect node influence across various higher-order cases. We first demonstrate that even within a fixed dynamical process, the influential nodes of higher-order systems can shift substantially as dynamical parameters change. This instability is analytically traced to the intrinsic parameter sensitivity of dynamical processes such as the high-order threshold model. Despite this, IPS effectively overcomes this sensitivity, maintaining superior accuracy across different parameter regimes. 
Further, we validate the effectiveness, robustness, and transferability of our approach within three classical higher-order spreading models and one opinion-evolution model on more than 20 empirical hypergraphs. 
We also emphasize that the results are explainable---anchored in clear physical reasoning---yet simple and efficient, positioning IPS as a practical tool for the reliable identification of influential nodes in real applications.

\section{Results}\label{sec2}

\subsection{Individual influence varies under different dynamics}\label{subsec2.1}
\begin{figure*}[h]
\centering
\includegraphics[]{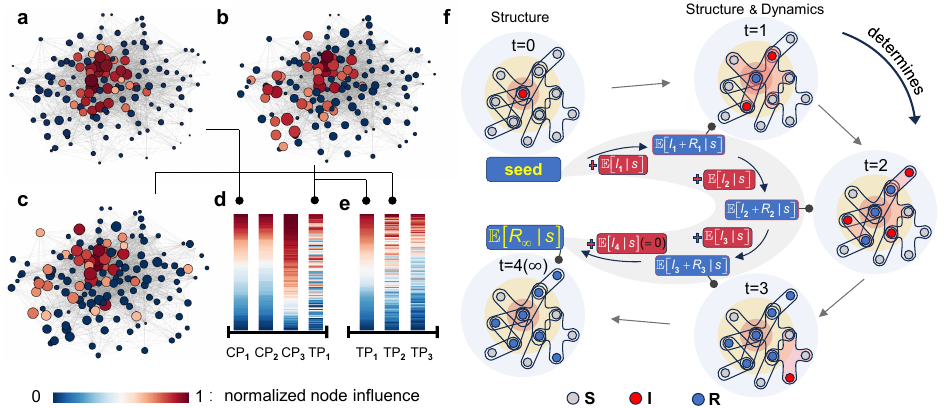}  
\caption{\textbf{Motivation and schematic of the IPS method.} 
\textbf{a-c}: The top 25\% of nodes Email-Enron are highlighted based on their normalized influence (using the bottom color map), under different spreading dynamics and parameter settings: \textbf{a} HCP model with parameter CP\textsubscript{1}, \textbf{b} HTC model with parameter TP\textsubscript{1}, \textbf{c} HTC model with parameter TP\textsubscript{2}. The remaining nodes are uniformly colored in dark blue, and the node size is proportional to its propagation range at the first time step. Results demonstrate that node influence is highly sensitive to the choice of spreading dynamics and parameter settings.
\textbf{d} and \textbf{e}: Complete comparison of node influence under different dynamical configurations. Each bar contains all the nodes in Email-Enron, and nodes are colored by their normalized node influence under the given model and dynamical parameters. The arrangement order of nodes is uniform to the first bar (CP\textsubscript{1} and TP\textsubscript{1}, respectively). Results show that dynamics affect node influential ranking, and the sensitivity to the parameter varies in different models. Node influence is averaged over 10,000 independent simulations, with dynamical parameters: CP\textsubscript{1}: $\nu=1$ and $\lambda=0.0115$, CP\textsubscript{2}: $\nu=2$ and $\lambda=0.0108$, CP\textsubscript{3}: $\nu=1$ and $\lambda=0.02$, TP\textsubscript{1}: $\theta=0.25$ and $\eta=0.05$, TP\textsubscript{2}: $\theta=0.5$ and $\eta=0.1$, TP\textsubscript{3}: $\theta=1/37$ and $\eta=0.027$, $\mu=1$ for all cases (see models and the meaning of dynamical parameters in Methods).
\textbf{f}: The idea and schematic of the IPS method. 
The importance of a node is essentially its range of influence. Taking spreading dynamics as a paradigmatic example, the influence of a node $s$ can be represented by the expected number of infected individuals given $s$ as the initial seed, i.e., $\mathbb{E}(R_\infty \mid s)$. For convenience, suppose $\mu=1$. $\mathbb{E}(R_\infty \mid s)$ can be decomposed as the accumulation of $\mathbb{E}(I_t\mid s)$, and approximated by $\mathbb{E}(I_t+R_t\mid s)$ for $\mathbb{E}(I_\infty\mid s)=0$. The toy hypergraphs in this panel give an example of a propagation chain.
} 
\label{fig1}
\end{figure*}

In this section, we will utilize two representative hypergraph-based propagation models to elucidate the complex effects of dynamical processes on node influence, a finding that motivates this study. After that, we propose a node influence evaluation framework that concisely and effectively integrates the underlying dynamics.

A hypergraph is generally given as $\mathcal{H}=(\mathcal{V},\mathcal{E})$, where $\mathcal{V}=\{v_1,v_2,...,v_N\}$ represents the set of $N$ nodes and $\mathcal{E}=\{h_1,h_2,...,h_M\}$ is the set of hyperedges. 
Here, a hyperedge $h$ contains a set of nodes, and $m=|h|$ is its size. 
Intuitively, a hyperedge denotes the interaction among a set of nodes (such as a chat group), which is a general extension of the normal pairwise edges where $m=2$~\cite{higherReview}.

To reveal the variance of node influence ranking under different dynamics, we consider two widespread propagation processes, 
the higher-order contagion model with power-law infection kernel (HCP model)~\cite{cpseeding,19nc} and the higher-order threshold contagion model (HTC model)~\cite{thres, Multistability} (see details in Methods). 
Using a widely used hypergraph constructed from the Email-Enron dataset as the underlying topology, we identify the ground-truth influential nodes under each of the two hypergraph dynamics models by measuring the total infection size through extensive simulations, following standard practice in prior works~\cite{kcore, hypercore}. 
The details of the simulations and the dataset can be seen in Methods. 
We evaluate three groups of propagation parameter settings for each model, denoted as CP\textsubscript{1}--CP\textsubscript{3} for the HCP model and TP\textsubscript{1}--TP\textsubscript{3} for the HTC model.
Figure~\ref{fig1}a-c highlight the top 25\% influential nodes under different dynamical mechanisms and parameters: Fig.\ref{fig1}a corresponds to the HCP mechanism with parameter setting CP\textsubscript{1}, while Fig.\ref{fig1}b and c correspond to the HTC dynamics with parameter settings TP\textsubscript{1} and TP\textsubscript{2}. Notably, the influential nodes and their importance significantly change under different dynamical settings. This indicates that both dynamical mechanisms and parameters play a pivotal role in identifying key nodes. 
For a more comprehensive and intuitive illustration, we arranged all nodes in a single sequence, using different colors to represent their influence. In Fig.~\ref{fig1}d and Fig.~\ref{fig1}e, CP\textsubscript{1} and TP\textsubscript{1} respectively serve as the baselines, where the nodes are ordered according to their influence scores. 
The ordering-similarity among CP\textsubscript{1}, CP\textsubscript{2}, and CP\textsubscript{3} reveals stable node influence patterns with no significant fluctuations. However, notable differences emerge when comparing CP\textsubscript{1} with TP\textsubscript{1}, as well as among TP\textsubscript{1}, TP\textsubscript{2}, and TP\textsubscript{3}.
Such observations indicate that the sensitivity of node influence to parameter changes varies across different higher-order dynamical processes. 
A natural question, then, is how to define a general measure of node influence that integrates dynamical information and remains effective under diverse higher-order interactions.

To begin with, we consider the standard $S$–$I$–$R$ notation, which has been widely used to describe various real-world spreading dynamics~\cite{higherReview,hypercore}. Generally, nodes that have not yet undergone propagation are denoted as $S$ (susceptible), nodes that have been activated and can propagate information, disease, or opinions are marked as $I$ (infected), and nodes that are immune to the propagation process and no longer participate in it are labeled as $R$ (recovered or removed).
Denote the number of nodes in state $S$, $I$ and $R$ at time $t$ as $S_{t}$, $I_{t}$, and $R_{t}$, respectively.
The essence of evaluating node influence lies in establishing a connection between the seed node and its eventual propagation outcome, i.e., the expected number of removed nodes at the end of the process, denoted as $\mathbb{E}[R_{\infty}|s]$, where $s$ is a given seed.
In fact, regardless of the specific dynamical mechanism, the final spreading range can be viewed as the accumulation of infected nodes over time, and is progressively approximated by the sequence $\{\mathbb{E}[I_t+R_t|s]\}$ (for $\lim_{t\to \infty}\mathbb{E}[I_t|s]=0$ in SIR models). For convenience of analysis, we suppose the recovery rate from $I$ to $R$ is $\mu=1$, and we have:
\begin{align}
\mathbb{E}[R_\infty \mid s] &= \sum_{t=0}^{\infty} \mathbb{E}[I_t \mid s], \nonumber \\
\mathbb{E}[I_t + R_t \mid s] &= \sum_{t'=0}^{t} \mathbb{E}[I_{t'} \mid s].
\label{eq0}
\end{align}
See Fig.~\ref{fig1}f for an instantiation.
For cases when $\mu \neq 1$, we only need to refine $I_t$ as the number of newly infected nodes at time $t$.
Note that each propagation step relies on the former step and ultimately traces back to the original seed node.
Intuitively, the early-stage propagation inherently encodes both the seed’s topological position and the system’s dynamical characteristics, which in turn, largely determine the final propagation outcome.
Therefore, we approximate the ultimate influence of each seed using its early-stage propagation range, forming the basis of our IPS (Initial Propagation Score) framework. 
Formally, given a dynamical process $P$ and a target time step $t$, the IPS measure is denoted as $IPS_{t}^{P}=\mathbb{E}[I_t + R_t \mid s]$ (also called the $t$-th order IPS).

\subsection{Effectiveness of IPS on diverse real-world hypergraphs}\label{subsec2.2}
\begin{figure*}[h]
\centering
\includegraphics[]{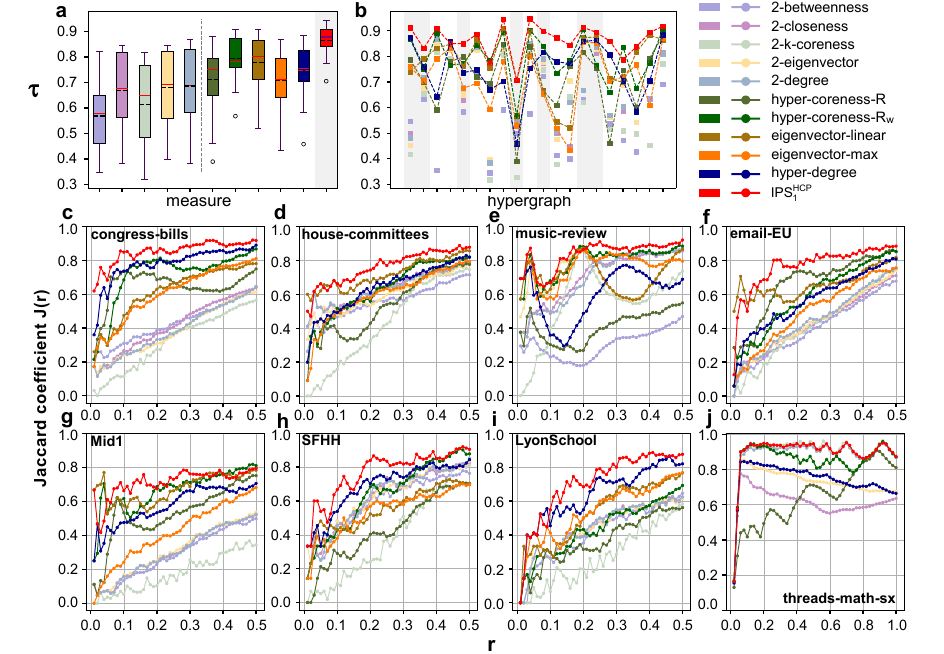}
\vspace{-8pt}
\caption{ 
\textbf{IPS successfully predicts node influence across diverse real-world hypergraphs.} IPS is compared with 10 commonly used hypergraph centrality measures, evaluated by Kendall's $\tau$ in \textbf{a}-\textbf{b} and Jaccard coefficient $J(r)$ in \textbf{c}-\textbf{j}.
\textbf{a} Each box summarizes Kendall's $\tau$ of a corresponding measure in 20 real hypergraphs. The solid and dashed lines in each box represent the median and the mean, respectively.  
\textbf{b} Values of Kendall's $\tau$ are plotted against the 20 hypergraphs, which are ordered by decreasing order of node count as follows: 
1. congress-bills, 2. house-committees, 3. music-review, 4. M\_PL\_062\_ins, 5. email-EU, 6. M\_PL\_015\_ins, 7. Mid1, 8. geometry-questions, 9. M\_PL\_062\_pl, 10. algebra-questions, 11. SFHH, 12. Elem1, 13. Thiers13, 14. senate-bills, 15. senate-committees, 16. LyonSchool, 17. InVS15, 18. email-Enron, 19. M\_PL\_015\_pl, 20. LH10.
Hyperstructure-based methods are connected by dashed lines for emphasis. 
Results show that IPS performs the best across almost all hypergraphs.
\textbf{c}--\textbf{j} show the Jaccard coefficient $J(r)$ of methods, which scans the ability of each method to identify key nodes at different scales. IPS (the red line) is the best in almost all cases. 
\textbf{c}--\textbf{i} reports results of the highlighted hypergraphs in \textbf{b}, while panel \textbf{j} (threads-math-sx) reports results of an additional large-scale hypergraph with 152,702 nodes. All colors in this figure are painted according to the upper-right legend. Simulation settings are detailed in Methods. Parameter settings are provided in Supplementary Table~S1.
}\label{fig2}
\end{figure*}

In this section, we adopt the HCP model to demonstrate the calculation process and show that even the first-order IPS ($t=1$) could exhibit strong predictive power in real-world systems.
At $t=1$, propagation can only occur within the first-order neighborhood of the seed node $s$, and the expected number of newly infected nodes is the sum of infection probabilities over its neighbors.
In HCP, the infection rate in a hyperedge with $i$ infected nodes is given as $\lambda i^{\nu}$, where $\lambda$ is a global propagation rate parameter and $\nu$ reflects the reinforcement effect.  
Since only the seed node is infected at $t=0$ and all other nodes are susceptible, the probability that a neighboring node becomes infected via a shared hyperedge is $1-e^{-{\lambda}}$ (see details in Methods). 
Due to the independence of infection events across hyperedges, the probability that a neighbor $i$ of $s$ becomes infected at $t=1$ is $p_{1}(i\mid s)=1-e^{-k_s(i){\lambda}} \approx k_s(i)\lambda$ ($\lambda$ is typically small), where $k_s(i)=|\{h\mid s, i\in h,h\in\mathcal{E}\}|$ is the number of hyperedges adjacent to $s$ that contain $i$.
Then the expected propagation range (number of infected nodes and recovered nodes) of the seed node $s$ at $t=1$ is:
\begin{align}
IPS_1^{HCP}&=\mathbb{E}(R_1+I_1|s) \nonumber \\
&=  1+\sum\limits_{i\in N_v^1(s)}(1-e^{-\lambda{k_s(i)}}) \nonumber \\ 
&\approx 1+\sum\limits_{i\in N_v^1(s)}\lambda{k_s(i)}  \nonumber \\  
&= 1 + \lambda\sum\limits_{h\in N_h^1(s)}(|h|-1),  
\label{eq1}
\end{align}
where $N_v^1(s)=\bigcup_{ s\in h, h \in \mathcal{E}}h \setminus\{s\}$ is the set of first-order neighbors of $s$, $N_h^1(s)=\{h|s\in h, h \in \mathcal{E}\}$ is the set of adjacent hyperedges of $s$. 
Eq.\eqref{eq1} indicates that the ranking of $IPS_1^{HCP}$ is solely determined by the sum of the sizes of its neighboring hyperedges, 
leading to a computational complexity of $\mathcal{O}(N\langle k\rangle$ that is highly efficient for large-scale applications (Here $\langle k\rangle$ is the average hyper-degree of the hypergraph, and the details for time complexity are provided in Supplementary Section~S1.3). We also notice that the ranking of $IPS_1^{HCP}$ is independent of the dynamical parameters, which is consistent with the previously observed robustness of influence rankings across different HCP parameters in Fig.~\ref{fig1}d.

We evaluate the predictive performance of $IPS_{1}^{HCP}$ on 20 real-world higher-order datasets from various domains, including human contact networks, email communications, legislative collaboration, online platforms, and ecological systems. These datasets vary in size, density, and functional context. 
A complete summary of the datasets is provided in Methods.

For comprehensive benchmarking, we include 10 existing centrality measures widely used in higher-order interactions, including the basic hyper-degree, the recently proposed hyper-coreness-$R$, hyper-coreness-$R_w$~\cite{hypercore}, the Node-and-edge-nonlinear-eigenvector centrality measures (abbreviated as eigenvector-linear and eigenvector-max), and the natural extension of five classic measures to the 2-projection networks of hypergraphs, containing degree, k-coreness, eigenvector, betweenness, and closeness. 
To distinguish these projection-based measures from their standard counterparts in pairwise networks, we prefix them with `2-' (e.g., 2-degree, 2-betweenness, etc.).

Here we use different indicators to evaluate and compare measures' performance. 
Specifically, we first use Kendall's $\tau$ to quantify the consistency between the node influence ranking based on different measures and the ground truth derived from over 300 times simulations, with a larger $\tau$ (closer to 1) indicating better performance of the measure. 
Figure~\ref{fig2}a clearly shows that the $IPS_1^{HCP}$ method, corresponding to the red box, has the best overall performance, which even outperforms the second-best algorithm by around 10\%. 
We then unfold the results on each hypergraph in Fig.~\ref{fig2}b. 
Notably, the $IPS_1^{HCP}$ method consistently performs the best in almost all datasets (red points), while other measures show large fluctuations across different hypergraphs.  
We also employ the Jaccard similarity, which quantifies the overlap between the predicted top  \(\lfloor rN \rfloor\) influential nodes by centrality measures and the ground truth, to further evaluate the prediction accuracy.
Figure~\ref{fig2}c--i present results on several representative hypergraphs (highlighted in gray in Fig.~\ref{fig2}b), and Fig.~\ref{fig2}j additionally shows a case on a large-scale hypergraph. Again, the $IPS_1^{HCP}$ method consistently achieves the highest similarity across nearly all settings. See the results for other hypergraphs in Supplementary Figures~S3,~S4, and S6.

\begin{figure}[h]
\centering
\includegraphics[]{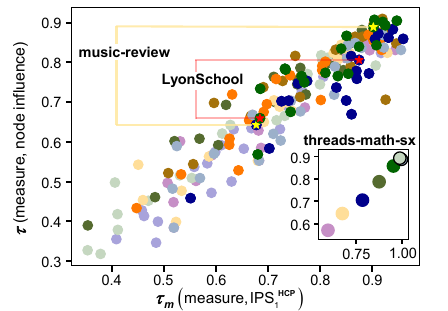}  
\caption{\textbf{Explanation for the variability of method performance.} 
This plot shows the relationship between each method’s identification performance and its rank correlation with IPS. The clear positive correlation indicates that stronger alignment with IPS leads to better identification performance. 
The inset shows the case of threads-math-sx (overlapped points are circled for clarity).  
}\label{fig3}
\end{figure}

Taken together, these results validate the superior stability and effectiveness of the $IPS_1^{HCP}$ method across a wide range of real-world hypergraphs. Other methods, by contrast, exhibit considerable variability, with performance often depending on both the network structure and the proportion $r$. For example, hyper-coreness-\(R_w\) and eigenvector-linear occasionally achieve high performance within certain ranges of $r$, but display sharp declines as $r$ increases. Moreover, while they often appear as the second-best performers (e.g., Fig.~\ref{fig2}e, f, g), they are outperformed by hyper-degree in several hypergraphs (e.g., Fig.~\ref{fig2}c, h, i). Furthermore, the general trend that hyperstructure-based methods outperform projection-based counterparts is reversed in the large-scale threads-math-sx hypergraph, where 2-k-coreness and 2-degree match the performance of $IPS_1^{HCP}$. 
Notably, several methods could not even be evaluated on this hypergraph due to the excessive computational cost.

In Fig.~\ref{fig3}, we further analyze the consistency between the $IPS_1^{HCP}$ method and other baseline methods.
We plot each method's Kendall correlation with $IPS_1^{HCP}$ ($\tau_m$) against its predictive performance ($\tau$, as defined in Fig.~\ref{fig2}a, b). Each point represents a specific method evaluated on a certain hypergraph, with color-coding consistent with the legend in Fig.~\ref{fig2}. 
We find a pronounced positive correlation, whereby methods that exhibit stronger agreement with $IPS_1^{HCP}$ tend to achieve higher accuracy.  
Such a clear alignment, which is not observed in other measures (see Supplementary Figure~S5, offers a unified explanation for the performance fluctuations of different algorithms across different higher-order structures. For instance, the highlighted nodes and the subgraphs in Fig.~\ref{fig3} illustrate why some baseline methods outperform others in Fig.~\ref{fig2}e, i, and j. 
More importantly, this suggests that our method may have captured the fundamental features governing node importance, namely, the coupling between local topology and early dynamics.

\subsection{Robustness of IPS towards different parameters}\label{subsec2.3}
We incorporate the second-order structural and dynamical information, and extend $IPS_1^{HCP}$ to $IPS_2^{HCP}$ under the HCP model, assessing whether such enrichment can substantially improve predictive accuracy. Furthermore, we will systematically analyze the robustness of the IPS methods under varying dynamical parameters.

Following Eq.~\eqref{eq0}, when considering second-order information ($t=2$), we have:
\begin{align}
    \mathbb{E}[R_2+I_2\mid s]=& \mathbb{E}[R_1+I_1\mid s] +  \mathbb{E}[I_{2} \mid s].
    \label{eq_t2}
\end{align}
Here $\mathbb{E}[R_1+I_1\mid s]$ is given by Eq.\eqref{eq1}, and $\mathbb{E}[I_{2} \mid s]$ denotes the number of newly infected nodes at time $t=2$. Let $p_{t,n}(i|s)$ represents the probability that the $n$-th order neighbor $i$ of the seed $s$ is infected at time $t$. Then, $\mathbb{E}[I_{2} \mid s]$ can be derived by 
\begin{align}
    \mathbb{E}[I_{2} \mid s]=& \sum\limits_{i\in N_v^1(s)}p_{2,1}(i|s)(1-p_{1,1}(i|s))   \nonumber \\
    &+ \sum\limits_{i\in N_v^2(s)}p_{2,2}(i|s),  
\label{eq_t2_e}
\end{align} 
where the first item is the expected number of first-order neighbors that are infected at time $t = 2$, and the remaining item records the expected number of second-order infected neighbors.  

We then need to compute $p_{t,n}(i|s)$. Let $N_h^1(s)$ and $N_h^2(s)=\bigcup _{i\in N_v^1(s)} N_h^1(i) \setminus N_h^1(s)$ represent the set of first-order adjacent hyperedges and second-order hyperedges, respectively. Moreover, the expected number of infected nodes within hyperedge $h$ can be approximated as follows: If $i\in N_v^1(s)$, $I_1(i,h) = \sum_{j\in h\cap  N_v^1(s),j\neq i}p_{1,1}(j|s)$; Otherwise, $I_2(h) = \sum_{j\in h \cap  N_v^1(s)}p_{1,1}(j|s)$. Finally, through enormous computation (see details in Supplementary Section~S1.1), we can obtain the expression of $p_{t,n}(i|s)$, i.e.,
\begin{align}
&p_{1,1}(i|s)=1-e^{-k_s(i)\lambda}, \qquad  \text{for }i \in N_v^1(s), \nonumber\\
&p_{2,1}(i|s)=
    \begin{aligned}[t]
    &1-
    \begin{aligned}[t]
        &\!\!\prod_{h\in  N_h^1(i)\cap N_h^1(s)\atop {I_1(i,h)}\geq1}\!\!\!\!f_1(i,h)
        \!\!\prod_{h\in  N_h^1(i)\cap N_h^1(s)\atop {I_1(i,h)}<1}\!\!\!\!f_2(i,h)   \nonumber\\
        &\!\!\prod_{h\in N_h^1(i)\cap N_h^2(s)\atop {I_1(i,h)}\geq1}\!\!\!\!f_3(i,h) 
        \!\!\prod_{h\in N_h^1(i)\cap N_h^2(s)\atop {I_1(i,h)}<1}\!\!\!\!f_4(i,h),  \nonumber\\
    \end{aligned} \nonumber\\
    &\text{for } i \in N_v^1(s), \nonumber\\
    \end{aligned} \nonumber\\
&p_{2,2}(i|s)=
    \begin{aligned}[t]
    &1-\!\prod_{h\in N_h^1(i)\cap N_h^2(s) \atop {I_2(h)}\geq1}\!\!\!f_5(h)
    \!\prod_{h\in N_h^1(i)\cap N_h^2(s) \atop {I_2(h)}<1}\!\!\! f_6(h), \nonumber\\
    &\text{for } i \in N_v^2(s), \nonumber\\
    \end{aligned} \nonumber\\
\label{eq_t2_p}
\end{align} 
with:
\begin{align}
&f_1(i,h)=\mu e^{-\lambda {{I_1(i,h)}}^\nu} +(1-\mu)e^{-\lambda ({I_1(i,h)}+1)^\nu},  \nonumber\\
&f_2(i,h)
    \begin{aligned}[t]
        &=\mu (1-{I_1(i,h)}(1-e^{-\lambda}))    \nonumber\\
        &+(1-\mu)e^{-\lambda ({I_1(i,h)}+1)^\nu},  \nonumber\\
    \end{aligned} \nonumber\\
&f_3(i,h) = e^{-\lambda {I_1(i,h)}^\nu}, \nonumber\\
&f_4(i,h) = 1-{I_1(i,h)}(1-e^{-\lambda}) , \nonumber\\
&f_5(h) = e^{-\lambda {I_2(h)}^\nu},  \nonumber\\
&f_6(h) = 1- {I_2(h)}(1-e^{-\lambda}). 
\label{eq_t2_f}
\end{align}
See Supplementary Section~S1.1 for an example in a hypergraph.

\begin{figure*}[h]
\centering
\includegraphics[]{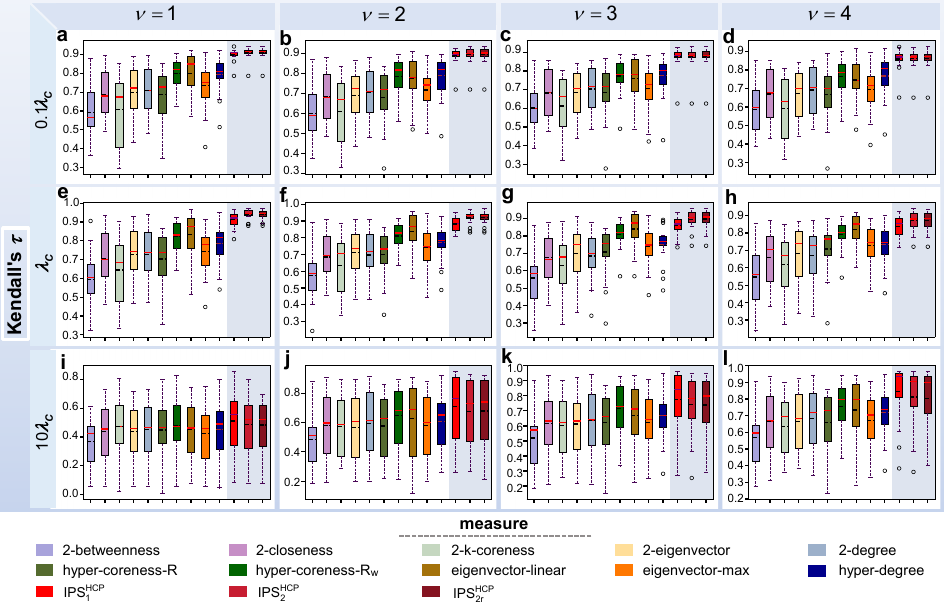}
\caption{\textbf{IPS methods consistently outperform other benchmarks under different dynamical parameters.}
We report Kendall's $\tau$ between measures and ground-truth under the HCP model in different dynamical parameters. Each column of panels corresponds to the $\nu$ given in the headline; each row corresponds to the $\lambda$ given on the left side. Here, $\lambda$ takes a uniform multiple of $\lambda_c$, while $\lambda_c$ may be different in each hypergraph.
Each panel contains the results in 20 real hypergraphs.
The solid line gives the median, and the dashed line gives the mean.
IPS methods achieve the highest values of Kendall's $\tau$ in all cases, indicating their outstanding performance and robustness.
Simulation details are given in Methods, and the specific values of dynamical parameters for each hypergraph are provided in Supplementary Table~S2.}\label{fig4}
\end{figure*}

It is noteworthy that $IPS_{2}^{HCP}$ exhibits a complexity of $\mathcal{O}(N\langle k\rangle k_{\max} m_{\max}^2)$, where $k_{\max}$ is the maximal hyper-degree and $m_{\max}=\max_{h\in \mathcal{E}}|h|$ is the maximal size of hyperedges. 
Similar to Eq.~\eqref{eq1}, through infinitesimal substitution that transforms node-level calculations into the hyperedge-level, we can make additional approximations and obtain a faster version with time complexity reduced to $\mathcal{O}(N \langle k\rangle k_{\max} m_{\max})$, denoted as $IPS_{2r}^{HCP}$. Details are provided in Supplementary Section~S1.2 and S1.3. When $k_{\max}$ and $m_{\max}$ are bounded (e.g., sparse hypergraphs), the complexity of both versions scales linearly with node number $N$.

We conduct extensive experiments with a large range of dynamical parameters to validate the performance and robustness of IPS methods systematically. 
We consider cases where the recovery probability $\mu = 0.1, 0.5, 1$, and the reinforcement parameter $\nu = 1, 2, 3, 4$, respectively. 
For the contagion parameter $\lambda$, we experimentally calculate the propagation threshold $\lambda_{c}$ for each network with given $\mu$ and $\nu$ (see Methods) and consider $0.1 \lambda_{c}$, $\lambda_{c}$, and $10  \lambda_{c}$ respectively. 
In summary, we examine $20$ real hypergraphs, each with $36$ groups of different dynamical parameters. 
We employ Kendall's $\tau$ as before to evaluate the effectiveness of different methods. 

Figure~\ref{fig4} presents the results for the case of $\mu=1$.
Overall, the IPS methods outperform all other baselines and exhibit strong robustness to parameter variations. Among them, the approximation $IPS_{2r}^{HCP}$ achieves performance comparable to $IPS_{2}^{HCP}$, 
while the relative ranking between second-order IPSs and $IPS_{1}^{HCP}$ depends on the propagation rate $\lambda$.
Not surprisingly, when $\lambda$ is set near the threshold, where the propagation process is highly sensitive to the choice of seed nodes and the accuracy of node ranking is therefore most crucial, $IPS_2^{HCP}$, which incorporates richer information, achieves the best performance (Fig.~\ref{fig4}e--h).
For small $\lambda$ (Fig.~\ref{fig4}a--d, where the propagation efficiency is limited), $IPS_{2}^{HCP}$ and $IPS_{1}^{HCP}$ exhibit similar performance, implying that one-step information is already sufficient to accurately assess node influence.
As for cases with extremely large $\lambda$ (Fig.~\ref{fig4}i--l), which are rare in the real world, $IPS_2^{HCP}$ shows no improvement over $IPS_1^{HCP}$. This is because, at such extremely high spread rates, propagation largely depends on the first step.
In addition, when $\mu \neq 1$, our IPS methods maintain superior performance across all parameter settings.
See results in Supplementary Section~S2.2.

To sum up, IPS methods consistently surpass other centrality measures while maintaining strong robustness towards different parameter regions. 
Its dynamics-informed formulation also provides clear physical interpretability, thereby enhancing its reliability.
Moreover, our results show that although the second-order IPSs perform better in certain scenarios, particularly near the critical threshold, the first-order IPS performs comparably and remains substantially superior to existing approaches. Considering the trade-off between computational efficiency and predictive accuracy, $IPS_1^{P}$ is more practical for real-world applications. We therefore focus on $IPS_1^{P}$ in the subsequent analyses.

\subsection{Transferability of IPS across various higher-order contagion dynamics.}\label{subsec2.4}
\begin{figure*}[h]
\centering
\includegraphics[]{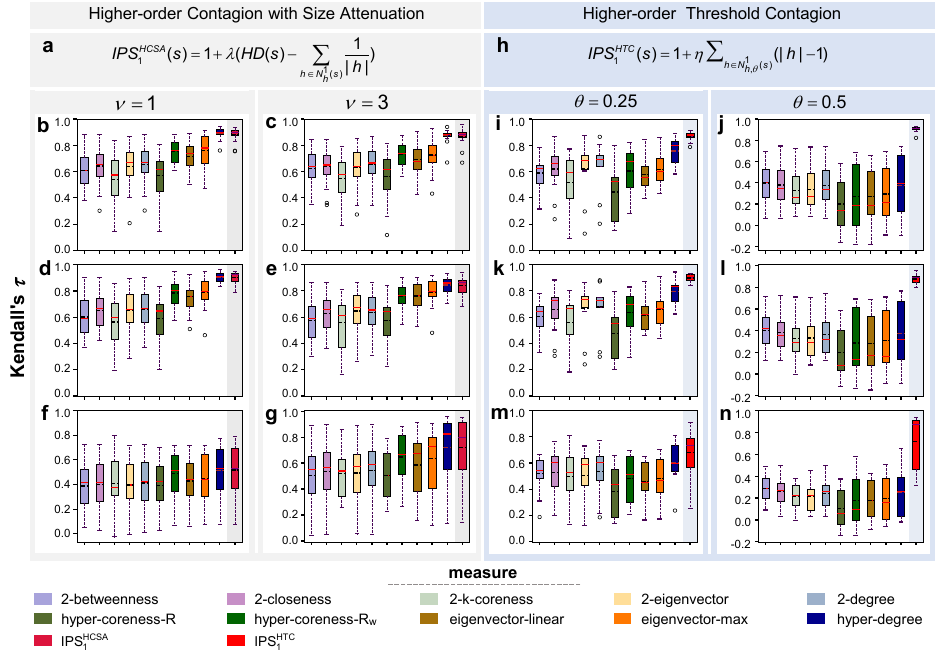}
\caption{\textbf{IPS performs the best across various high-order contagion dynamics.}
\textbf{a},\textbf{h}: First-order IPS under the HCSA and HTC models, respectively.
Box plots provide Kendall's $\tau$ between the ground-truth influence ranking under the corresponding dynamics and the ranking induced by each measure, evaluated on multiple real-world hypergraphs. \textbf{b}--\textbf{g} aggregate results under HCSA. The column headers indicate $\nu$ and the rows indicate $\lambda$: \textbf{b}, \textbf{c}: $0.1\lambda_c$, \textbf{d}, \textbf{e}: $\lambda_c$, \textbf{f}, \textbf{g}: $10\lambda_c$. 
Panels \textbf{i}--\textbf{n} aggregate results over 11 real-world hypergraphs for HTC; column headers indicate $\theta$ and rows indicate $\eta$: \textbf{i}, \textbf{j}: $0.1\eta_c$, \textbf{k}, \textbf{l}: $\eta_c$, \textbf{m}, \textbf{n}: $10\eta_c$ (or $2\eta_c$ instead if $10\eta_c \geq1$). 
In all panels, $\mu=1$. 
The specific dynamical parameters are given in Supplementary Table~S5 and Table~S7. 
See Methods for simulation settings.
}
\label{fig5}
\end{figure*}

IPS builds on the intuition that a node’s early propagation range in a hypergraph serves as an accurate proxy for its eventual influence, thereby enabling natural generalization across diverse higher-order contagion dynamics.
In this section, we further assess its transferability under a size-attenuation variant of the HCP model and the HTC model.

We consider a phenomenon where the interaction intensity and communication efficiency within groups may decline as the group size increases, particularly in face-to-face interactions~\cite{cpseeding,variant1}.
We model this phenomenon by introducing an attenuation term: define the infection rate for susceptible nodes within a hyperedge $h$ containing $i$ infected nodes as $\frac{\lambda}{|h|}i^{\nu}$.
This model is a variant of the basic HCP model, and we call it the higher-order contagion model with size attenuation (HCSA). 
Following the computational methodology of $IPS^{HCP}_{1}$, we derive: 
\begin{align}
    IPS^{HCSA}_{1}(s)&=\! 1+\!\!\!\sum_{i \in N_v^1(s)}(1\! -\!e\!^{-\sum_{h_j \in N_h^1(s),i \in h_j}\! \frac{\lambda}{|h_j|}})    \nonumber \\
    & \approx \! 1+\lambda\sum_{h\in N_h^1(s)}(1-\frac{1}{|h|})  \nonumber \\
    &=\! 1+\lambda (HD(s)- \!\!\! \sum_{h\in N_h^1(s)} \!\! \frac{1}{|h|}),
\end{align}
where HD$(s)$ represents the hyper-degree value of node $s$.

We evaluate the performance of $IPS_1^{HCSA}$ in the same way as before.
Results in Fig.~\ref{fig5}b--g show that $IPS_1^{HCSA}$ exhibits robust and superior performance across all scenarios, while the hyper-degree method, which is formally similar to $IPS_1^{HCSA}$, exhibits comparable performance.
However, this arises from the “weak monotonicity” of hyper-degree, which assigns identical values to many nodes and thereby induces a spurious bias in Kendall's $\tau$. Once the spurious gain is neutralized, $IPS^{HCSA}_{1}$ again maintains a consistent advantage over hyper-degree (see details in Supplementary Section~S4.1.2).
Additionally, contrary to the case of the previous HCP model (Fig.~\ref{fig2}a), eigenvector-max performs better than eigenvector-linear in the HCSA model.
This is because eigenvector-linear assigns higher centrality to nodes belonging to larger hyperedges and possessing more connections, while, eigenvector-max disregards the size of hyperedges and focuses more on the number of connections~\cite{hypereigen-ne}. 
That is, methods that are inherently more similar to the corresponding IPS approach tend to achieve better predictive performance, which is in line with the earlier observations.
Above results are corresponding to the cases when $\mu=1$, we also provide more results for cases when $\mu=0.1$ in Supplementary Section~S4.1.2.

We also consider the higher-order threshold contagion (HTC) model, 
where infection occurs collectively, i.e., if the number of infectious nodes in a hyperedge exceeds (or equals) a fraction $\theta$ of its size, all susceptible nodes in it get infected with probability $\eta$. Note that the propagation only occurs in hyperedges that satisfy the threshold condition, we have:
\begin{align}
IPS^{HTC}_{1}(s) &= 1+\sum_{i \in N_{v,\theta}^1(s)}[1-(1-\eta)^{k_s^\theta(i)}] \nonumber\\
&\approx 1+ \eta \sum_{i \in N_{v,\theta}^1(s)} k_s^\theta(i)  \nonumber\\
&=1+\eta \sum_{h\in N_{h,\theta}^1(s)}(|h|-1),
\label{htc}
\end{align}
where $N_{v,\theta}^1(s)=\{i\mid i\in h\in N_h^1(s),|h|\leq 1/\theta\}$, ${k_s^\theta(i)}=|\{h\mid i\in h\in N_h^1(s),|h|\leq 1/\theta\}|$, and $N_{h,\theta}^1(s)=\{h\mid s \in h, |h|\leq1/\theta\}$.

We consider $\theta \in \{0.25,\, 0.5\}$, and $\eta \in \{0.1\eta_{c},\,\eta_{c},\,10\eta_{c}\,(\text{or} \, \,2\eta_c\, \, \text{instead if}\, \, 10\eta_c \geq1)\}$ to evaluate $IPS^{HTC}_{1}$.
Here, $\eta_c$ is the threshold of $\eta$ for the given $\theta$ and recovery probability $\mu$ (see the calculation of $\eta_c$ in Methods).
The performance of different measures is shown in Fig.~\ref{fig5}i-n.
We find that the approximations of $IPS^{HTC}_{1}$ attain superior predictive accuracy compared to all other baselines.
Besides, contrary to most cases under HCP and HCSA, the projection-based 2-eigenvector and 2-degree outperform hyper-coreness-$R_w$ and eigenvector-linear. 
The performance fluctuation among baselines, together with the consistent superiority of $IPS_1^{HTC}$, once again underscores the importance of incorporating dynamical information into node ranking in the context of complex higher-order interactions. Above results are the cases when $\mu=1$; results when $\mu=0.1$ are in Supplementary Section~S4.2.2.

\begin{figure*}[h]
\centering
\includegraphics[scale=0.98]{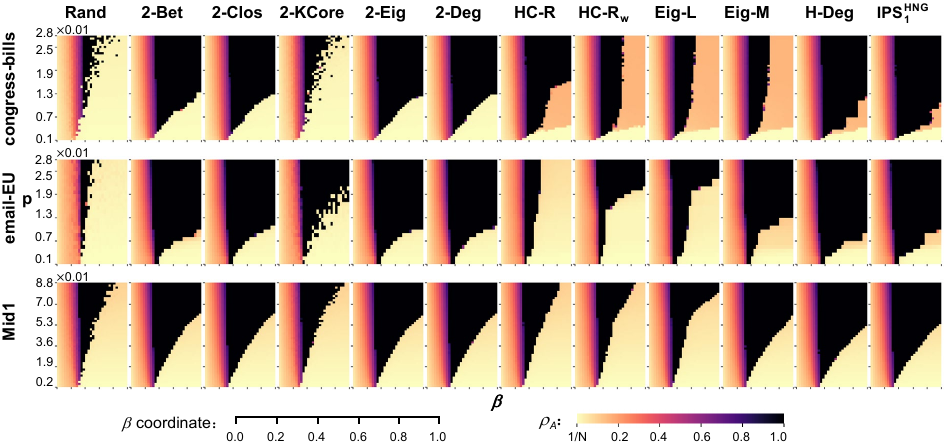}
\caption{\textbf{IPS exhibits strong cross-paradigm transferability.}
We evaluate the performance of $IPS^{HNG}_{1}$ when it serves as a committed nodes selection strategy on three real-world hypergraphs, and compare it with other methods. 
Each heatmap shows the steady-state abundance of name $A$ ($\rho_A$) in cases of different $(\beta,p)$. 
In each grid cell, the $\lfloor pN \rfloor$ committed nodes are selected according to the measure indicated in the panel header, where $N$ is the number of nodes in the hypergraph.
Larger black regions (i.e., $\rho_A=1$) correspond to the lower critical mass $p_c$, indicating more efficient minority takeover. 
Results shows that $IPS^{HNG}_{1}$ effectively reduces the required $p_c$.
Measures in panel headers are abbreviations: Rand: Random, 2-Bet: 2-betweenness, 2-Clos: 2-closeness, 2-KCore: 2-k-coreness, 2-Eig: 2-eigenvector, 2-Deg: 2-degree, HC-R: hyper-coreness-$R$, HC-$R_w$: hyper-coreness-$R_w$, Eig-L: eigenvector-linear, Eig-M: eigenvector-max, H-Deg: hyper-degree, $IPS^{HNG}_{1}$: $IPS^{HNG}_{1}$. 
Details of the agreement rules and simulations are provided in Methods.
}\label{fig6}
\end{figure*}
From a broader perspective, the first-order IPS formulation theoretically reveals the parameter sensitivity of node influence ranking.
Under the HTC model, $IPS^{HTC}_{1}$ depends explicitly on the dynamical parameters through the threshold condition, i.e, the indicator of whether $|h|\leq 1/\theta$. 
Consequently, $IPS^{HTC}_{1}$ at $\theta=0.25$ differs from that at $\theta=0.5$, accounting for the substantial variation in influence rankings across HTC parameter settings (Fig.~\ref{fig1}e).
By contrast, for the HCP model, $IPS^{HCP}_{1}$ is independent of the dynamical parameters, which explains the high concordance of rankings across different HCP parameter choices (Fig.~\ref{fig1}d).
Taken together, the outstanding predictive performance and clear interpretability of IPS demonstrate its strong transferability across modeling dynamics and parameter regimes.

\subsection{Beyond contagion: performance of IPS on higher-order naming game}\label{subsec2.5}
Finally, we assess cross-paradigm transferability of IPS by moving beyond contagion to an opinion dynamics—the naming game (NG) model.
The NG model simulates the emergence of convention under competitive opinion evolution, which is widely used to investigate the critical mass of committed minority opinion to overturn the majority~\cite{ng2,ng4}. 
Here, we adopt the higher-order naming game model (HNG) in the hypergraph framework~\cite{ng5}, and select committed minority agents based on different influence ranking algorithms to examine which approach most effectively promotes the reversal of minority opinions.

In the HNG model, there are two names (denoted as $A$ and $B$), representing different social norms.
Initially, a fraction $p$ of committed nodes hold opinion vocabulary $\{A\}$, whereas the remaining (uncommitted) majority holds $\{B\}$. 
At each following step, a hyperedge is sampled, and two processes occur: (1) broadcast: a random speaker proposes a name from its vocabulary and shares it with other members; (2) negotiation: all members update their vocabulary (except committed nodes) based on a particular agreement rule and the intensity of social influence $\beta$. 
See details of the model in Methods.
With the different setup of the $\beta$ and $p$, the names evolve to different abundances in steady state.
Particularly, as $p$ increases, the minority can take over the system completely, where the minimal $p$ required is the critical mass $p_c$~\cite{hypercore,ng5}.

Following the above mechanism, we focus on its broadcast stage, which is similar to the propagation process. For a committed agent $s$, its expected broadcast range at the initial moment is naturally given as: 
\begin{align}
    IPS^{HNG}_{1}(s)&=1+\sum_{h \in N_h^1(s)}\frac{1}{M}\frac{1}{|h|}(|h|-1) \nonumber \\
    &=1 + \frac{1}{M}(HD(s) -\!\!\!\! \sum_{h \in N_h^1(s)}\frac{1}{|h|}).
\end{align}

We evaluate the effectiveness of $IPS^{HNG}_{1}$ on reducing the critical mass $p_c$ in real-world hypergraphs and compare it with 11 baselines, including a random selection strategy. 
We select the top $\lfloor pN\rfloor$ nodes and record the steady-state abundance of name $A$ (the fraction of nodes holding only $A$, denoted as $\rho_A$). 
Figure~\ref{fig6} shows heatmaps of $\rho_A$ over the $(\beta,p)$ grid for each measure. 
Black regions indicate successful minority takeover ($\rho_A=1$), whereas colored regions correspond to $\rho_A<1$; for each fixed $\beta$, the smallest $p$ at which the heatmap turns black gives the critical mass $p_c$. 
Thus, a larger black region implies a lower $p_c$ and hence more efficient reversal.
Compared with random selection, centrality-based selection substantially lowers $p_c$, although effectiveness varies across measures. 
Among them, $IPS^{HNG}_{1}$---and, as expected, the formally similar hyper-degree (HD)—outperform the alternatives by a noticeable margin. 
By contrast, hyper-coreness-$R_w$ and eigenvector-linear, which perform well under other dynamics, are less effective under HNG. 
Here we set the agreement rule as the unanimity rule; see results in more hypergraphs and results for the union rule in Supplementary Section~S5.
These results highlight the potential of the IPS method for broader applicability across multiple classes of dynamical processes.

\section{Discussion}\label{sec3}
This work proposes the IPS centrality measure, a general framework to evaluate node influence in hypergraphs by the initial propagation information.
We conduct extensive experiments to evaluate IPS on over 20 real-world hypergraphs and multiple propagation dynamics (including two fundamental models and a variant), and validate that IPS can effectively predict node influence across different dynamics (i.e., cross-model transferability). 
We show that IPS exhibits strong robustness in different parameter regions, and theoretically addresses the different dependence of node influence rankings on dynamical parameters.
Beyond propagation processes, IPS effectively assists minority norm dominance in consensus progress (the higher-order naming game model), which further shows its cross-paradigm transferability and broadens its applicability. We also evaluate IPS for the downstream task, epidemic containment~\cite{imm1,imm2}, and the results show that IPS outperforms other centrality measures overall (see Supplementary Section~S6). 
Moreover, IPS requires only local hypergraph information and is computationally lightweight, making it scalable to large-scale hypergraphs. 
Overall, IPS integrates the structural and dynamical characteristics of high-order interactions in a natural way, and mitigates the robustness and transferability issues arising from the nonlinearity and complexity of high-order dynamical systems.

By virtue of a clear physical meaning that is directly connected to the spreading context, IPS attains strong interpretability and constitutes the basis for its reliability in practical applications and allows the incorporation of other real-world mechanisms (e.g., activity levels and competition) conveniently~\cite{ac2,njp}.  
Furthermore, the consistent effectiveness of IPS also suggests that the integration of local topology with early-time dynamics may constitute an essential determinant of node influence.
This finding not only advances research on node influence prediction but also, to some extent, alleviates its reliance on low-reliability heuristics.

Nowadays, emerging machine learning approaches provide a complementary pathway for obtaining dynamical descriptions of real systems from data, such as the discovery of explicit, interpretable equations (e.g., neural symbolic regression and scalable symbolic enumeration)~\cite{real1, ml-f-sym, ml-learning} as well as the inference of higher-order interaction structures from time series (e.g., higher-order Granger–based frameworks)~\cite{ml-f-high}. These advances naturally integrate with IPS, enabling a physics-informed data-driven framework for identifying influential nodes in complex systems~\cite{ml-discovering}.
Future research can also apply IPS to more downstream tasks~\cite{app}, including political propaganda~\cite{Twitterinfluencers, political2025}, rumor suppression~\cite{rumor}, and cybersecurity~\cite{sc}, among others.

\section*{Methods}\label{sec4}
\subsection*{Propagation dynamics and simulation details}
\subsubsection*{Higher-order contagion model with power-law infection kernel and its size attenuation variant}
In the higher-order model with power-law infection kernel (HCP model), within a hyperedge that contains $i$ infected nodes, each infected node attempts to spread at a rate of $r = \lambda i^\nu$. In the size-attenuation variant (HCSA), the infection rate is $r=\frac{\lambda}{|h|}i^\nu$, where $|h|$ is the size of hyperedge $h$.
In both models, we control the dynamical parameters $\nu \geq 1$ and $\lambda >0$, where $\nu$ characterizes the strength of higher-order reinforcement. In practice, $\nu$ captures effects such as cumulative exposure in biological processes, social reinforcement, etc. When $\nu=1$, there is no higher-order reinforcement, and the model reduces to the conventional SIR model on hypergraphs.

We convert the infection rate into a per-contact infection probability.
Assuming infection events are independent and follow a Poisson process, the number of infection events $X$ occurring within one time unit in a hyperedge follows $P(X=k)=\frac{r^k e^{-r}}{k!}$, where $r$ is the propagation rate.
The probability that a susceptible node becomes infected through a hyperedge $h$ during one time step is therefore $1-P(X=0)=1-e^{-r}$. If there are $i$ infected nodes in $h$, the infection probability of a susceptible node in $h$ at one time step is $1-e^{-\lambda i^{\nu}}$ for HCP and $1-e^{-\frac{\lambda}{|h|}i^\nu}$ for HCSA.
The recovery probability $\mu$ is taken in the interval $[0,1]$.

\subsubsection*{Higher-order threshold contagion model}
The higher-order threshold contagion (HTC) model describes social contagions driven by group influence, such as public opinion shifts or collective behavior adoption.
At each time step, select one hyperedge $h$ randomly. If the number of infected nodes within it ($i_h$) satisfies $i_h \geq \lceil {\theta}|h| \rceil$, all susceptible nodes in $h$ become infected with probability $\eta$. 
Infected nodes recover independently with probability $\mu \in [0,1]$. 

\subsubsection*{Simulation details} 
We use discrete-time simulations to estimate the propagation range for given dynamical parameters and an initial infection seed. Specifically: 
Set the seed node to the $I$ state and all remaining nodes in the hypergraph to the $S$ state; In the following times, $S$-state nodes become infected (convert to the $I$ state) according to the transition rules of the corresponding model (HCP, HCSA, or HTC); Each $I$-state node recovers independently with probability $\mu$; All state updates are performed synchronously, i.e., nodes that become infected (or recovered) at time $t$ enter the $I$ (or $R$) state at time $t+1$.
When there is no $I$-state node in the hypergraph, the propagation terminates. We record the number of nodes in the $R$ state as the propagation range.
The ground-truth influence of a node is defined as the average propagation range obtained when that node is used as the initial seed, averaged over multiple Monte Carlo realizations.

To improve simulation efficiency, we randomly sample (without replacement) a subset of nodes and evaluate their influence instead of testing all nodes: we sample $10\%$ of nodes in Fig.~\ref{fig2}j (threads-math-sx) and $50\%$ of nodes in Fig.~\ref{fig4} and Fig.~\ref{fig5}. 
To enhance randomness and ensure reliable results, we re-sample a new set of candidate initial nodes independently for each group of dynamical parameters. Note that simulations are always performed on the full hypergraph structure, and Fig.~\ref{fig2}a--i reports results for all nodes in each hypergraph (no partial sampling).
For all results under the HCP, HCSA, and HTC models, we conduct 1000 simulations per seed when $\mu=1$ or $0.5$, and 300 simulations per seed when $\mu=0.1$.
The specific dynamical parameters for each hypergraph are provided in the Supplementary Information.

\subsection*{Higher-order naming game model and its simulations}
In the higher-order naming game (HNG) model, there are two names (denoted as $A$ and $B$), and each agent (i.e., the hypergraph node) can hold one or both names in its vocabulary. 
For each pair of $\beta$ and $p$, we initialize the system as follows. First, we rank nodes according to a given measure and select the top $\lfloor{pN}\rfloor$ ranked nodes as the committed minority. These nodes have a fixed vocabulary $\{A\}$ and never change it. 
All remaining nodes start with vocabulary $\{B\}$.
The subsequent name evolution proceeds in discrete time steps: 
At each time step, randomly select a hyperedge, and then randomly select a node in this hyperedge as the speaker. 
The speaker samples a name uniformly at random from its vocabulary and broadcasts this name to all other nodes in the hyperedge (hearers).
The hearers in the hyperedge jointly determine whether an agreement is reached.
Under the unanimity rule, agreement is achieved if all hearers already have the broadcast name in their vocabularies.
Also, we can adopt a more relaxed rule, called the union rule, under which agreement is achieved if at least one hearer in the group already holds the broadcast name.
If agreement is achieved, then with probability $\beta \in [0,1]$ all non-committed nodes in the hyperedge erase all other names and keep only the broadcast name. Committed nodes always keep $\{A\}$.
If agreement is not achieved, then every non-committed hearer that does not already have the broadcast name adds it to its vocabulary.
Then, the system goes to the next time step.
The process continues until it reaches an absorbing state with $\rho _{A}=1$ (all non-committed holds the name $A$ only) or until a maximum time step $t_{\max}=1\times 10^6$ is reached, by which time the abundance of names has converged to a stationary regime.
In the latter case, the stationary value of $\rho_{A}$ is obtained by averaging over 100 samples taken in the last $1\times 10^5$ time steps. 
For each pair of $\beta$ and $p$ in the main text, the reported $\rho_{A}$ is the median value of 100 simulations (50 simulations for the congress-bills hypergraph).

\subsection*{Calculation of propagation threshold}
To numerically estimate the critical threshold in our SIR-type propagation models, we adopt the susceptibility-based method commonly used to detect phase transitions in finite systems~\cite{Multistability,sus3,sus4}.
The susceptibility $\chi$ is defined as $\chi = \frac{\langle R_{\infty}^2\rangle - \langle R_{\infty}\rangle ^2}{\langle R_{\infty}\rangle}$, where $R_{\infty}$ is the final propagation range in one simulation (i.e., the number of recovered nodes at the end of each simulation ).
This quantity measures the relative magnitude of fluctuations in outbreak size and typically peaks near the transition point, thus providing a reliable numerical estimate of the propagation threshold when an analytical solution is not available.
For the HCP and HCSA model, we fix $\nu$ and $\mu$, and scan the parameter $\lambda$ on a logarithmic grid.
For each $\lambda$, we perform 300 realizations when $\mu=0.1$ and 1,000 realizations when $\mu=1$, using randomly selected initial seeds.
We then take the value of $\lambda$ that maximizes $\chi(\lambda)$ as the threshold $\lambda_c$.
Similarly, for the HTC model, we fix $\theta$ and $\mu$ and scan the parameter $\eta$. 
The critical threshold $\eta_c$ is determined as the value of $\eta$ that maximizes $\chi(\eta)$ following the same procedure.

\subsection*{Datasets}
In this work, we adopt several widely used higher-order networks spanning diverse domains, comprising 20 medium-sized hypergraphs and one large-scale hypergraph (Table~\ref{tab}).
The medium-sized hypergraphs contain topics of email communications, political affiliations, online interactions, ecological datasets, and face-to-face interactions. We follow the previous work~\cite{hypercore} to preprocess the medium-sized hypergraphs and follow the previous work~\cite{datas} for the large-scale hypergraph threads-math-sx.

Email communication datasets contain email-EU and email-Enron~\cite{datas,email-eu,email-enron},  nodes are the email addresses within a European research and Enron corporation, respectively, while each hyperedge includes the senders and receivers of an email.

Political affiliation datasets contain congress-bills, senate-bills, house-committees, and senate-committees~\cite{house,bills1,datas,housesenata}. Nodes in these hypergraphs are members of the U.S. Congress or the U.S. Senate, and each hyperedge includes the members in a committee or cosponsoring a bill.

Online interaction datasets contain music-review, geometry-questions, and algebra-questions~\cite{music, questions}. Nodes in music-review are Amazon users, and each hyperedge involves nodes who review a specific product belonging to the category of blues music.
Nodes in geometry-questions and algebra-questions are users in MathOverflow, and each hyperedge contains users who have answered the same question on the topic of algebra or geometry, respectively.

Ecological systems datasets contain M\_PL\_062\_ins, M\_PL\_015\_ins, M\_PL\_062\_pl, and M\_PL\_015\_pl~\cite{flowers62,flowers15,flowerweb}.
Nodes in M\_PL\_062\_ins and M\_PL\_015\_ins are insects, and each hyperedge contains insects that pollinate the same plant.
By contrast, nodes in M\_PL\_062\_pl and M\_PL\_015\_pl are plants, and each hyperedge connects nodes pollinated by the same insect species.

Face-to-face interaction datasets are aggregated from time-resolved face-to-face human interaction databases, containing middle-school (Mid1~\cite{Elem1_Mid1}), primary-schools (LyonSchool~\cite{Lyon,datas}, Elem1~\cite{Elem1_Mid1}), high-school (Thiers~\cite{Thiers13}), conference (SFHH~\cite{InVS15_2_SFHH}), workplace (InVS15~\cite{InVS15_1,InVS15_2_SFHH}), and hospital (LH10~\cite{LH10}). Nodes in these datasets are humans. The hyperedges are the maximal cliques of the time-resolved networks, which are aggregated over a time window of 15 minutes.

In the large-scale dataset threads-math-sx~\cite{datas}, nodes represent the users of a forum, and each hyperedge denotes a set of users answering a question.
\begin{table}[h]
\vspace{-2em}
\centering 
\caption{Datasets} 
\setlength{\tabcolsep}{1pt} 
\renewcommand{\arraystretch}{1.2} 
\label{tab:my_table} 
\begin{tabular}{l|llllll}
\toprule 
  Dataset & $N$ & $M$ &  $m_{\max}$ & $\langle m\rangle$ & $k_{\max}$ & $\langle k\rangle$ \\ 
\midrule 
    congress-bills &1718 & 83105 & 25 & 8.81 & 3964 & 48.37  \\ 
    house-committees &1290 & 335 & 81 & 35.25  & 44 & 0.26  \\ 
    music-review  &1104 & 685 & 83 & 15.30  & 127 & 0.62   \\
    M\_PL\_062\_ins  &1044 & 456 & 58 & 33.45  & 157 & 0.44   \\
    email-EU  &979 & 24399 & 25 & 3.49  & 910 & 24.92   \\
    M\_PL\_015\_ins  &663 & 126 & 124 & 23.08  & 103 & 0.19    \\
    Mid1  &591 & 61521 & 13 & 3.94  & 1208 & 104.10    \\
    geometry-questions  &580 & 888 & 230 & 13.00  & 227 & 1.53   \\
    M\_PL\_062\_pl &456 & 866 & 157 & 17.41  & 57 & 1.90    \\
    algebra-questions  &420 & 979 & 107 & 7.57  & 328 & 2.33  \\
    SFHH  &403 & 6398 & 10 & 2.72  & 222 & 15.88   \\
    Elem1 &339 & 20940 & 16 & 4.73  & 1052 & 61.77    \\
    Thiers13 &327 & 4795 & 7 & 3.09  & 131 & 14.66   \\
    senate-bills  &294 & 21721 & 99 & 9.90  & 3222 & 73.88    \\
    senate-committees  &282 & 301 & 31 & 17.57  & 61 & 1.07    \\
    LyonSchool  &242 & 10848 & 10 & 4.05  & 438 & 44.83   \\
    InVS15  &217 & 3279 & 10 & 2.77  & 141 & 15.11   \\
    email-Enron  &143 & 1459 & 37 & 3.13  & 117 & 10.20  \\
    M\_PL\_015\_pl &130 & 401 & 104 & 6.60  & 101 & 3.08   \\
    LH10  &76 & 1102 & 7 & 3.45  & 205 & 14.50 \\
    threads-math-sx & 152702 & 534768 & 21 & 2.61 & 11358 & 3.50  \\
\bottomrule
\end{tabular}
\label{tab}
\end{table}

\newpage

\backmatter

\bmhead{Supplementary information}
Supplementary information accompanies this paper in a separate file.

\bmhead{Acknowledgements}
This work is supported by National Science and Technology Major Project (2022ZD0116800), Program of National Natural Science Foundation of China (12425114, 12201026, 12501702, 12501718, 62441617), the Fundamental Research Funds for the Central Universities, and Beijing Natural Science Foundation (Z230001).

\section*{Declarations}


\bmhead{Conflict of interest}
The authors declare no competing interests.




\bmhead{Data availability}
The study uses publicly available datasets. All raw data are accessible from the original sources cited in the Methods. The processed datasets used in this work are obtained from previously published studies, as described in the Methods section.


\bmhead{Code availability}
All code used in this study is publicly available at: \url{https://github.com/Haoyajing/IPS-codes}.

\bmhead{Author contribution}
Y.H., S.T., and Z.Z. conceived the study. Y.H., L.L., and X.W. designed the methodology and implemented the algorithm. Y.H., Z.H., and M.W. performed the experiments and analyzed the data. S.T. and Z.Z. supervised the project. All authors contributed to discussions and to the writing and revision of the manuscript.

\bibliography{sn-bib}

\end{document}


\renewcommand{\thesection}{S\arabic{section}}
\renewcommand{\thesubsection}{S\arabic{section}.\arabic{subsection}}
\renewcommand{\theequation}{S\arabic{equation}}
\renewcommand{\thefigure}{S\arabic{figure}}
\renewcommand{\thetable}{S\arabic{table}}

\setcounter{section}{0}
\setcounter{equation}{0}
\setcounter{figure}{0}
\setcounter{table}{0}

\title[Article Title]{Supplementary Material for "A unified framework for identifying influential nodes in hypergraphs"}



\author[1,4]{\fnm{Yajing} \sur{Hao}}

\author*[2,4,5,6,7]{\fnm{Longzhao} \sur{Liu}}\email{longzhao@buaa.edu.cn}

\author*[2,4,5,6,7]{\fnm{Xin} \sur{Wang}}\email{wangxin\_1993@buaa.edu.cn}

\author[2,4]{\fnm{Zhihao} \sur{Han}}

\author[1,4]{\fnm{Ming} \sur{Wei}}

\author[2,3,4,5,6,7,8,9]{\fnm{Zhiming} \sur{Zheng}}

\author*[2,3,4,5,6,7,8,9]{\fnm{Shaoting} \sur{Tang}}\email{tangshaoting@buaa.edu.cn}

\affil[1]{\orgdiv{School of Mathematical Sciences}, \orgname{Beihang University}, \orgaddress{\city{Beijing}, \postcode{100191}, \country{China}}}

\affil[2]{\orgdiv{School of Artificial Intelligence}, \orgname{Beihang University}, \orgaddress{\city{Beijing}, \postcode{100191}, \country{China}}}

\affil[3]{\orgdiv{Hangzhou International Innovation Institute}, \orgname{Beihang University}, \orgaddress{\city{Hangzhou}, \postcode{311115}, \country{China}}}

\affil[4]{\orgdiv{Key laboratory of Mathematics, Informatics and Behavioral Semantics}, \orgname{Beihang University}, \orgaddress{\city{Beijing}, \postcode{100191}, \country{China}}}

\affil[5]{\orgdiv{Beijing Advanced Innovation Center for Future Blockchain and Privacy Computing}, \orgname{Beihang University}, \orgaddress{\city{Beijing}, \postcode{100191}, \country{China}}}

\affil[6]{\orgdiv{Zhongguancun Laboratory}, \orgaddress{\city{Beijing}, \postcode{100094}, \country{China}}}

\affil[7]{\orgdiv{State Key Laboratory of Complex \& Critical Software Environment}, \orgname{Beihang University}, \orgaddress{\city{Beijing}, \postcode{100191}, \country{China}}}

\affil[8]{\orgdiv{Institute of Trustworthy Artificial Intelligence}, \orgname{Zhejiang Normal University}, \orgaddress{\city{Hangzhou}, \postcode{310012}, \country{China}}}

\affil[9]{\orgdiv{Beijing Academy of Blockchain and Edge Computing}, \orgaddress{\city{Beijing}, \postcode{100085}, \country{China}}}





\maketitle
This Supplementary Information provides detailed derivations, algorithmic analyses, and extended experimental results supporting the main text “A unified framework for identifying influential nodes in hypergraphs”.
It includes explicit formulations and calculation examples of the proposed IPS measure, complexity analyses, and validations on real hypergraphs.
Comprehensive parameter settings and full simulation results are presented for three higher-order propagation models (HCP, HCSA, and HTC) and an additional higher-order naming game model, verifying the robustness and transferability of the proposed framework.
Furthermore, we report the effectiveness of IPS in epidemic containment tasks, demonstrating its consistent advantages.

\newpage
\tableofcontents

\newpage
\section{Method and time complexity}\label{S1}
\subsection{Calculation for $IPS_2^{HCP}$}\label{S1.1}
When considering second-order information ($t=2$), we have:
\begin{align}
    \mathbb{E}[R_2+I_2\mid s]=& \mathbb{E}[R_1+I_1\mid s] +  \mathbb{E}[I_{2} \mid s].
    \label{eq_t2}
\end{align}
Here $\mathbb{E}[R_1+I_1\mid s]$ is the expected number of infected and recovered nodes at time $t=1$, and $\mathbb{E}[I_{2} \mid s]$ denotes the expected number of newly infected nodes at time $t=2$.
Let $p_{t,n}(i|s)$ represent the probability that the $n$-th order neighbor $i$ of the seed $s$ is infected at time $t$. Then,  $\mathbb{E}[R_1+I_1\mid s]$ and $\mathbb{E}[I_{2} \mid s]$ can be derived by 
\begin{align}
    \mathbb{E}[R_1+I_1\mid s]=& 1+\sum\limits_{i\in N_v^1(s)}p_{1,1}(i|s)   \nonumber \\
    \mathbb{E}[I_{2} \mid s]=& \sum\limits_{i\in N_v^1(s)}p_{2,1}(i|s)(1-p_{1,1}(i|s))   
    + \sum\limits_{i\in N_v^2(s)}p_{2,2}(i|s),  
\label{eq_t2_e}
\end{align} 
where the first item in $\mathbb{E}[I_{2} \mid s]$ is the expected number of first-order neighbors that are infected at time $t = 2$, and the remaining item records the expected number of second-order infected neighbors.
To compute $p_{t,n}(i|s)$, we first let $N_h^1(s)$ and $N_h^2(s)=\bigcup _{i\in N_v^1(s)} N_h^1(i) \setminus N_h^1(s)$ represent the set of first-order adjacent hyperedges and second-order hyperedges, respectively. Then, the expected number of infected nodes within hyperedge $h$ can be approximated as follows: If $i\in N_v^1(s)$, $I_1(i,h) = \sum_{j\in h\cap  N_v^1(s),j\neq i}p_{1,1}(j|s)$; Otherwise, $I_2(h) = \sum_{j\in h \cap  N_v^1(s)}p_{1,1}(j|s)$. With these notations, we can obtain the expression of $p_{t,n}(i|s)$, i.e.,
\begin{align}
&p_{1,1}(i|s)=1-e^{-k_s(i)\lambda}, \nonumber\\
&p_{2,1}(i|s)=
    \begin{aligned}[t]
    &1-
    \begin{aligned}[t]
        &\!\!\prod_{h\in  N_h^1(i)\cap N_h^1(s)\atop {I_1(i,h)}\geq1}\!\!\!\!f_1(i,h)
        \!\!\prod_{h\in  N_h^1(i)\cap N_h^1(s)\atop {I_1(i,h)}<1}\!\!\!\!f_2(i,h)  
        &\!\!\prod_{h\in N_h^1(i)\cap N_h^2(s)\atop {I_1(i,h)}\geq1}\!\!\!\!f_3(i,h) 
        \!\!\prod_{h\in N_h^1(i)\cap N_h^2(s)\atop {I_1(i,h)}<1}\!\!\!\!f_4(i,h),  \nonumber\\
    \end{aligned} \nonumber\\
    \end{aligned} \nonumber\\
&p_{2,2}(i|s)=
    \begin{aligned}[t]
    &1-\!\prod_{h\in N_h^1(i)\cap N_h^2(s) \atop {I_2(h)}\geq1}\!\!\!f_5(h)
    \!\prod_{h\in N_h^1(i)\cap N_h^2(s) \atop {I_2(h)}<1}\!\!\! f_6(h), 
    \end{aligned} 
\label{eq_t2_p}
\end{align} 
with:
\begin{align}
&f_1(i,h)=\mu e^{-\lambda {{I_1(i,h)}}^\nu} +(1-\mu)e^{-\lambda ({I_1(i,h)}+1)^\nu},  \nonumber\\
&f_2(i,h)
    \begin{aligned}[t]
        &=\mu (1-{I_1(i,h)}(1-e^{-\lambda}))    \nonumber\\
        &+(1-\mu)e^{-\lambda ({I_1(i,h)}+1)^\nu},  \nonumber\\
    \end{aligned} \nonumber\\
&f_3(i,h) = e^{-\lambda {I_1(i,h)}^\nu}, \nonumber\\
&f_4(i,h) = 1-{I_1(i,h)}(1-e^{-\lambda}) , \nonumber\\
&f_5(h) = e^{-\lambda {I_2(h)}^\nu},  \nonumber\\
&f_6(h) = 1- {I_2(h)}(1-e^{-\lambda}). 
\label{eq_t2_f}
\end{align}
Here $f_1(i,h)$ is the probability that a first-order neighbor does not get infected through a first-order adjacent hyperedge $h$ of $s$. The two terms in $f_1(i,h)$ separately deal with situations where the seed is recovered at $t=2$ or not. If the seed has recovered (with probability $\mu$), the infection probability of $i$ is given as $1-(1-e^{-\lambda I_1(i,h)^\nu})=e^{-\lambda I_1(i,h)^\nu}$; otherwise, if the seed is still in the $I$ state (with probability $1-\mu$), the infected nodes should be modified to $I_1(i,h)+1$, and the infection probability of $i$ is given as $1-(1-e^{-\lambda (I_1(i,h)+1)^\nu})= e^{-\lambda (I_1(i,h)+1)^\nu}$.
For the case of $I_1(i,h)<1$, the power calculation may be problematic. In this case, $I_1(i,h)$ can be regarded as the probability that there is one infected node in $h$. Thus, the probability of $i$ not being infected through the first-order adjacent hyperedge, denoted by $f_2(i,h)$, can be approximated by $\mu (1-{I_1(i,h)}(1-e^{-\lambda}))+(1-\mu)e^{-\lambda ({I_1(i,h)}+1)^\nu}$.
Similarly, functions $f_3(i,h)$ and $f_4(i,h)$ denote the probabilities that a first-order neighbor does not get infected through second-order adjacent hyperedges.
Finally, the multiplication of these probabilities gives the probability of $i$ not getting infected at $t=2$ (i.e., $1-p_{2,1}(i|s)$). 
Similarly, $f_5(h)$ and $f_6(h)$ calculate the not-infected probability of the second-order neighbors. 
Substituting Eq.~\eqref{eq_t2_e}--\eqref{eq_t2_f} into Eq.~\eqref{eq_t2}, we finally obtain $IPS_2^{HCP}=\mathbb{E}[R_2+I_2\mid s]$. 

\begin{figure}[H]
    \centering
    \includegraphics[scale=0.85]{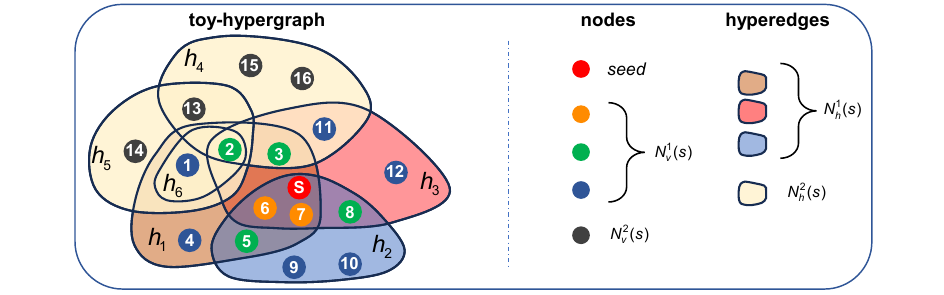}
    \captionsetup{justification=raggedright, singlelinecheck=false, width=\textwidth}
    \caption{\textbf{A toy-hypergraph for instantiation calculation of $IPS_{2}^{HCP}$.} There are 17 nodes and 6 hyperedges in this hypergraph, where $h_1=\{s,1,2,3,4,5,6,7\}$, $h_2=\{s,5,6,7,8,9,10\}$, $h_3=\{s,2,3,6,7,8,11,12\}$, $h_4=\{2,3,11,13,15,16\}$, $h_5=\{1,2,13,14\}$, $h_6=\{1,2\}$. The red node $s$ is assigned as the seed, and we have $N_v^1(s)=\{1,2,3,4,5,6,7,8,9,10,11,12\}$, $N_v^2(s)=\{13,14,15,16\}$, $N_h^1(s)=\{h_1,h_2,h_3\}$, and $N_h^2(s)=\{h_4,h_5,h_6\}$. }
    \label{toynet}
\end{figure}

Figure~\ref{toynet} presents the schematic diagram of $IPS_2^{HCP}$'s calculation, which has 17 nodes and 6 hyperedges.  We choose $s$ (red node) as the seed. Denote $N_v^1(s)$ the set of $s$' neighbors, and $N_v^2(s)$ the set of its second-order neighbors. Then, we have:
\begin{align}
     IPS_{2}^{HCP}(s)&=\mathbb{E}[R_{2}+I_{2}\mid s] 
     = \mathbb{E}[R_1+I_1\mid s] +  \mathbb{E}[I_{2} \mid s]   \nonumber \\
     &=1 + \sum\limits_{i\in N_v^1(s)}p_{1,1}(i|s) + \sum\limits_{i\in N_v^1(s)}p_{2,1}(i|s)(1-p_{1,1}(i|s))  + \sum\limits_{i\in N_v^2(s)}p_{2,2}(i|s),  
\label{maineq}
\end{align}
where $p_{t,n}(i|s)$ is the infection probability of the $n$-th order neighbor $i$ of the seed $s$ at time step $t$. For $t=1$, we have $p_{1,1}(i|s)=1-e^{-\lambda k_s(i) }$ for each node in $N_v^1(s)$. Take node 1 in the toy hypergraph as an example to illustrate the calculation of $p_{2,1}(i|s)$. According to Fig.~\ref{toynet}, $p_{2,1}(1|s)$ can be written as:
\begin{align}
    p_{2,1}(1|s)&=1-
        \!\!\prod_{h\in  N_h^1(1)\cap N_h^1(s)\atop {I_1(1,h)}\geq1}\!\!\!\!f_1(1,h)
        \!\!\prod_{h\in  N_h^1(1)\cap N_h^1(s)\atop {I_1(1,h)}<1}\!\!\!\!f_2(1,h)   
        \!\!\prod_{h\in N_h^1(1)\cap N_h^2(s)\atop {I_1(1,h)}\geq1}\!\!\!\!f_3(1,h) 
        \!\!\prod_{h\in N_h^1(1)\cap N_h^2(s)\atop {I_1(1,h)}<1}\!\!\!\!f_4(1,h)  \nonumber \\
        &=1- 
        \prod_{h\in  \{h_1\} \atop {I_1(1,h)}\geq1}f_1(1,h)
         \prod_{h\in  \{h_1\} \atop {I_1(1,h)}<1}f_2(1,h)
         \prod_{h\in  \{h_5,h_6\} \atop {I_1(1,h)}\geq1}f_3(1,h)
         \prod_{h\in  \{h_5,h_6\} \atop {I_1(1,h)}<1}f_4(1,h)\nonumber \\
        &=1-
        \begin{aligned}[t]
            &(\mathds{1}_{\{I_1(1,h_1)\geq 1\}}f_1(1,h_1)+\mathds{1}_{\{I_1(1,h_1)< 1\}}f_2(1,h_1))    \\
            &\cdot(\mathds{1}_{\{I_1(1,h_5)\geq 1\}}f_3(1,h_5)+\mathds{1}_{\{I_1(1,h_5)< 1\}}f_4(1,h_5))   \\
            &\cdot(\mathds{1}_{\{I_1(1,h_6)\geq 1\}}f_3(1,h_6)+\mathds{1}_{\{I_1(1,h_6)< 1\}}f_4(1,h_6)),  
        \end{aligned}  
    \label{node1}
\end{align}
where $f_j(1,h)$ ($j=1,2,3,4$) calculates the probabilities of node 1 not being infected through a hyperedge $h$.
Denote the number of infected nodes in $h$ (except for node 1 itself) as $I_1(1,h)$. Then, if $I_1(1,h_1)\geq1$, the probability of node 1 not being infected through $h_1$, i.e., $f_1(1, h_1)$, can be written as: 
\begin{align}
    f_1(1,h_1)&=\mu [1-(1-e^{-\lambda {{I_1(1,h_1)}}^\nu})] +(1-\mu)[1-(1-e^{-\lambda ({I_1(1,h_1)}+1)^\nu})] \nonumber  \\ 
    &=\mu e^{-\lambda {{I_1(1,h_1)}}^\nu} +(1-\mu)e^{-\lambda ({I_1(1,h_1)}+1)^\nu},
\end{align}
where $I_1(1,h_1)=\sum_{i\in h_1\cap N_v^1(s),i\neq1}p_{1,1}(i)=\sum_{i\in h_1}p_{1,1}(i) - p_{1,1}(1)$.
Otherwise, the probability of not being infected is:
\begin{align}
    f_2(1,h_1)=\mu (1-I_1(1,h_1)(1-e^{-\lambda})) +(1-\mu)e^{-\lambda ({I_1(1,h_1)}+1)^\nu}.
\end{align}
For the adjacent hyperedge $h_5$, which does not contain the seed node $s$ directly (i.e., $h_5 \in N_h^2(s)$), we do not need to consider the state of the seed. If $I_1(1,h_5) \geq 1$, the probability of not being infected is:
\begin{align}
    f_3(1,h_5)=e^{-\lambda {I_1(1,h_5)}^\nu};
\end{align}
if $I_1(1,h_5) < 1$, the probability can be approximated by: 
\begin{align}
    f_4(1,h_5)=1-{I_1(1,h_5)}(1-e^{-\lambda}),
\end{align}
where $I_1(1,h_5)=\sum_{i\in h_5\cap N_v^1(s),i\neq1}p_{1,1}(i)=p_{1,1}(2)$.
Similarly, for $h_6$: if $I_1(1,h_6) \geq 1$, the probability is 
if $I_1(1,h_6) < 1$, the probability is 
The infection probability of the remaining nodes in $N_v^1(s)$ can be calculated in a similar procedure.

Then, we take node 13 as an example to show the calculations of $p_{2,2}(i|s)$. According to Fig.~\ref{toynet}, we have:
\begin{align}
    p_{2,2}(13|s)
    &=1-\!\prod_{h\in N_h^1(13)\cap N_h^2(s) \atop {I_2(h)}\geq1}\!\!\!f_5(h)
    \!\prod_{h\in N_h^1(13)\cap N_h^2(s) \atop {I_2(h)}<1}\!\!\! f_6(h)  \nonumber  \\
    &=1-\prod_{h\in \{h_4,h_5\} \atop {I_2(h)}\geq1}f_5(h)
        \prod_{h\in \{h_4,h_5\} \atop {I_2(h)}<1}f_6(h)    \nonumber  \\
    &=1- \begin{aligned}[t]
        &(\mathds{1}_{\{I_2(h_4)\geq 1\}}f_5(h_4)\!+\!\mathds{1}_{\{I_2(h_4)< 1\}}f_6(h_4))\!\cdot\!(\mathds{1}_{\{I_2(h_5)\geq 1\}}f_5(h_5)\!+\!\mathds{1}_{\{I_2(h_5)< 1\}}f_6(h_5)).
        \end{aligned}   
\end{align} 
Similarly, $f_5(h)$ and $f_6(h)$ calculate the probabilities of a node within it not being infected through this hyperedge $h$. Since second-order neighbors do not affect the expected infected nodes in their adjacent hyperedges, we have the number of infected nodes in $h$ given as $I_2(h)=\sum_{j\in h \cap  N_v^1(s)}p_{1,1}(j|s)$ (for $h \in N_h^{1}(i)$, $i \in N_v^2(s)$). Then, if $I_2(h_4) \geq 1$, the probability of not being infected is: 
\begin{align}
    f_5(h_4)=e^{-\lambda {I_2(1,h_4)}^\nu},
\end{align}
where $I_2(h_4)=p_{1,1}(2)+p_{1,1}(3)+p_{1,1}(11)$. If $I_2(h_4) < 1$, the probability is 
\begin{align}
    f_6(h_4)=1-{I_2(h_4)}(1-e^{-\lambda}).
\end{align}
Similarly, for $h_5$: if $I_2(h_5) \geq 1$, the probability is 
if $I_2(h_5) < 1$, the probability can be approximated by: 
where $I_2(h_5)=p_{1,1}(1)+p_{1,1}(2)$. The infection probability of the other node in $N_v^2(s)$ at $t=2$ can be calculated similarly. Finally, take all of the infection probabilities together into Eq.~\eqref{maineq}, and we get $IPS_{2}^{HCP}(s)$.

\subsection{Details of $IPS_{2r}^{HCP}$}\label{S1.2}
By transforming from node-level calculation to hyperedge-level, we can obtain a faster version of the algorithm, i.e., $IPS_{2r}^{HCP}$. 
Here, we use $I(h)=\lambda\sum_{i \in h\cap N_v^1(s)}k_s(i)$ to estimate the number of infected nodes in each $h$.
Then, each susceptible node $i$ gets infected with the following probability:
\begin{align}
    p(i)=1-\prod_{h \in N_h^1(i)}[1-(1-e^{-\lambda {I(h)}^{\nu}})]=1-\prod_{h \in N_h^1(i)}e^{-\lambda {I(h)}^{\nu}}=1-e^{-\sum_{h\in N_h^1(i)}\lambda {I(h)}^{\nu}}
    \label{eq_2_r_1}
\end{align}
Applying the first-order Taylor approximation to the above equation, we can obtain the approximate: $\hat{p}(i)\approx \sum_{h\in N_h^1(i)}\lambda {I(h)}^{\nu}$.
Thus, we get:
\begin{equation}
    \mathbb{E}[I_2\mid s] \approx \sum_{i\in N_{v}^1(s)\cup N_{v}^2(s)}\hat{p}(i) \approx \sum_{i\in N_{v}^1(s)\cup N_{v}^2(s)}\sum_{h\in N_h^1(i)}\lambda {I(h)}^{\nu} =\sum_{h\in N_h^1(i),i\in N_{v}^1(s)\cup N_{v}^2(s)}\lambda {I(h)}^{\nu}|h|
    \label{eq_2_r_2}
\end{equation}
Furthermore, we do some rectifications to the above approximation to reduce errors:
1)~The term $\lambda {I(h)}^{\nu}$ is the agent of the infection probability, thus we rectify it to $\lambda I(h)$ (i.e., $\lambda I(h)\cdot 1^{\nu}$) when $I(h)<1$ (similar to before) and set the upper bound 1. 
2)~Use $|h|-I(h)$ (or $|h|-I(h)-1$ for hyperedges in $N_h^{1}(s)$) to rectify the size of hyperedges, i.e., reconsider the infection has happened at $t=1$ to rectify the potential propagation range. 
3)~Separately calculate cases of $N_{h}^{1}(s)$ and $N_{h}^{2}(s)$ (i.e., seed status related or not). We have:
\begin{align}
&IPS_{2r}^{HCP}(s)
    \begin{aligned}[t]
    =& \mathbb{E}[R_{1}+I_{1}\mid s] + \mathbb{E}[I_{2}\mid s] \nonumber \\
    \approx &1+\sum_{i\in N_{_v}^1(s)}k_s(i)+\mu \sum_{h \in N_h^{1}(s)} g_1(h) \cdot  H(|h|-I(h)-1) \nonumber \\
    &+(1-\mu )\sum_{h \in N_h^{1}(s)} g_2(h) \cdot  H(|h|-I(h)-1)  \nonumber \\
    &+\sum_{h \in N_h^{2}(s)} g_1(h) \cdot  H(|h|-I(h)), \nonumber \\
    \end{aligned}
\nonumber \\
&g_1(h) = 
\left\{
\begin{aligned} 
    &\lambda I(h),          && \text{if } I(h)<1 \nonumber\\
    &\min(\lambda {I(h)}^\nu,1),          && \text{if }I(h)\geq1,  \nonumber\\
\end{aligned},
\right. \nonumber\\
&g_2(h)=\min(\lambda( I(h)+1)^\nu,1),  \nonumber\\
&H(x)=\max(x,0)
\label{eqt7}
\end{align}  
where $g_1(h)$ and $g_2(h)$ estimate the infection probability in each hyperedge, and the terms of the function $H$ estimate the number of susceptible nodes in each hyperedge. Here, we use a step function to control the estimation range to avoid unreasonable negative values.

\subsection{Time complexity analysis}\label{S1.3}
From Eq.~(2) and Eq.~(7)--(8) in the main text, the first-order IPS methods, such as $IPS_1^{HCP}$, $IPS_1^{HCSA}$ and $IPS_1^{HTC}$, only require information of the first-order adjacent hyperedges. 
Thus, the time complexity of $IPS_1^{HCP}$ , $IPS_1^{HCSA}$ and $IPS_1^{HTC}$ is given as: 
$\sum_{s\in\mathcal{V}}HD(s)=N\langle k\rangle$. 

According to Eq.~(3)--(6) and Eq.~\eqref{eq_2_r_1}--\eqref{eqt7}, calculating $IPS_2^{HCP}$ and $IPS_{2r}^{HCP}$ requires traversing both the first- and second-order structures. For a seed $s$, the number of its second-order hyperedges satisfies: 
\begin{align}
  |N_h^2(s)|\leq k_{\max}|N_v^1(s)|\leq k_{\max}HD(s)m_{\max}.
\end{align}
Summing over all candidate seeds yields:
\begin{align}
  \sum_{s \in \mathcal{V}}|N_h^2(s)| \leq N\langle k\rangle k_{\max}m_{\max} 
\end{align} 
That is, the time complexity of $IPS_{2r}^{HCP}$ is $\mathcal{O}(N\langle k\rangle k_{\max} m_{\max})$. 
Compared to $IPS_{2r}^{HCP}$, the calculation of $IPS_2^{HCP}$ additionally requires $\mathcal{O}(|h|)$ calculations in each
second-order adjacent hyperedge.
Accordingly, the time complexity of $IPS_2^{HCP}$ is $\mathcal{O}(N\langle k\rangle k_{\max}m_{\max}^2)$.

\subsection{Validation of IPS methods}\label{S1.4}
\begin{figure}[H]
    \centering
    \includegraphics[scale=1]{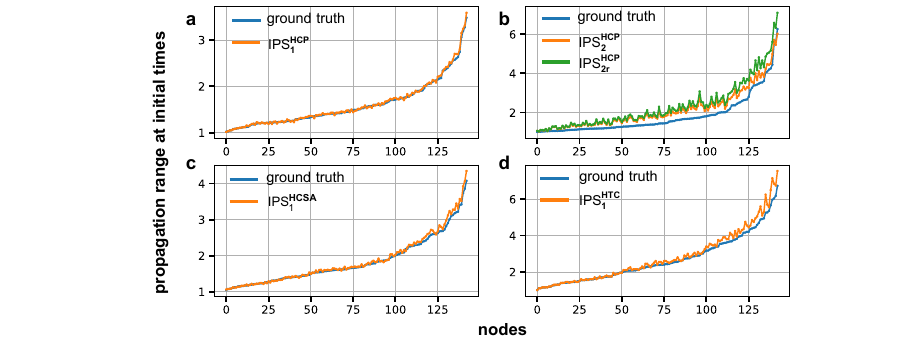}
    \captionsetup{justification=raggedright, singlelinecheck=false, width=\textwidth}
    \caption{\textbf{Validation of IPS methods in a real-world hypergraph.} 
    We present the ground truth propagation range of each node and theoretical values based on the IPS method under different dynamical conditions. \textbf{a} and \textbf{b}: Shown are cases of HCP with parameters $\nu=3$, $\lambda=0.005$ and $\mu=1$ when $t=1$ and $t=2$, respectively. \textbf{c}: We present the case under HCSA with parameters $\nu=3$, $\lambda=0.04$ and $\mu=1$. \textbf{d}: We present the case under HTC with parameters $\theta=0.25$, $\eta = 0.05$, and $\mu=1$. The ground truths are obtained by taking the average of 1000 simulations.}
    \label{vali}
\end{figure}

We use the real-world hypergraph, email-Enron, to validate the accuracy of the above calculations. 
Fig.~\ref{vali} compares the ground truth propagation range (the blue line) and the values derived from the IPS methods.
Results demonstrate that the theoretical values, including $IPS_1$ and $IPS_2$, predict the ground truth well under different dynamics, including HCP (Fig.~\ref{vali}a, b),  HCSA (Fig.~\ref{vali}c), and HTC (Fig.~\ref{vali}d).

\newpage
\section{Complementary results and parameter settings for Fig.~2-3}\label{S2}
This section shows parameter settings and complementary results of effectiveness evaluation under the higher-order contagion model with power-law infection kernel (HCP, corresponding to Sec.~2.2 and Sec.~2.3 in the main text). Subsection~\ref{S2.1} provides the parameter settings in Fig.2 and Fig.3 in the main text; Subsection~\ref{S2.2} and Sec.~\ref{S2.3} show the complementary results of Fig.2 and Fig.3 in the main text; Subsection~\ref{S2.4} provides more experiments conducted on the large-scale hypergraph.

\subsection{Parameter settings in Figs.~2--3}\label{S2.1}
Table~\ref{Pfig23} gives the parameter settings for Fig.~2 and Fig.~3 in the main text (other simulation results are shown in Fig.~\ref{jaccard}--\ref{fig3all}). 
Consistent with previous work~\cite{hypercore}, we use the same dynamical parameters to conduct  Monte Carlo simulations on hypergraphs numbered 1-20, and use the average propagation range of nodes to represent the ground truth of their influence. To accurately reflect the truth, propagation range of each node is averaged over 300 simulations.   
Due to the extremely large size of threads-math-sx (hypergraph 21), we randomly sample $10\%$ nodes to simulate their influence. In other datasets, all nodes are included.
\vspace{-1em}
\begin{table}[h]
\centering
\setlength{\tabcolsep}{6pt}
\renewcommand{\arraystretch}{1.6}
\begin{tabular}{@{}c@{\hspace{4em}}c@{}} 
\begin{tabular}[t]{|l|l|lll|}   
\hline
No. & Dataset & $\nu$ & $\lambda$ & $\mu$ \\
\hline
1   & congress-bills      & 1.5   & $5\times10^{-5}$ & 0.1 \\
\hline
2   & house-committees    & 4.0  & $5\times10^{-5}$  & 0.1  \\
\hline
3   & music-review        & 3.0  & $5\times10^{-4}$  & 0.1  \\
\hline
4   & M\_PL\_062\_ins     & 4.0  & $5\times10^{-5}$  & 0.1  \\
\hline
5   & email-EU            & 4.0   & $5\times10^{-5}$ & 0.1  \\
\hline
6   & M\_PL\_015\_ins     & 1.25  & $5\times10^{-4}$ & 0.1 \\
\hline
7   & Mid1                & 4.0   & $5\times10^{-5}$ & 0.1  \\
\hline
8   & geometry-questions  & 4.0   & $5\times10^{-4}$ & 0.1 \\
\hline
9   & M\_PL\_062\_pl      & 4.0   & $5\times10^{-5}$ & 0.1 \\
\hline
10   & algebra-questions   & 4.0   & $1\times10^{-3}$ & 0.1 \\
\hline
11   & SFHH                & 4.0   & $1\times10^{-2}$ & 0.1  \\
\hline
\end{tabular}
&
\begin{tabular}[t]{|l|l|lll|}
\hline
No. & Dataset & $\nu$ & $\lambda$ & $\mu$ \\
\hline
12   & Elem1              & 4.0   & $1\times10^{-4}$ & 0.1  \\
\hline
13   & Thiers13           & 4.0   & $1\times10^{-3}$ & 0.1   \\
\hline
14   & senate-bills       & 4.0   & $5\times10^{-5}$ & 0.1  \\
\hline
15   & senate-committees  & 4.0   & $1\times10^{-4}$ & 0.1  \\
\hline
16   & LyonSchool         & 4.0   & $1\times10^{-3}$ & 0.1  \\
\hline
17   & InVS15             & 4.0   & $1\times10^{-3}$ & 0.1  \\
\hline
18   & email-Enron        & 4.0   & $5\times10^{-4}$ & 0.1  \\
\hline
19   & M\_PL\_015\_pl     & 2.0   & $5\times10^{-4}$ & 0.1 \\
\hline
20   & LH10               &  1.5   & $1\times10^{-2}$ & 0.1   \\
\hline
21   & threads-math-sx    & 1.0  & $1\times10^{-1}$ & 1  \\
\hline
\end{tabular}
\end{tabular}
\captionsetup{justification=raggedright, singlelinecheck=false, width=\textwidth}
\caption{\textbf{Parameter settings in Fig.~2--3. }
Parameter values of $\nu$, $\lambda$, and $\mu$ are shown under different hypergraphs. 
}
\label{Pfig23}

\end{table}

\vspace{-3em}
\subsection{Complementary results of  Fig.~2}\label{S2.2}
\begin{figure}[H]
    \centering
    \includegraphics[]{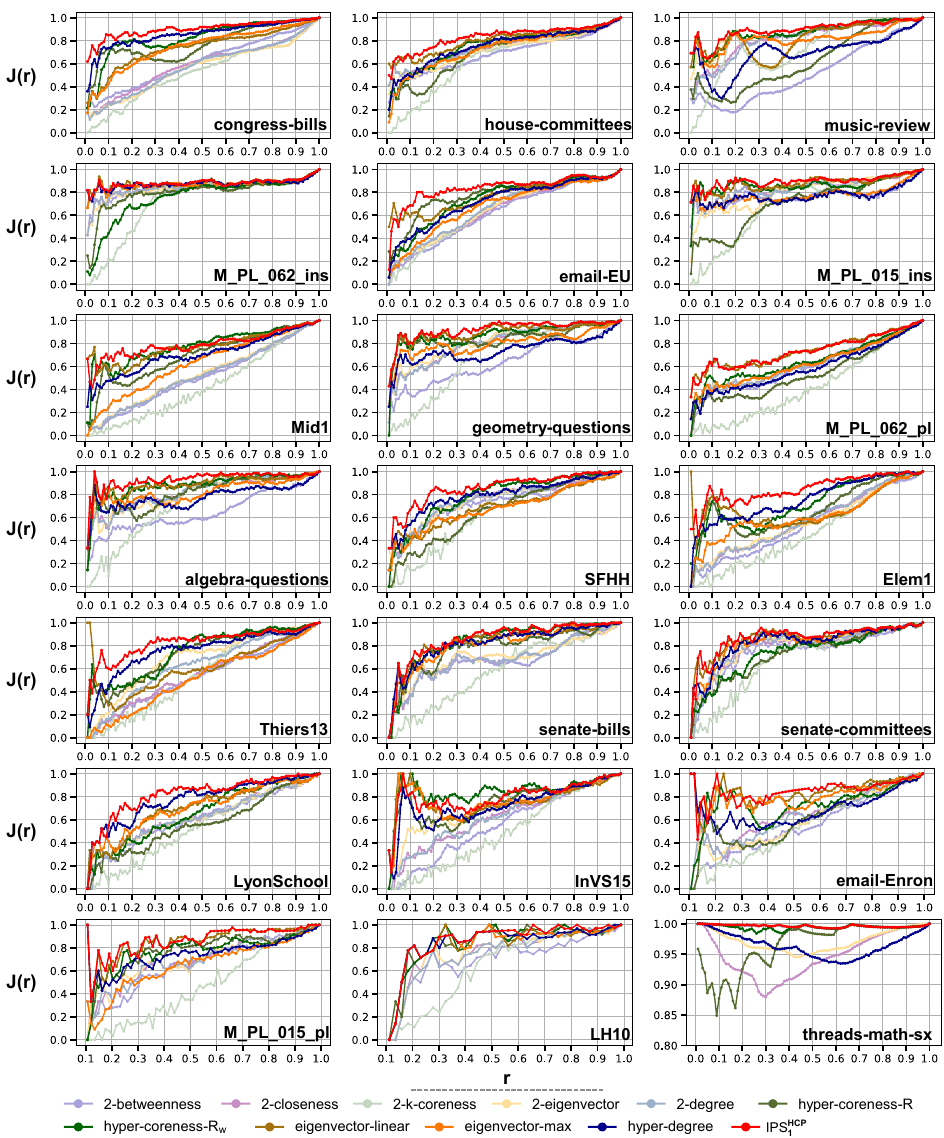}
    \captionsetup{justification=raggedright, singlelinecheck=false, width=\textwidth}
    \caption{\textbf{Results of the Jaccard coefficient.} Each subplot compares the Jaccard coefficient $J(r)$ among 10 classical methods and IPS method under 21 real hypergraphs.
    }
    \label{jaccard}
\end{figure}

\begin{figure}[H]
    \centering
    \includegraphics[]{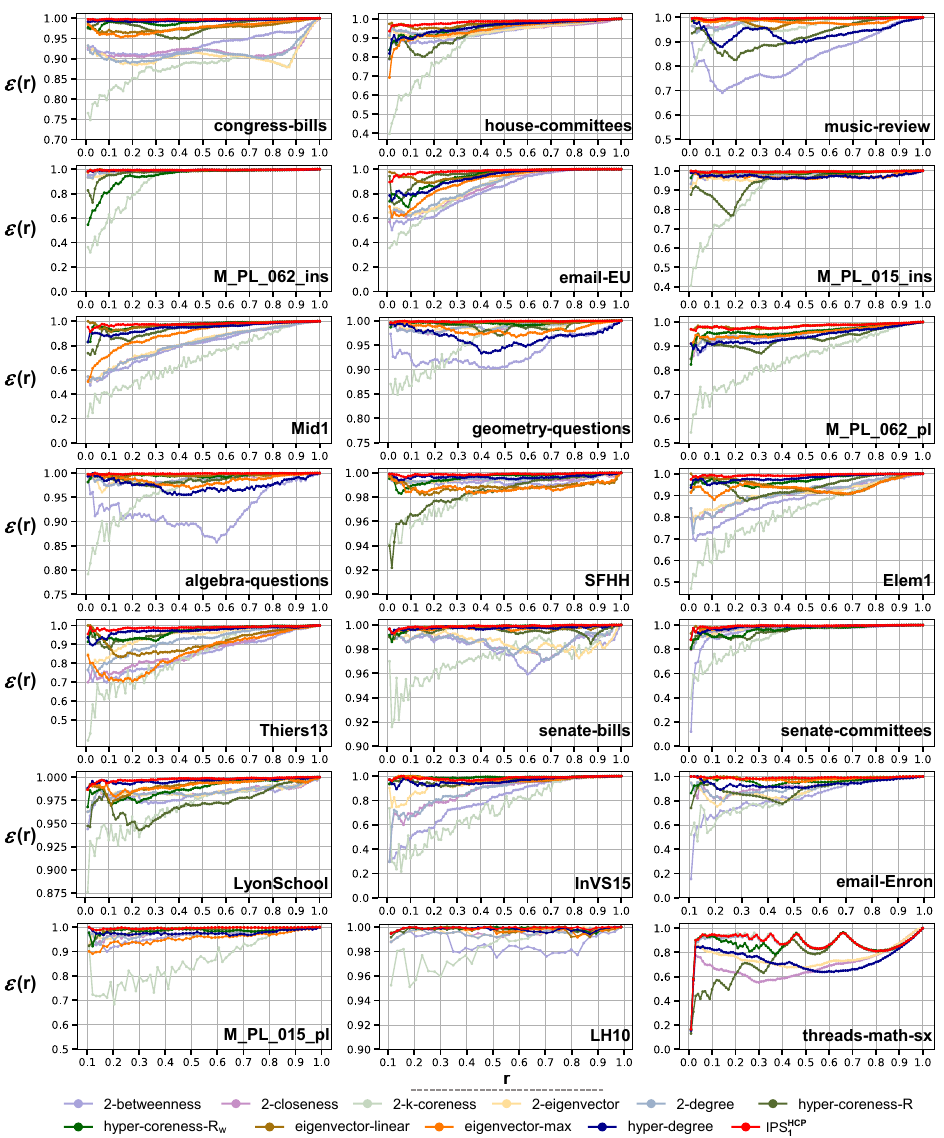}
    \captionsetup{justification=raggedright, singlelinecheck=false, width=\textwidth}
    \caption{\textbf{Results of the imprecision function.} For different real hypergraphs, we compare the imprecision function of IPS method and other existing methods. Results demonstrate that IPS has the best performance.}
    \label{unex}
\end{figure}

Figure~\ref{jaccard} shows results of the Jaccard coefficient $J(r)$ under more hypergraphs, as a complementary part of Fig.~2. 
The Jaccard coefficient J(r) is calculated by $J(r)=\frac{|T_g \cap T_m|}{|T_g\cup T_m|}$, 
where $T_g$ and $T_m$, respectively, represent the set of top $rN$ ($r\in [0,1]$) nodes based on ground truth and a particular measure of node's importance. 
We find that IPS has the highest Jaccard coefficient under almost all scenarios, which means that the proposed measure performs the best.
Figure~\ref{unex} shows another widely-used index, i.e., the imprecision function $\epsilon(r)$, which can be written as $\epsilon(r) = 1-\frac{R_{T_m}}{R_{T_g}}$, 
where $R_{T_g}$ and $R_{T_m}$ represent the average propagation range of nodes in $T_g$ and $T_m$.
A larger value of $\epsilon(r)$ reflects better performance of the centrality measure. 
The imprecision function also shows that the IPS method performs better than other measures in most scenarios.

\subsection{Complementary results for Fig.~3}\label{S2.3}
\begin{figure}[H]
    \centering
    \includegraphics[]{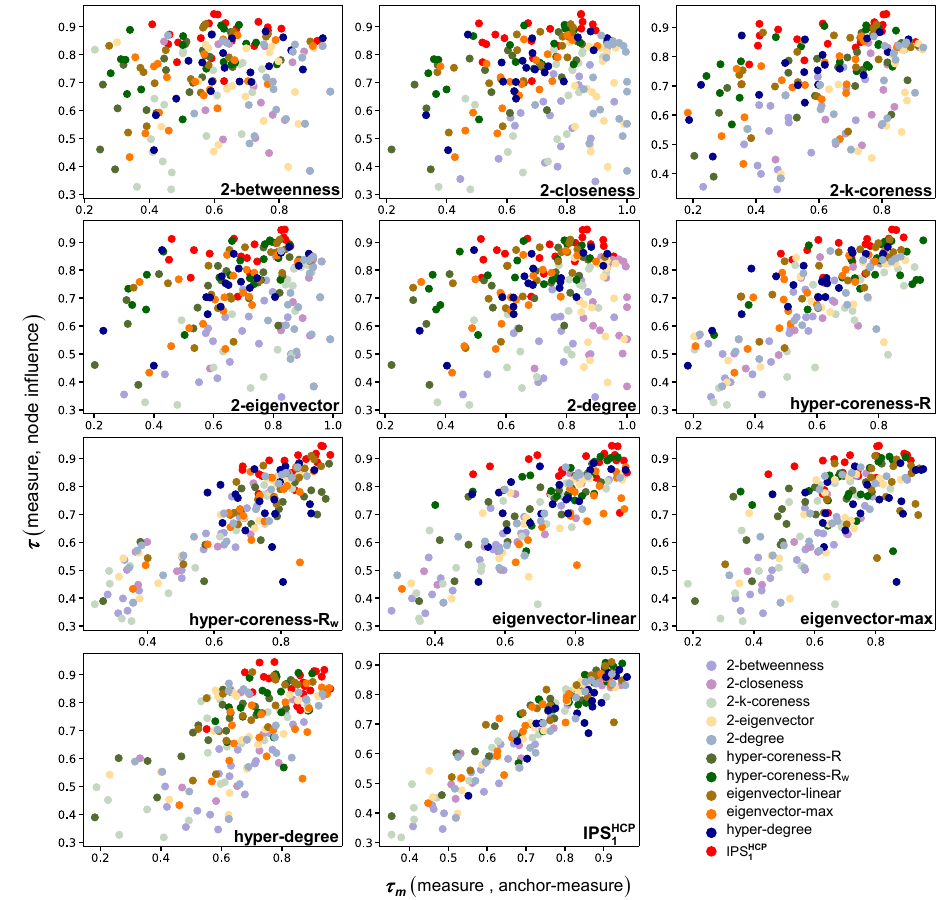}
    \captionsetup{justification=raggedright, singlelinecheck=false, width=\textwidth}
    \caption{\textbf{Correlations between performance of measures and their similarity with the anchor measure.} 
    Each subplot corresponds to an anchor measure (right bottom). Points represent the correlations under different real hypergraphs, and different colors indicate different measures.}
    \label{fig3all}
\end{figure}

We explore the correlation between the performance of methods (quantified by Kendall's $\tau$) and their similarity to an anchor measure. IPS is used as the anchor measure in Fig.~3 of the main text. Figure~\ref{fig3all} presents situations where other methods serve as the anchor measure.
In each subplot, the anchor measure is indicated at the bottom right corner. The horizontal axis represents the correlation between a measure (indicated by its color) and the current anchor measure in a hypergraph (denoted as $\tau_m(\text{measure, anchor measure})$, measure $\neq$ anchor measure). The vertical axis is the performance of the measure. 
Results demonstrate that there exists a positive correlation only when IPS is the anchor measure. This means that IPS can be used to explains the performance variation of other methods.

\subsection{More results on the large-scale real-world dataset}\label{S2.4}
\begin{figure}[H]
    \centering
    \includegraphics[]{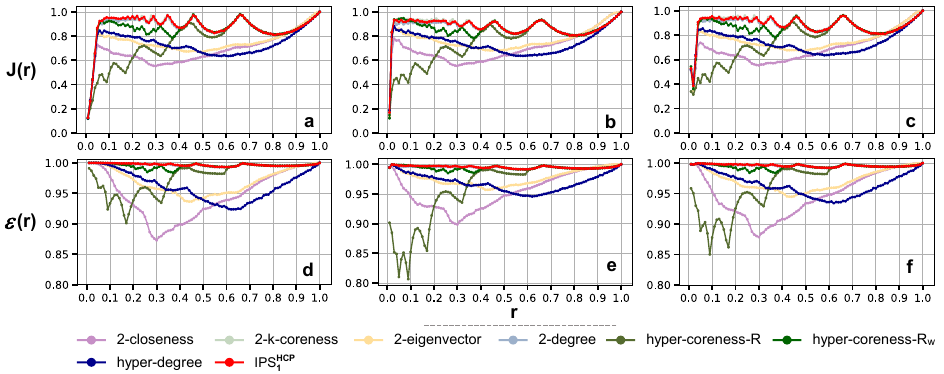}
    \captionsetup{justification=raggedright, singlelinecheck=false, width=\textwidth}
    \caption{\textbf{More experiments on a large-scale network.} 
    We show \textbf{a}--\textbf{c} the Jaccard coefficient and \textbf{d}--\textbf{f} imprecision function under different dynamical settings: in \textbf{a}, \textbf{d} $(\nu,\lambda,\mu )=(1, 0.2, 1)$, $(3,0.05,1)$ in \textbf{b}, \textbf{e}, and $(3,0.1,1)$ in \textbf{c}, \textbf{f}. In all parameters, IPS shows outstanding performance.
    }
    \label{bignet}
\end{figure}

We conduct more experiments on the large-scale hypergraph threads-math-sx. Figure~\ref{bignet} shows the Jaccard coefficient $J(r)$ and imprecision function $\epsilon(r)$ to explore the performance of different measures on a large-scale network. Both indices suggest that IPS consistently achieves outstanding performance.

\newpage
\section{Robustness of IPS under different parameters}
In this section, we validate the robustness of IPS's performance under different real-world hypergraphs and dynamical parameters based on the HCP model. We present the performance of IPS on 20 real-world hypergraphs under 36 groups of parameters (3 values of $\mu$, 4 values of $\nu$, and 3 values of $\lambda$).
Subsection.~\ref{S2.5} provides parameter settings, and Subsection.~\ref{S2.6} shows the results.
\subsection{Parameter settings}\label{S2.5}
\begin{figure}[H]
    \vspace{-3em}
    \centering
    \includegraphics[]{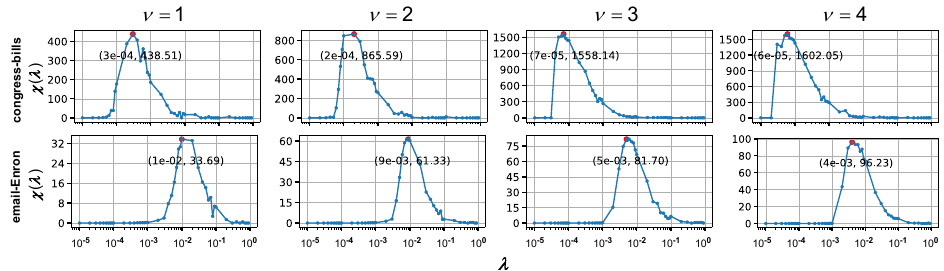}
    \captionsetup{justification=raggedright, singlelinecheck=false, width=\textwidth}
    \caption{\textbf{Determination of $\lambda_c$ of HCP when $\mu=1$ based on susceptibility.} 
    Each subplot corresponds to a parameter $v$ and a real hypergraph, where susceptibility is shown against $\lambda$. In the subplots, each point is obtained from 1000 independent simulations, each starting from a randomly selected seed. The peak, indicated by a red point, represents the threshold ($\lambda_c$).
    }
    \label{p-hnc-10-1}
\end{figure}

We study the robustness of the IPS method by varying dynamical parameters ($\mu$, $\nu$, $\lambda$). We consider three values of recovery probability $\mu$, including 0.1, 0.5, and 1.0; three representative values of $\lambda$, including an under-threshold value ($0.1\lambda_c$), the threshold value ($\lambda_c$), and an over-threshold value ($10\lambda_c$); and four values of $\nu$, including 1, 2, 3, and 4, indicating increasing nonlinear enhancement effectiveness. 
Here, we estimate the spreading thresholds ($\lambda_c$) under different hypergraphs and parameters ($\nu$ and $\mu$) by using the susceptibility $\chi$, which is given as: 
\begin{align}
    \chi = \frac{\langle R_{\infty}^2\rangle - \langle R_{\infty}\rangle ^2}{\langle R_{\infty}\rangle},
    \label{chi}
\end{align}
where $R_{\infty}$ represents the stationary spreading sizes of simulations.
For the HCP model, Fig.~\ref{p-hnc-10-1} presents examples of the susceptibility against $\lambda$,  where each point is the result of 1000 simulations. Complying with prior works, the peak in susceptibility corresponds to the spreading threshold ($\lambda_c$).  Table~\ref{hnc-mu10} presents the estimated thresholds under different settings (corresponding to Fig.~4 in the main text).

\newpage
\begin{table}[h]
\centering
\setlength{\tabcolsep}{3pt}
\renewcommand{\arraystretch}{1.6}  
\begin{tabular}{@{}c@{\hspace{0.8em}}c@{}} 
\begin{tabular}[t]{|l|cccc|}   
\hline
\multirow{2}{*}{Dataset}  & \multicolumn{4}{c|}{$\lambda_c$} \\
\hhline{~|----|}
 & $\nu=1$ & $\nu=2$ & $\nu=3$ & $\nu=4$  \\
\hline
congress-bills      & $3\!\times\!10^{-4}$   & $2\!\times\!10^{-4}$ & $7\!\times\!10^{-5}$&  $6\!\times\!10^{-5}$\\
\hline
house-committees    & $4\!\times\!10^{-3}$  & $1\!\times\!10^{-3}$  & $9\!\times\!10^{-4}$ & $7\!\times\!10^{-4}$  \\
\hline
music-review        & $6\!\times\!10^{-3}$  & $2\!\times\!10^{-3}$  & $2\!\times\!10^{-3}$ & $1\!\times\!10^{-3}$  \\
\hline
M\_PL\_062\_ins     & $4\!\times\!10^{-3}$  & $1\!\times\!10^{-3}$  & $6\!\times\!10^{-4}$ & $5\!\times\!10^{-4}$  \\
\hline
email-EU            & $7\!\times\!10^{-3}$   & $2\!\times\!10^{-3}$ & $2\!\times\!10^{-3}$ & $7\!\times\!10^{-4}$  \\
\hline
M\_PL\_015\_ins     & $9\!\times\!10^{-3}$  & $3\!\times\!10^{-3}$ & $2\!\times\!10^{-3}$ & $2\!\times\!10^{-3}$ \\
\hline
Mid1                & $1\!\times\!10^{-3}$   & $9\!\times\!10^{-4}$ & $7\!\times\!10^{-4}$ & $5\!\times\!10^{-4}$  \\
\hline
geometry-questions  & $5\!\times\!10^{-3}$   & $1\!\times\!10^{-3}$ & $5\!\times\!10^{-4}$ & $3\!\times\!10^{-4}$ \\
\hline
M\_PL\_062\_pl      & $1\!\times\!10^{-3}$   & $6\!\times\!10^{-4}$ & $4\!\times\!10^{-4}$ & $3\!\times\!10^{-4}$ \\
\hline
algebra-questions   & $7\!\times\!10^{-3}$   & $2\!\times\!10^{-3}$ & $2\!\times\!10^{-3}$ & $1\!\times\!10^{-3}$ \\
\hline
\end{tabular}
&
\begin{tabular}[t]{|l|cccc|}
\hline
\multirow{2}{*}{Dataset}  & \multicolumn{4}{c|}{$\lambda_c$} \\
\hhline{~|----|}
 & $\nu=1$ & $\nu=2$ & $\nu=3$ & $\nu=4$  \\
\hline
SFHH                & $2\!\times\!10^{-2}$   & $2\!\times\!10^{-2}$ & $1\!\times\!10^{-2}$ & $1\!\times\!10^{-2}$\\
\hline
Elem1              & $1\!\times\!10^{-3}$   & $1\!\times\!10^{-3}$ & $8\!\times\!10^{-4}$ & $5\!\times\!10^{-4}$  \\
\hline
Thiers13           & $2\!\times\!10^{-2}$   & $2\!\times\!10^{-2}$ &  $1\!\times\!10^{-2}$  &  $9\!\times\!10^{-3}$  \\
\hline
senate-bills       & $8\!\times\!10^{-5}$   & $4\!\times\!10^{-5}$ &  $2\!\times\!10^{-5}$  &  $2\!\times\!10^{-5}$ \\
\hline
senate-committees  & $3\!\times\!10^{-3}$   & $2\!\times\!10^{-3}$ &  $2\!\times\!10^{-3}$  &  $1\!\times\!10^{-3}$ \\
\hline
LyonSchool         & $3\!\times\!10^{-3}$   & $2\!\times\!10^{-3}$ &  $2\!\times\!10^{-3}$  &  $2\!\times\!10^{-3}$ \\
\hline
InVS15             & $2\!\times\!10^{-2}$   & $2\!\times\!10^{-2}$ &  $2\!\times\!10^{-2}$  &  $2\!\times\!10^{-2}$ \\
\hline
email-Enron        & $1\!\times\!10^{-2}$   & $9\!\times\!10^{-3}$ &  $5\!\times\!10^{-3}$  &  $4\!\times\!10^{-3}$ \\
\hline
M\_PL\_015\_pl     & $5\!\times\!10^{-3}$   & $2\!\times\!10^{-3}$ &  $2\!\times\!10^{-3}$  &  $9\!\times\!10^{-4}$\\
\hline
LH10               &  $1\!\times\!10^{-2}$   & $1\!\times\!10^{-2}$ &  $1\!\times\!10^{-2}$  &  $8\!\times\!10^{-3}$  \\
\hline
\end{tabular}
\end{tabular}
\captionsetup{justification=raggedright, singlelinecheck=false, width=\textwidth}
\caption{\textbf{Spreading thresholds ($\lambda_c$) in the HCP model when $\mu=1$.}
}
\label{hnc-mu10}
\end{table}

\clearpage

\begin{figure}[H]
    \centering
    \includegraphics[]{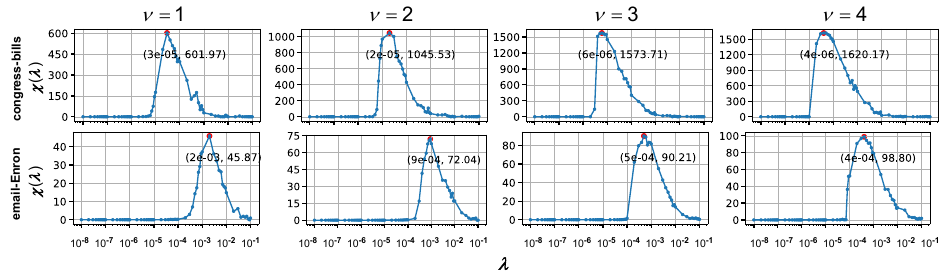}
    \captionsetup{justification=raggedright, singlelinecheck=false, width=\textwidth}
    \caption{\textbf{Determination of $\lambda_c$ of HCP when $\mu=0.1$ based on susceptibility.} Each subplot corresponds to a parameter $v$ and a real hypergraph, where susceptibility is shown against $\lambda$. In the subplots, each point is obtained from 300 independent simulations, each starting from a randomly selected seed. The peak, indicated by  a red point, represents the threshold ($\lambda_c$).
    }
    \label{p-hnc-1-1}
\end{figure}

\begin{figure}[H]
    \centering
    \includegraphics[]{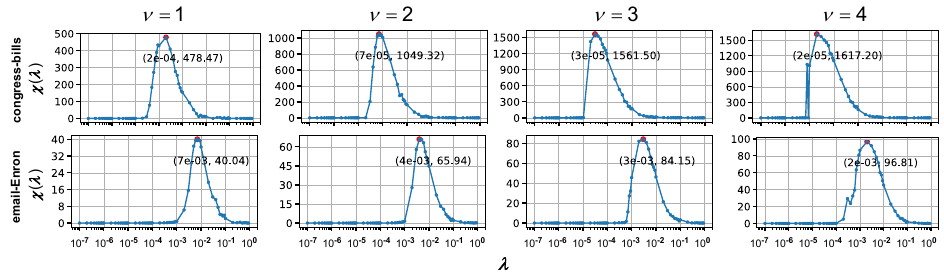}
    \captionsetup{justification=raggedright, singlelinecheck=false, width=\textwidth}
    \caption{\textbf{Determination of $\lambda_c$ of HCP when $\mu=0.5$ based on susceptibility.} Each subplot corresponds to a parameter $v$ and a real hypergraph, where susceptibility is shown against $\lambda$. In the subplots, each point is obtained from 1000 independent simulations, each starting from a randomly selected seed. The peak, indicated by a red point, represents the threshold ($\lambda_c$).}
    \label{p-hnc-5-1}
\end{figure}

Similarly, we explore the threshold of $\lambda$ under more cases ($\mu \neq 1$) by using the susceptibility $\chi$. 
Specifically, Tab.~\ref{hnc-mu1} and Tab.~\ref{hnc-mu5} present the spreading thresholds of HCP when $\mu=0.1$ and $\mu=0.5$, respectively. Fig.~\ref{p-hnc-1-1} and Fig.~\ref{p-hnc-5-1} show examples of locating the threshold of $\eta$ by susceptibility.
\begin{table}[!h]
\centering
\setlength{\tabcolsep}{3pt}
\renewcommand{\arraystretch}{1.6}  
\begin{tabular}{@{}c@{\hspace{0.8em}}c@{}} 
\begin{tabular}[t]{|l|cccc|}  
\hline
\multirow{2}{*}{Dataset}  & \multicolumn{4}{c|}{$\lambda_c$} \\
\hhline{~|----|}
 & $\nu=1$ & $\nu=2$ & $\nu=3$ & $\nu=4$  \\
\hline
congress-bills      & $3\!\times\!10^{-5}$   & $2\!\times\!10^{-5}$ & $6\!\times\!10^{-6}$&  $4\!\times\!10^{-6}$\\
\hline
house-committees    & $4\!\times\!10^{-4}$  & $1\!\times\!10^{-4}$  & $5\!\times\!10^{-5}$ & $5\!\times\!10^{-5}$  \\
\hline
music-review        & $9\!\times\!10^{-4}$  & $2\!\times\!10^{-4}$  & $9\!\times\!10^{-5}$ & $6\!\times\!10^{-5}$  \\
\hline
M\_PL\_062\_ins     & $6\!\times\!10^{-4}$  & $1\!\times\!10^{-4}$  & $4\!\times\!10^{-5}$ & $4\!\times\!10^{-5}$  \\
\hline
email-EU            & $6\!\times\!10^{-4}$   & $5\!\times\!10^{-4}$ & $1\!\times\!10^{-4}$ & $6\!\times\!10^{-5}$  \\
\hline
M\_PL\_015\_ins     & $1\!\times\!10^{-3}$  & $4\!\times\!10^{-4}$ & $2\!\times\!10^{-4}$ & $9\!\times\!10^{-5}$ \\
\hline
Mid1                & $1\!\times\!10^{-4}$   & $1\!\times\!10^{-4}$ & $8\!\times\!10^{-5}$ & $4\!\times\!10^{-5}$  \\
\hline
geometry-questions  & $4\!\times\!10^{-4}$   & $9\!\times\!10^{-5}$ & $4\!\times\!10^{-5}$ & $2\!\times\!10^{-5}$ \\
\hline
M\_PL\_062\_pl      & $1\!\times\!10^{-4}$   & $5\!\times\!10^{-5}$ & $3\!\times\!10^{-5}$ & $2\!\times\!10^{-5}$ \\
\hline
algebra-questions   & $1\!\times\!10^{-3}$   & $2\!\times\!10^{-4}$ & $2\!\times\!10^{-4}$ & $8\!\times\!10^{-5}$ \\
\hline
\end{tabular}
&
\begin{tabular}[t]{|l|cccc|}
\hline
\multirow{2}{*}{Dataset}  & \multicolumn{4}{c|}{$\lambda_c$} \\
\hhline{~|----|}
 & $\nu=1$ & $\nu=2$ & $\nu=3$ & $\nu=4$  \\
\hline
SFHH                & $2\!\times\!10^{-3}$   & $2\!\times\!10^{-3}$ & $2\!\times\!10^{-3}$ & $1\!\times\!10^{-3}$\\
\hline
Elem1              & $2\!\times\!10^{-4}$   & $1\!\times\!10^{-4}$ & $8\!\times\!10^{-5}$ & $4\!\times\!10^{-5}$  \\
\hline
Thiers13           & $2\!\times\!10^{-3}$   & $2\!\times\!10^{-3}$ &  $1\!\times\!10^{-3}$  &  $1\!\times\!10^{-3}$  \\
\hline
senate-bills       & $1\!\times\!10^{-5}$   & $4\!\times\!10^{-6}$ &  $2\!\times\!10^{-6}$  &  $9\!\times\!10^{-7}$ \\
\hline
senate-committees  & $6\!\times\!10^{-4}$   & $2\!\times\!10^{-4}$ &  $8\!\times\!10^{-5}$  &  $7\!\times\!10^{-5}$ \\
\hline
LyonSchool         & $3\!\times\!10^{-4}$   & $3\!\times\!10^{-4}$ &  $2\!\times\!10^{-4}$  &  $1\!\times\!10^{-4}$ \\
\hline
InVS15             & $3\!\times\!10^{-3}$   & $2\!\times\!10^{-3}$ &  $2\!\times\!10^{-3}$  &  $2\!\times\!10^{-3}$ \\
\hline
email-Enron        & $2\!\times\!10^{-3}$   & $9\!\times\!10^{-4}$ &  $5\!\times\!10^{-4}$  &  $4\!\times\!10^{-4}$ \\
\hline
M\_PL\_015\_pl     & $5\!\times\!10^{-4}$   & $2\!\times\!10^{-4}$ &  $9\!\times\!10^{-5}$  &  $9\!\times\!10^{-5}$\\
\hline
LH10               &  $4\!\times\!10^{-3}$   & $9\!\times\!10^{-4}$ &  $7\!\times\!10^{-4}$  &  $6\!\times\!10^{-4}$  \\
\hline
\end{tabular}
\end{tabular}
\captionsetup{justification=raggedright, singlelinecheck=false, width=\textwidth}
\caption{\textbf{Spreading thresholds ($\lambda_c$) in the HCP model when $\mu=0.1$.}
}
\label{hnc-mu1}
\end{table}

\begin{table}[!h]
\centering
\setlength{\tabcolsep}{3pt}
\renewcommand{\arraystretch}{1.6}  
\begin{tabular}{@{}c@{\hspace{0.8em}}c@{}} 
\begin{tabular}[t]{|l|cccc|} 
\hline
\multirow{2}{*}{Dataset}  & \multicolumn{4}{c|}{$\lambda_c$} \\
\hhline{~|----|}
 & $\nu=1$ & $\nu=2$ & $\nu=3$ & $\nu=4$  \\
\hline
congress-bills      & $2\!\times\!10^{-4}$   & $7\!\times\!10^{-5}$ & $3\!\times\!10^{-5}$&  $2\!\times\!10^{-5}$\\
\hline
house-committees    & $2\!\times\!10^{-3}$  & $6\!\times\!10^{-4}$  & $3\!\times\!10^{-4}$ & $2\!\times\!10^{-4}$  \\
\hline
music-review        & $4\!\times\!10^{-3}$  & $9\!\times\!10^{-4}$  & $6\!\times\!10^{-4}$ & $4\!\times\!10^{-4}$  \\
\hline
M\_PL\_062\_ins     & $2\!\times\!10^{-3}$  & $5\!\times\!10^{-4}$  & $3\!\times\!10^{-4}$ & $2\!\times\!10^{-4}$  \\
\hline
email-EU            & $4\!\times\!10^{-3}$   & $1\!\times\!10^{-3}$ & $7\!\times\!10^{-4}$ & $4\!\times\!10^{-4}$  \\
\hline
M\_PL\_015\_ins     & $6\!\times\!10^{-3}$  & $2\!\times\!10^{-3}$ & $8\!\times\!10^{-4}$ & $5\!\times\!10^{-4}$ \\
\hline
Mid1                & $6\!\times\!10^{-4}$   & $5\!\times\!10^{-4}$ & $4\!\times\!10^{-4}$ & $2\!\times\!10^{-4}$  \\
\hline
geometry-questions  & $1\!\times\!10^{-3}$   & $4\!\times\!10^{-4}$ & $2\!\times\!10^{-4}$ & $1\!\times\!10^{-4}$ \\
\hline
M\_PL\_062\_pl      & $5\!\times\!10^{-4}$   & $3\!\times\!10^{-4}$ & $2\!\times\!10^{-4}$ & $1\!\times\!10^{-4}$ \\
\hline
algebra-questions   & $4\!\times\!10^{-3}$   & $1\!\times\!10^{-3}$ & $8\!\times\!10^{-4}$ & $5\!\times\!10^{-4}$ \\
\hline
\end{tabular}
&
\begin{tabular}[t]{|l|cccc|}
\hline
\multirow{2}{*}{Dataset}  & \multicolumn{4}{c|}{$\lambda_c$} \\
\hhline{~|----|}
 & $\nu=1$ & $\nu=2$ & $\nu=3$ & $\nu=4$  \\
\hline
SFHH                & $9\!\times\!10^{-3}$   & $9\!\times\!10^{-3}$ & $8\!\times\!10^{-3}$ & $6\!\times\!10^{-3}$\\
\hline
Elem1              & $7\!\times\!10^{-4}$   & $5\!\times\!10^{-4}$ & $4\!\times\!10^{-4}$ & $2\!\times\!10^{-4}$  \\
\hline
Thiers13           & $8\!\times\!10^{-3}$   & $8\!\times\!10^{-3}$ &  $6\!\times\!10^{-3}$  &  $5\!\times\!10^{-3}$  \\
\hline
senate-bills       & $5\!\times\!10^{-5}$   & $2\!\times\!10^{-5}$ &  $1\!\times\!10^{-5}$  &  $8\!\times\!10^{-6}$ \\
\hline
senate-committees  & $2\!\times\!10^{-3}$   & $9\!\times\!10^{-4}$ &  $5\!\times\!10^{-4}$  &  $4\!\times\!10^{-4}$ \\
\hline
LyonSchool         & $1\!\times\!10^{-3}$   & $1\!\times\!10^{-3}$ &  $9\!\times\!10^{-4}$  &  $6\!\times\!10^{-4}$ \\
\hline
InVS15             & $9\!\times\!10^{-3}$   & $1\!\times\!10^{-2}$ &  $1\!\times\!10^{-2}$  &  $7\!\times\!10^{-3}$ \\
\hline
email-Enron        & $7\!\times\!10^{-3}$   & $4\!\times\!10^{-3}$ &  $3\!\times\!10^{-3}$  &  $2\!\times\!10^{-3}$ \\
\hline
M\_PL\_015\_pl     & $3\!\times\!10^{-3}$   & $1\!\times\!10^{-3}$ &  $7\!\times\!10^{-4}$  &  $6\!\times\!10^{-4}$\\
\hline
LH10               &  $8\!\times\!10^{-3}$   & $5\!\times\!10^{-3}$ &  $4\!\times\!10^{-3}$  &  $3\!\times\!10^{-3}$  \\
\hline
\end{tabular}
\end{tabular}
\captionsetup{justification=raggedright, singlelinecheck=false, width=\textwidth}
\caption{\textbf{Spreading thresholds ($\lambda_c$) in the HCP model when $\mu=0.5$. }
}
\label{hnc-mu5}
\end{table}


\newpage
\subsection{Results for robustness evaluation}\label{S2.6}
\begin{figure}[!h]
    \centering
    \includegraphics[]{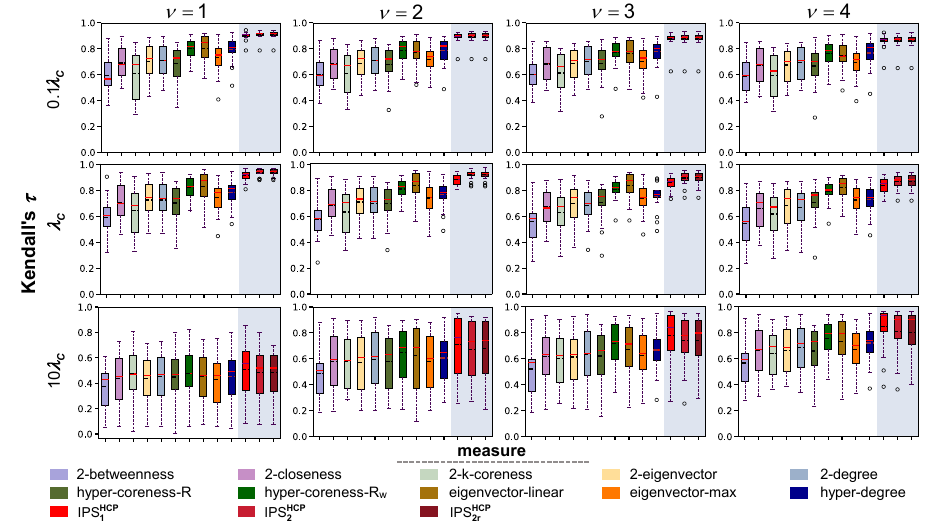}
    \captionsetup{justification=raggedright, singlelinecheck=false, width=\textwidth}
    \caption{\textbf{Robustness analysis when $\mu=1.0$.} 
    Results of Kendall's $\tau$ between node measure and node influence are shown as boxplots (containing 20 real hypergraphs). 
    The solid line in each box is the median value, and the dashed line is the mean value. Node influence is obtained through 1000 simulations.
    }
    \label{hnc-10}
\end{figure}

\begin{figure}[H]
    \centering
    \includegraphics[]{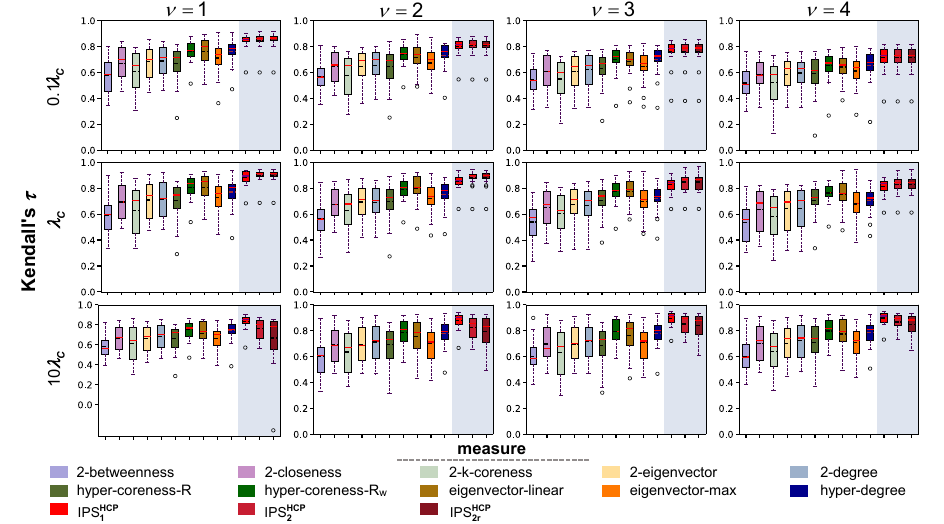}
    \captionsetup{justification=raggedright, singlelinecheck=false, width=\textwidth}
    \caption{\textbf{Robustness analysis when $\mu=0.1$.}
    Results of Kendall's $\tau$ between node measure and node influence are shown as boxplots (containing 20 real hypergraphs). 
    The solid line in each box is the median value, and the dashed line is the mean value. Node influence is obtained through 300 simulations.
    }
    \label{hnc-1}
\end{figure}

\begin{figure}[H]
    \vspace{-15pt}
    \centering
    \includegraphics[]{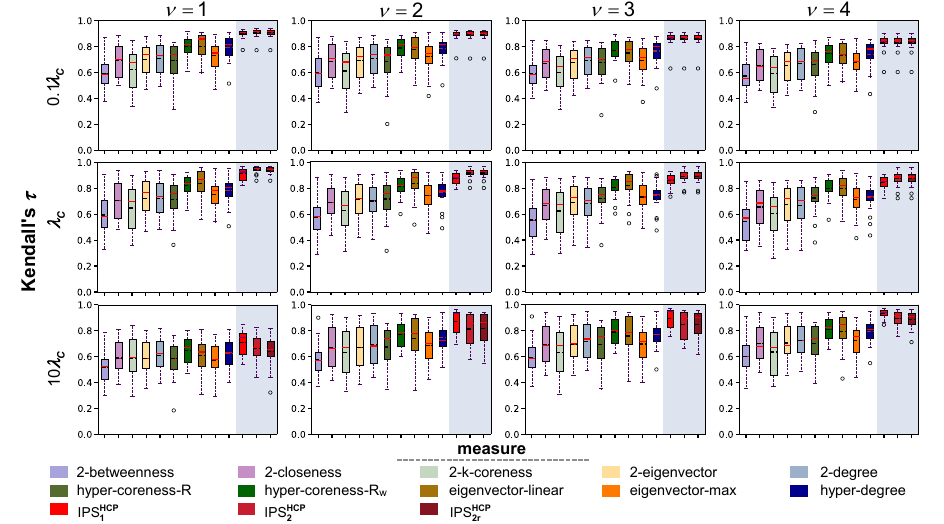}
    \captionsetup{justification=raggedright, singlelinecheck=false, width=\textwidth}
    \caption{\textbf{Robustness analysis when $\mu=0.5$.}
    Results of Kendall's $\tau$ between node measure and node influence are shown as boxplots (containing 20 real hypergraphs). 
    The solid line in each box is the median value, and the dashed line is the mean value. Node influence is obtained through 1000 simulations.
    }
    \label{hnc-5}
\end{figure}

For each group of parameters and each hypergraph shown in Tab.~\ref{hnc-mu10} (also \ref{hnc-mu1} and \ref{hnc-mu5}), we take $50\%$ nodes from it to serve as the seed in turn, and then calculate Kendall's $\tau$ for measures. 
Finally, results of Kendall's $\tau$ 
are shown in Figure~\ref{hnc-10}--\ref{hnc-5} (corresponding to cases with $\mu=1$, $\mu=0.1$, and $\mu=0.5$, respectively); in each figure, subplots are arranged by different settings of $\lambda$ and $\nu$.
Results show that the IPS methods, whatever $IPS_1^{HCP}$, $IPS_2^{HCP}$ and $IPS_{2r}^{HCP}$, outperform existing methods in all dynamical parameters. It highlights that the performance of IPS is robust to parameters. 
Note that the parameterized measures $IPS_{2}^{HCP}$ and $IPS_{2r}^{HCP}$ require pre-rescaling when $\mu \neq 1$ (i.e., $\lambda \gets \frac{\lambda}{\mu}$, $\mu \gets 1$).


\newpage
\section{Transferability of IPS under different spreading dynamics}
We study the effectiveness of the IPS framework in other spreading dynamics, including the higher-order contagion model with size attenuation (HCSA) and the higher-order threshold contagion model (HTC), to evaluate its transferability.
\subsection{Transferability evaluation under HCSA}\label{S3}
\subsubsection{Parameter settings for HCSA}\label{S3.1}
\begin{figure}[H]
    \centering
    \includegraphics[]{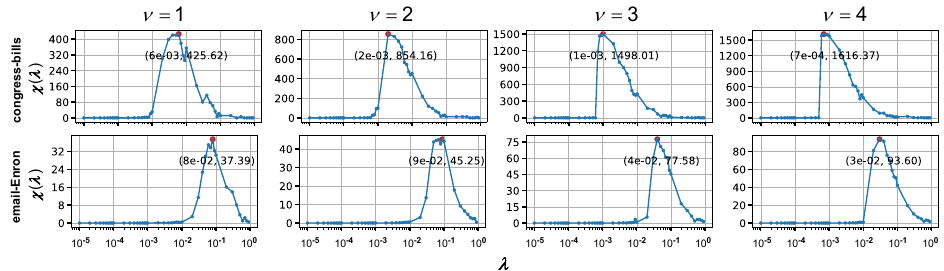}
    \captionsetup{justification=raggedright, singlelinecheck=false, width=\textwidth}
    \caption{\textbf{Determination of $\lambda_c$ based on susceptibility when $\mu =1$ in the HCSA model.} 
    Each subplot corresponds to a parameter $\nu$ and a real hypergraph, where susceptibility is shown against $\lambda$. 
    In the subplots, each point is obtained from 1000 independent simulations, each starting from a randomly selected seed.  The peak, indicated by a red point, represents the threshold ($\lambda_c$).}
    \label{hncsa-10-1}
\end{figure}

\begin{figure}[H]
    \centering
    \includegraphics[]{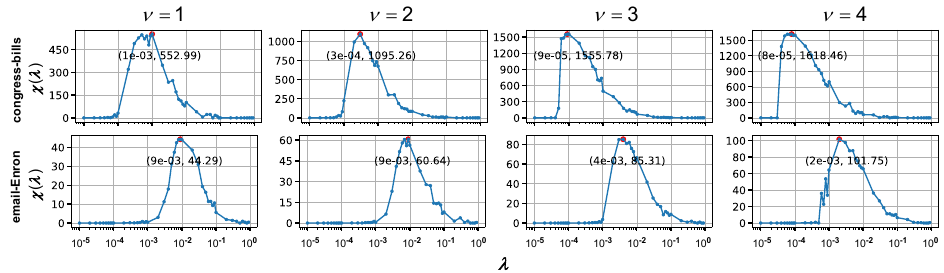}
    \captionsetup{justification=raggedright, singlelinecheck=false, width=\textwidth}
    \caption{\textbf{Determination of $\lambda_c$ based on susceptibility when $\mu =0.1$ in the HCSA model.} 
    Each subplot corresponds to a parameter $\nu$ and a real hypergraph, where susceptibility is shown against $\lambda$. 
    In the subplots, each point is obtained from 300 independent simulations, each starting from a randomly selected seed.  The peak, indicated by a red point, represents the threshold ($\lambda_c$).}
    \label{hncsa-1-1}
\end{figure}

To ensure the robustness of the results, we consider situations of different parameters under the HCSA model, i.e., $\mu=1,0.1$; $\nu=1,2,3,4$; $\lambda=0.1\lambda_c, \lambda_c, 10\lambda_c$, respectively. 
Similarly, the spreading threshold ($\lambda_c$) under given $\nu$ and $\mu$ in each hypergraph is obtained by using susceptibility in Eq.\eqref{chi} (see examples for the case of $\mu=1$ in Fig.~\ref{hncsa-10-1}, and $\mu=0.1$ in Fig.~\ref{hncsa-1-1}).
The values of $\lambda_c$ in each hypergraph are presented in Tab.~\ref{hncsa-mu10} ($\mu=1$, also corresponding to Fig.~5 in the main text) and Tab.\ref{hncsa-mu1} ($\mu=0.1$).
\vspace{-1em}
\begin{table}[!h]
\centering
\setlength{\tabcolsep}{3pt}
\renewcommand{\arraystretch}{1.6}  
\begin{tabular}{@{}c@{\hspace{0.8em}}c@{}} 
\begin{tabular}[t]{|l|cccc|}  
\hline
\multirow{2}{*}{Dataset}  & \multicolumn{4}{c|}{$\lambda_c$} \\
\hhline{~|----|}
 & $\nu=1$ & $\nu=2$ & $\nu=3$ & $\nu=4$  \\
\hline
congress-bills      & $6\!\times\!10^{-3}$   & $2\!\times\!10^{-3}$ & $1\!\times\!10^{-3}$&  $7\!\times\!10^{-4}$\\
\hline
house-committees    & $2\!\times\!10^{-1}$  & $6\!\times\!10^{-2}$  & $5\!\times\!10^{-2}$ & $4\!\times\!10^{-2}$  \\
\hline
music-review        & $3\!\times\!10^{-1}$  & $7\!\times\!10^{-2}$  & $5\!\times\!10^{-2}$ & $3\!\times\!10^{-2}$  \\
\hline
M\_PL\_062\_ins     & $9\!\times\!10^{-2}$  & $4\!\times\!10^{-2}$  & $2\!\times\!10^{-2}$ & $2\!\times\!10^{-2}$  \\
\hline
email-EU            & $3\!\times\!10^{-2}$   & $3\!\times\!10^{-2}$ & $1\!\times\!10^{-2}$ & $8\!\times\!10^{-3}$  \\
\hline
M\_PL\_015\_ins     & $3\!\times\!10^{-1}$  & $1\!\times\!10^{-1}$ & $7\!\times\!10^{-2}$ & $6\!\times\!10^{-2}$ \\
\hline
Mid1                & $4\!\times\!10^{-3}$   & $4\!\times\!10^{-3}$ & $4\!\times\!10^{-3}$ & $3\!\times\!10^{-3}$  \\
\hline
geometry-questions  & $9\!\times\!10^{-2}$   & $3\!\times\!10^{-2}$ & $2\!\times\!10^{-2}$ & $1\!\times\!10^{-2}$ \\
\hline
M\_PL\_062\_pl      & $5\!\times\!10^{-2}$   & $3\!\times\!10^{-2}$ & $2\!\times\!10^{-2}$ & $2\!\times\!10^{-2}$ \\
\hline
algebra-questions   & $9\!\times\!10^{-2}$   & $5\!\times\!10^{-2}$ & $3\!\times\!10^{-2}$ & $2\!\times\!10^{-2}$ \\
\hline
\end{tabular}
&
\begin{tabular}[t]{|l|cccc|}
\hline
\multirow{2}{*}{Dataset}  & \multicolumn{4}{c|}{$\lambda_c$} \\
\hhline{~|----|}
 & $\nu=1$ & $\nu=2$ & $\nu=3$ & $\nu=4$  \\
\hline
SFHH                & $5\!\times\!10^{-2}$   & $5\!\times\!10^{-2}$ & $6\!\times\!10^{-2}$ & $4\!\times\!10^{-2}$\\
\hline
Elem1              & $6\!\times\!10^{-3}$   & $5\!\times\!10^{-3}$ & $4\!\times\!10^{-3}$ & $3\!\times\!10^{-3}$  \\
\hline
Thiers13           & $5\!\times\!10^{-2}$   & $4\!\times\!10^{-2}$ &  $4\!\times\!10^{-2}$  &  $4\!\times\!10^{-2}$  \\
\hline
senate-bills       & $2\!\times\!10^{-3}$   & $1\!\times\!10^{-3}$ &  $7\!\times\!10^{-4}$  &  $6\!\times\!10^{-4}$ \\
\hline
senate-committees  & $1\!\times\!10^{-1}$   & $4\!\times\!10^{-2}$ &  $3\!\times\!10^{-2}$  &  $2\!\times\!10^{-2}$ \\
\hline
LyonSchool         & $1\!\times\!10^{-2}$   & $1\!\times\!10^{-2}$ &  $9\!\times\!10^{-3}$  &  $6\!\times\!10^{-3}$ \\
\hline
InVS15             & $6\!\times\!10^{-2}$   & $5\!\times\!10^{-2}$ &  $6\!\times\!10^{-2}$  &  $5\!\times\!10^{-2}$ \\
\hline
email-Enron        & $8\!\times\!10^{-2}$   & $9\!\times\!10^{-2}$ &  $4\!\times\!10^{-2}$  &  $3\!\times\!10^{-2}$ \\
\hline
M\_PL\_015\_pl     & $9\!\times\!10^{-2}$   & $5\!\times\!10^{-2}$ &  $3\!\times\!10^{-2}$  &  $3\!\times\!10^{-2}$\\
\hline
LH10               &  $4\!\times\!10^{-2}$   & $4\!\times\!10^{-2}$ &  $4\!\times\!10^{-2}$  &  $3\!\times\!10^{-2}$  \\
\hline
\end{tabular}
\end{tabular}
\captionsetup{justification=raggedright, singlelinecheck=false, width=\textwidth}
\caption{\textbf{Spreading thresholds ($\lambda_c$) in the HCSA model when $\mu=1.0$.}}
\label{hncsa-mu10}
\end{table}

\vspace{-2em}
\begin{table}[!h]
\centering
\setlength{\tabcolsep}{3pt}
\renewcommand{\arraystretch}{1.6}  
\begin{tabular}{@{}c@{\hspace{0.8em}}c@{}} 
\begin{tabular}[t]{|l|cccc|}   
\hline
\multirow{2}{*}{Dataset}  & \multicolumn{4}{c|}{$\lambda_c$} \\
\hhline{~|----|}
 & $\nu=1$ & $\nu=2$ & $\nu=3$ & $\nu=4$  \\
\hline
congress-bills      & $1\!\times\!10^{-3}$   & $3\!\times\!10^{-4}$ & $9\!\times\!10^{-5}$&  $8\!\times\!10^{-5}$\\
\hline
house-committees    & $2\!\times\!10^{-2}$  & $6\!\times\!10^{-3}$  & $3\!\times\!10^{-3}$ & $2\!\times\!10^{-3}$  \\
\hline
music-review        & $4\!\times\!10^{-2}$  & $7\!\times\!10^{-3}$  & $4\!\times\!10^{-3}$ & $2\!\times\!10^{-3}$  \\
\hline
M\_PL\_062\_ins     & $1\!\times\!10^{-2}$  & $6\!\times\!10^{-3}$  & $2\!\times\!10^{-3}$ & $8\!\times\!10^{-4}$  \\
\hline
email-EU            & $3\!\times\!10^{-3}$   & $3\!\times\!10^{-3}$ & $1\!\times\!10^{-3}$ & $7\!\times\!10^{-4}$  \\
\hline
M\_PL\_015\_ins     & $5\!\times\!10^{-2}$  & $1\!\times\!10^{-2}$ & $5\!\times\!10^{-3}$ & $4\!\times\!10^{-3}$ \\
\hline
Mid1                & $6\!\times\!10^{-4}$   & $6\!\times\!10^{-4}$ & $4\!\times\!10^{-4}$ & $2\!\times\!10^{-4}$  \\
\hline
geometry-questions  & $1\!\times\!10^{-2}$   & $4\!\times\!10^{-3}$ & $2\!\times\!10^{-3}$ & $8\!\times\!10^{-4}$ \\
\hline
M\_PL\_062\_pl      & $5\!\times\!10^{-3}$   & $3\!\times\!10^{-3}$ & $2\!\times\!10^{-3}$ & $8\!\times\!10^{-4}$ \\
\hline
algebra-questions   & $2\!\times\!10^{-2}$   & $8\!\times\!10^{-3}$ & $2\!\times\!10^{-3}$ & $2\!\times\!10^{-3}$ \\
\hline
\end{tabular}
&
\begin{tabular}[t]{|l|cccc|}
\hline
\multirow{2}{*}{Dataset}  & \multicolumn{4}{c|}{$\lambda_c$} \\
\hhline{~|----|}
 & $\nu=1$ & $\nu=2$ & $\nu=3$ & $\nu=4$  \\
\hline
SFHH                & $9\!\times\!10^{-3}$   & $5\!\times\!10^{-3}$ & $5\!\times\!10^{-3}$ & $4\!\times\!10^{-3}$\\
\hline
Elem1              & $9\!\times\!10^{-4}$   & $7\!\times\!10^{-4}$ & $4\!\times\!10^{-4}$ & $2\!\times\!10^{-4}$  \\
\hline
Thiers13           & $6\!\times\!10^{-3}$   & $6\!\times\!10^{-3}$ &  $5\!\times\!10^{-3}$  &  $5\!\times\!10^{-3}$  \\
\hline
senate-bills       & $3\!\times\!10^{-4}$   & $1\!\times\!10^{-4}$ &  $5\!\times\!10^{-5}$  &  $4\!\times\!10^{-5}$ \\
\hline
senate-committees  & $9\!\times\!10^{-3}$   & $4\!\times\!10^{-3}$ &  $2\!\times\!10^{-3}$  &  $2\!\times\!10^{-3}$ \\
\hline
LyonSchool         & $1\!\times\!10^{-3}$   & $1\!\times\!10^{-3}$ &  $8\!\times\!10^{-4}$  &  $5\!\times\!10^{-4}$ \\
\hline
InVS15             & $7\!\times\!10^{-3}$   & $6\!\times\!10^{-3}$ &  $9\!\times\!10^{-3}$  &  $5\!\times\!10^{-3}$ \\
\hline
email-Enron        & $9\!\times\!10^{-3}$   & $9\!\times\!10^{-3}$ &  $4\!\times\!10^{-3}$  &  $2\!\times\!10^{-3}$ \\
\hline
M\_PL\_015\_pl     & $1\!\times\!10^{-2}$   & $5\!\times\!10^{-3}$ &  $3\!\times\!10^{-3}$  &  $2\!\times\!10^{-3}$\\
\hline
LH10               &  $6\!\times\!10^{-3}$   & $6\!\times\!10^{-3}$ &  $4\!\times\!10^{-3}$  &  $4\!\times\!10^{-3}$  \\
\hline
\end{tabular}
\end{tabular}
\captionsetup{justification=raggedright, singlelinecheck=false, width=\textwidth}
\caption{\textbf{Spreading thresholds ($\lambda_c$) in the HCSA model when $\mu=0.1$.}}
\label{hncsa-mu1}
\end{table}

\newpage
\subsubsection{Results for transferability evaluation under HCSA}\label{S3.2}
\begin{figure}[H]
    \centering
    \includegraphics[]{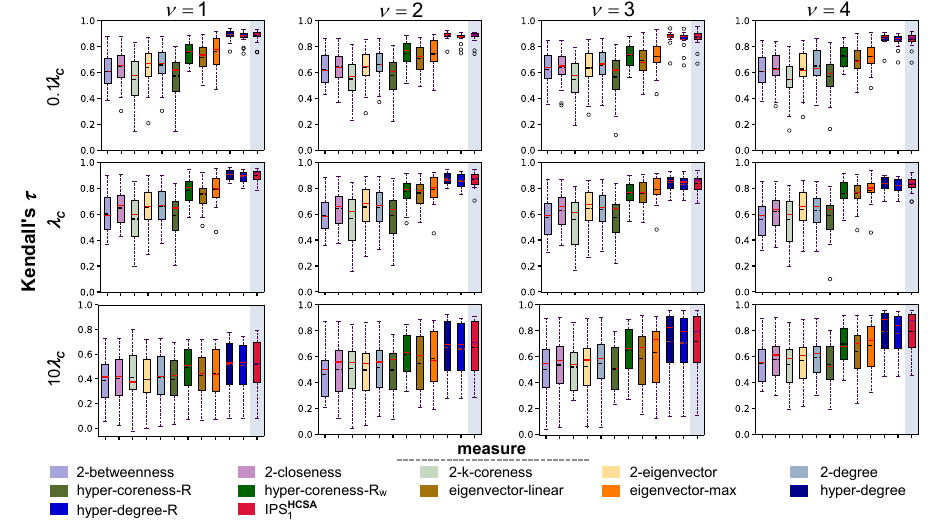}
    \captionsetup{justification=raggedright, singlelinecheck=false, width=\textwidth}
    \caption{\textbf{Performance of measures in the HCSA model when $\mu=1.0$.}
    Results of Kendall's $\tau$ between node measure and node influence are shown as boxplots (containing 20 real hypergraphs). The solid line in each box is the median value, and the dashed line is the mean value. Node influence is obtained through 1000 simulations.}
    \label{hncsa-10}
\end{figure}

\begin{figure}[H]
    \centering
    \includegraphics[]{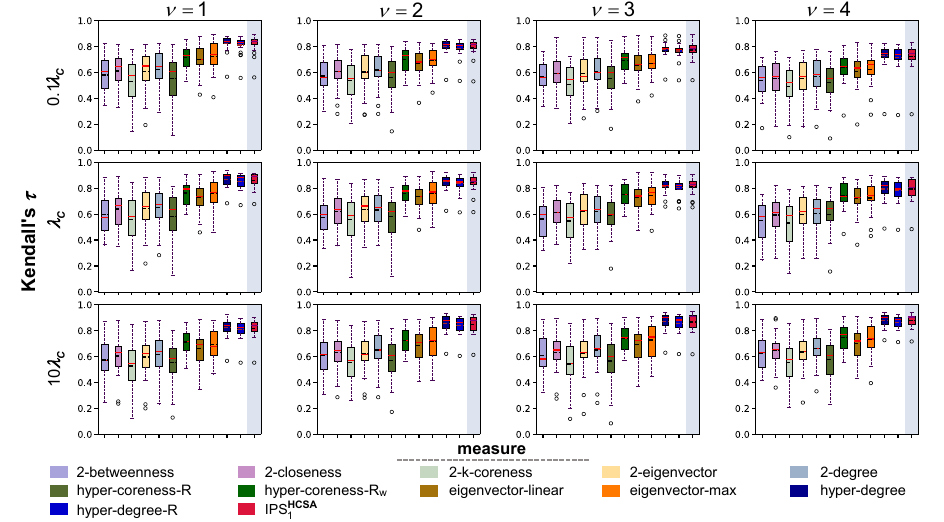}
    \captionsetup{justification=raggedright, singlelinecheck=false, width=\textwidth}
    \caption{\textbf{Performance of measures in the HCSA model when $\mu=0.1$.}
    Results of Kendall's $\tau$ between node measure and node influence are shown as boxplots (containing 20 real hypergraphs). The solid line in each box is the median value, and the dashed line is the mean value. Node influence is obtained through 300 simulations.}
    \label{hncsa-1}
\end{figure}

For each group of parameters and each hypergraph shown in Tab.~\ref{hncsa-mu10} (also \ref{hncsa-mu1}), we take $50\%$ nodes from it to serve as the seed in turn, and then calculate Kendall's $\tau$ for measures. Finally, results of Kendall's $\tau$ are shown in Fig.~\ref{hncsa-10} and Fig.\ref{hncsa-1} (corresponding to cases with $\mu=1$ and $\mu=0.1$, respectively); in each figure, subplots are arranged by different settings of $\lambda$ and $\nu$.

We find that IPS ($IPS_1^{HCSA}$) achieves the best performance in most parameter settings.
This verifies the transferability of IPS to the HCSA model. 
Besides, the hyper-degree method exhibits comparable performance, which can be explained from two perspectives.
On the one hand, its calculation is similar to $IPS^{HCSA}_{1}$.
On the other hand, hyper-degree gains an artificial advantage in Kendall's $\tau$ due to its “weak monotonicity”, i.e., assigning identical values to many nodes. 
That is, the hyper-degree has a smaller number of centrality values, causing Kendall's $\tau$ to be biased towards larger values. 
To verify this, we add a uniformly distributed random variable $\delta \in (0,1)$ to each node's hyper-degree value. This randomized hyper-degree (noted as hyper-degree-R) expands the measure values while preserving the original ordinal relationships among nodes with different hyper-degree values. Results in the following Fig.~\ref{hncsa-10} and \ref{hncsa-1} show that the $\tau$ values of hyper-degree-R under different parameters are consistently lower than those of IPS methods.

\clearpage

\subsection{Transferability evaluation under HTC}\label{S4}
\subsubsection{Parameter settings for HTC}\label{S4.1}
\begin{figure}[H]
    \centering
    \includegraphics[]{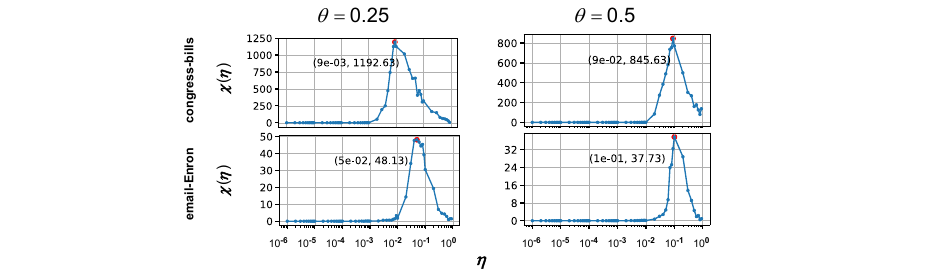}
    \captionsetup{justification=raggedright, singlelinecheck=false, width=\textwidth}
    \caption{\textbf{Determination of $\eta_c$ based on susceptibility when $\mu =1$ in the HTC model.} 
    Each subplot corresponds to a parameter $\nu$ and a real hypergraph, where susceptibility is shown against $\eta$. 
    In the subplots, each point is obtained from 1000 independent simulations, each starting from a randomly selected seed. The peak, indicated by a red point, represents the threshold ($\eta_c$).}
    \label{htc-10-1}
\end{figure}

\begin{figure}[H]
    \centering
    \includegraphics[]{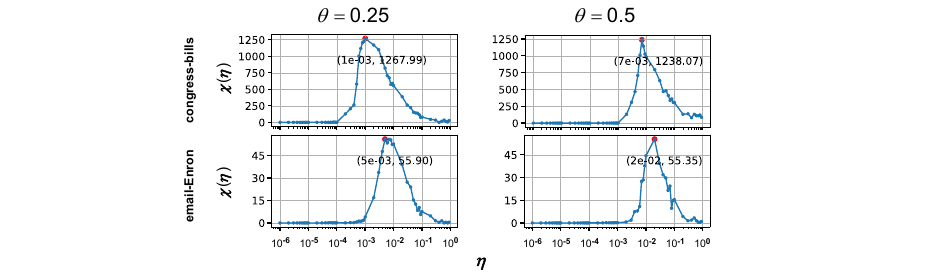}
    \captionsetup{justification=raggedright, singlelinecheck=false, width=\textwidth}
    \caption{\textbf{Determination of $\eta_c$ based on susceptibility when $\mu =0.1$ in the HTC model.} 
    Each subplot corresponds to a parameter $\nu$ and a real hypergraph, where susceptibility is shown against $\eta$. 
    In the subplots, each point is obtained from 300 independent simulations, each starting from a randomly selected seed.  The peak, indicated by a red point, represents the threshold ($\eta_c$).}
    \label{htc-1-1}
\end{figure}

Furthermore, we test the transferability of the IPS method under the HTC model, where we consider different parameters to ensure the robustness of the results, including $\mu=1,0.1$; $\theta=0.25,\,0.5$; $\eta=0.1\eta_c$, $\eta_c$, $10\eta_c$ (or $2\eta_c$ instead when $10\eta_c \geq1$), respectively.
In this model, we also determine the threshold of $\eta$ (i.e., $\eta_c$) under given $\theta$ and $\mu$ through the calculation of susceptibility in Eq.~\eqref{chi}, as exemplified in Fig.~\ref{htc-10-1} for cases when $\mu=1$, and Fig.~\ref{htc-1-1} for cases when $\mu=0.1$. 
Here, for subsequent analysis, we retain 11 hypergraphs where the peak of $\chi$ can be clearly observed. 
The values of $\eta_c$ are given in the following Tab.~\ref{htc-mu10} ($\mu=1$, corresponding to Fig.~5 in the main text) and Tab.~\ref{htc-mu1} ($\mu=0.1$), respectively.

\begin{table}[h]
\centering
\setlength{\tabcolsep}{3pt}
\renewcommand{\arraystretch}{1.6} 
\begin{tabular}{@{}c@{\hspace{3em}}c@{}} 
\begin{tabular}[t]{|l|cc|}  
\hline
\multirow{2}{*}{Dataset}  & \multicolumn{2}{c|}{$\eta_c$} \\
\hhline{~|--|}
 & $\theta=0.25$ & $\theta=0.5$   \\
\hline
congress-bills        & $9\!\times\!10^{-3}$ & $9\!\times\!10^{-2}$ \\
\hline
email-EU              & $2\!\times\!10^{-2}$ & $9\!\times\!10^{-2}$   \\
\hline
Mid1                  & $3\!\times\!10^{-3}$ & $5\!\times\!10^{-2}$   \\
\hline
SFHH                  & $2\!\times\!10^{-2}$ & $7\!\times\!10^{-2}$ \\
\hline 
Elem1                & $6\!\times\!10^{-3}$ & $2\!\times\!10^{-1}$   \\
\hline
Thiers13             & $2\!\times\!10^{-2}$ &  $2\!\times\!10^{-1}$    \\
\hline
\end{tabular}
&
\begin{tabular}[t]{|l|cc|}
\hline
\multirow{2}{*}{Dataset}  & \multicolumn{2}{c|}{$\eta_c$} \\
\hhline{~|--|}
 & $\theta=0.25$ & $\theta=0.5$   \\
\hline
senate-bills        & $5\!\times\!10^{-3}$ &  $5\!\times\!10^{-2}$   \\
\hline
LyonSchool          & $6\!\times\!10^{-3}$ &  $2\!\times\!10^{-1}$   \\
\hline
InVS15              & $3\!\times\!10^{-2}$ &  $1\!\times\!10^{-1}$   \\
\hline
email-Enron         & $5\!\times\!10^{-2}$ &  $1\!\times\!10^{-1}$   \\
\hline
LH10                 & $2\!\times\!10^{-2}$ &  $2\!\times\!10^{-1}$    \\
\hline
\end{tabular}
\end{tabular}
\captionsetup{justification=raggedright, singlelinecheck=false, width=\textwidth}
\caption{\textbf{Threshold of parameter $\eta$ in the HTC model when $\mu=1.0$.}}
\label{htc-mu10}
\end{table}

\begin{table}[h]
\centering
\setlength{\tabcolsep}{3pt}
\renewcommand{\arraystretch}{1.6}  
\begin{tabular}{@{}c@{\hspace{3em}}c@{}} 
\begin{tabular}[t]{|l|cc|}   
\hline
\multirow{2}{*}{Dataset}  & \multicolumn{2}{c|}{$\eta_c$} \\
\hhline{~|--|}
 & $\theta=0.25$ & $\theta=0.5$   \\
\hline
congress-bills        & $1\!\times\!10^{-3}$ & $7\!\times\!10^{-3}$ \\
\hline
email-EU              & $2\!\times\!10^{-3}$ & $6\!\times\!10^{-3}$   \\
\hline
Mid1                  & $3\!\times\!10^{-4}$ & $5\!\times\!10^{-3}$   \\
\hline
SFHH                  & $3\!\times\!10^{-3}$ & $1\!\times\!10^{-2}$ \\
\hline 
Elem1                & $7\!\times\!10^{-4}$ & $1\!\times\!10^{-2}$   \\
\hline
Thiers13             & $3\!\times\!10^{-3}$ &  $2\!\times\!10^{-2}$    \\
\hline
\end{tabular}
&
\begin{tabular}[t]{|l|cc|}
\hline
\multirow{2}{*}{Dataset}  & \multicolumn{2}{c|}{$\eta_c$} \\
\hhline{~|--|}
  & $\theta=0.25$ & $\theta=0.5$   \\
\hline
senate-bills        & $6\!\times\!10^{-4}$ &  $5\!\times\!10^{-3}$   \\
\hline
LyonSchool          & $1\!\times\!10^{-3}$ &  $2\!\times\!10^{-2}$   \\
\hline
InVS15              & $3\!\times\!10^{-3}$ &  $1\!\times\!10^{-2}$   \\
\hline
email-Enron         & $5\!\times\!10^{-3}$ &  $2\!\times\!10^{-2}$   \\
\hline
LH10                 & $4\!\times\!10^{-3}$ &  $2\!\times\!10^{-2}$    \\
\hline
\end{tabular}
\end{tabular}
\captionsetup{justification=raggedright, singlelinecheck=false, width=\textwidth}
\caption{\textbf{Threshold of parameter $\eta$ in the HTC model when $\mu=0.1$.} }
\label{htc-mu1}
\end{table}

\newpage
\subsubsection{Results for transferability evaluation unde HTC}\label{S4.2}
\begin{figure}[H]
    \centering
    \includegraphics[]{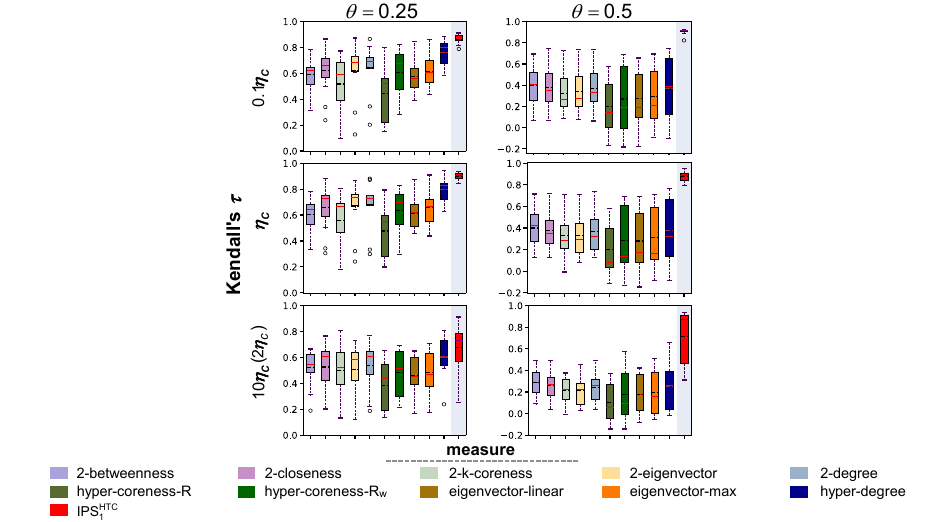}
    \captionsetup{justification=raggedright, singlelinecheck=false, width=\textwidth}
    \caption{\textbf{Performance of measures in the HTC model when $\mu=1.0$.}
    Results of Kendall's $\tau$ between node measure and node influence are shown as boxplots (containing 11 real hypergraphs). The solid line in each box is the median value, and the dashed line is the mean value. Node influence is obtained through 1000 simulations.}
    \label{htc-10}
\end{figure}

\begin{figure}[H]
    \centering
    \includegraphics[]{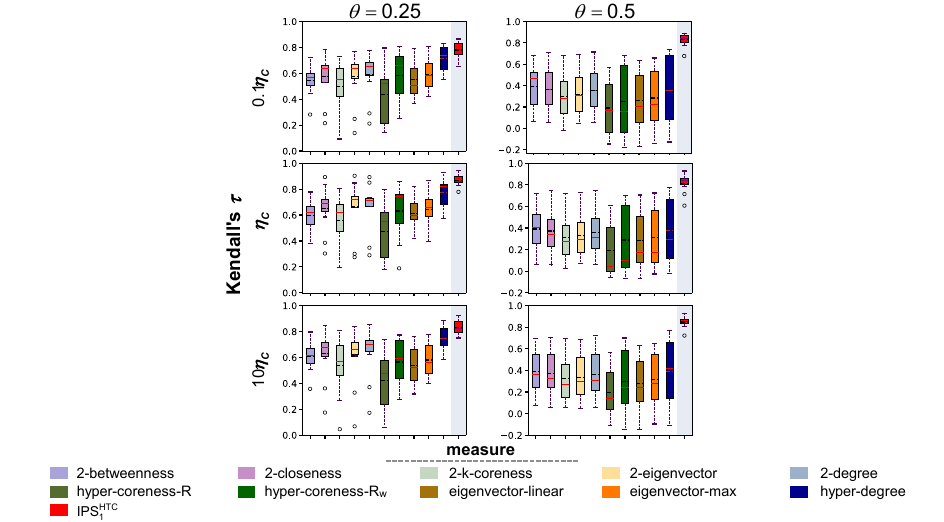}
    \captionsetup{justification=raggedright, singlelinecheck=false, width=\textwidth}
    \caption{\textbf{Performance of measures in the HTC model when $\mu=0.1$.}
    Results of Kendall's $\tau$ between node measure and node influence are shown as boxplots (containing 11 real hypergraphs). The solid line in each box is the median value, and the dashed line is the mean value. Node influence is obtained through 300 simulations.}
    \label{htc-1}
\end{figure}

For each group of parameters and each hypergraph shown in Tab.~\ref{htc-mu10} (also \ref{htc-mu1}), we take $50\%$ nodes from it to serve as the seed in turn, and then calculate Kendall's $\tau$ for measures. Finally, results of Kendall's $\tau$ are shown in Fig.~\ref{htc-10} and Fig.\ref{htc-1} (corresponding to cases with $\mu=1$ and $\mu=0.1$, respectively); in each figure, subplots are arranged by different settings of $\theta$ and $\eta$.
Notably, $IPS_{1}^{HTC}$ significantly outperforms other methods in all parameter settings.
This finding validates the effectiveness of IPS in the HTC model and further verifies the transferability of the IPS framework.

\newpage
\section{Cross-paradigm transferability of IPS} 
\begin{figure}[H]
    \centering
    \includegraphics[]{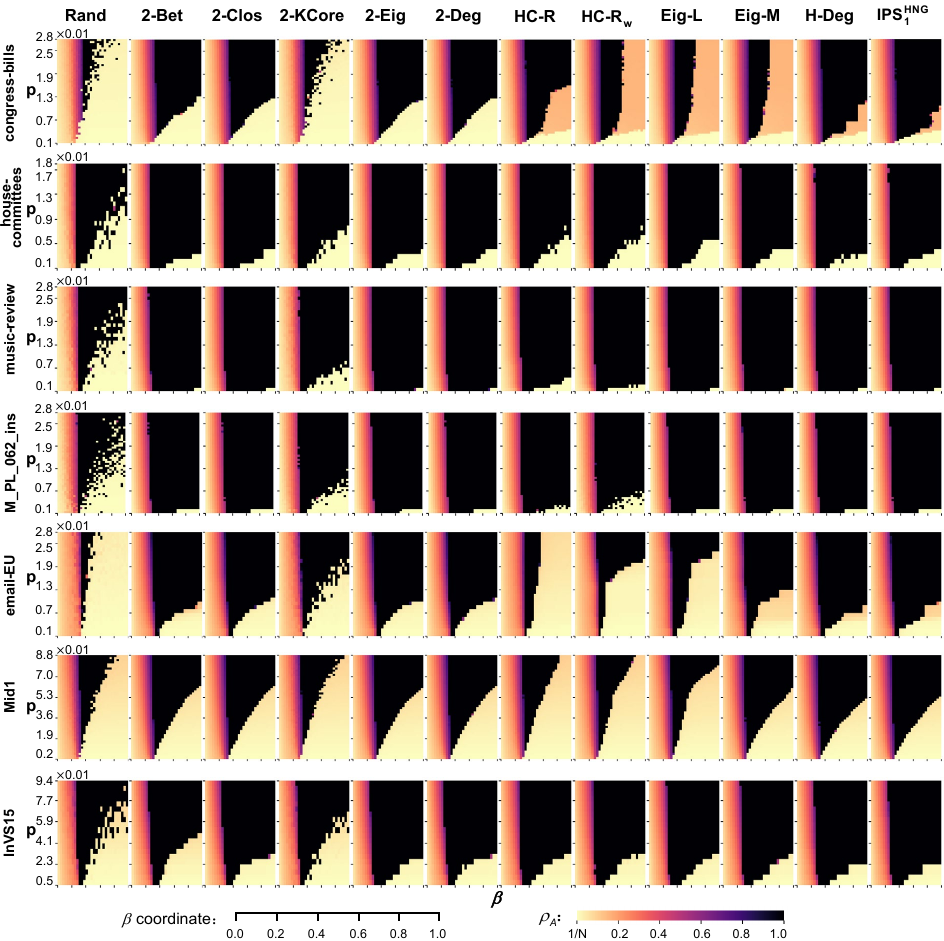}
    \captionsetup{justification=raggedright, singlelinecheck=false, width=\textwidth}
    \caption{\textbf{The steady-state abundance of name $A$ under the unanimity rule.}
    The vertical axis $p$ corresponds to the top $pN$ of nodes identified by different measures, which is abbreviated at the top of the column. Specifically, Rand: Random, 2-Bet: 2-betweenness, 2-Clos: 2-closeness, 2-KCore: 2-k-coreness, 2-Eig: 2-eigenvector, 2-Deg: 2-degree, HC-R: hyper-coreness-$R$, HC-$R_w$: hyper-coreness-$R_w$, Eig-L: eigenvector-linear, Eig-M: eigenvector-max, H-Deg: hyper-degree, $ IPS^{HNG}_{1}$: $ IPS^{HNG}_{1}$.}
    \label{unanimity}
\end{figure}

\begin{figure}[H]
    \centering
    \includegraphics[]{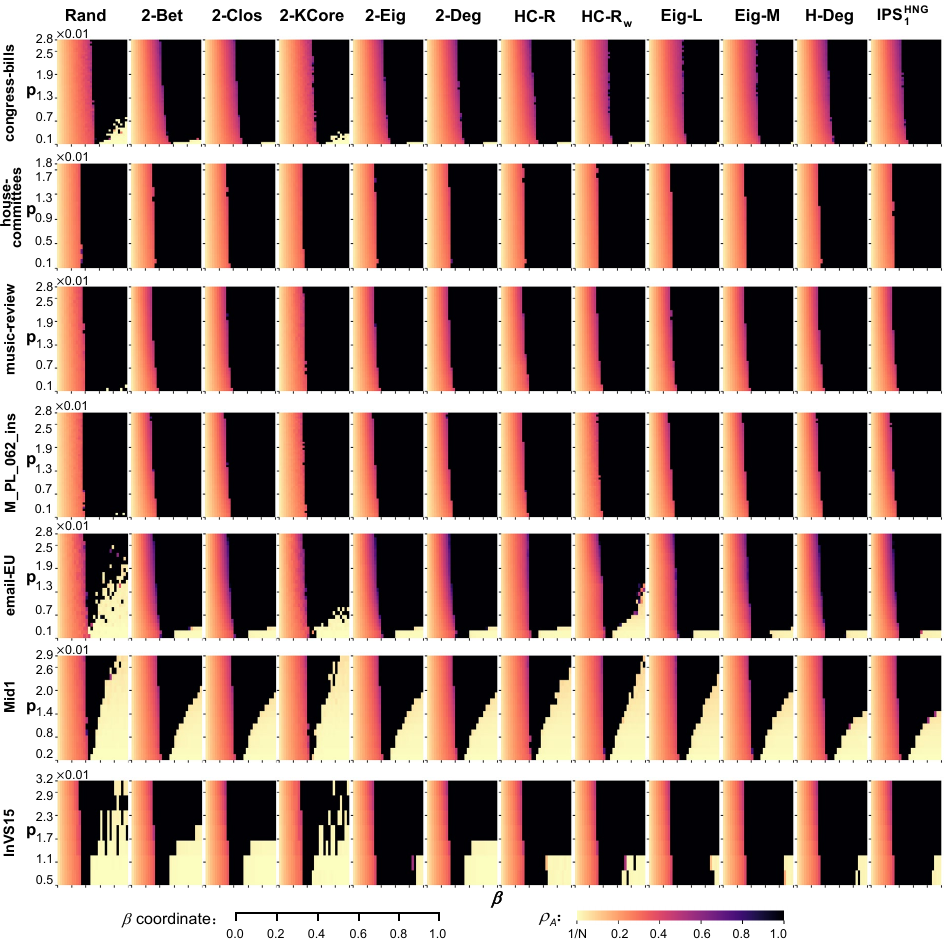}
    \captionsetup{justification=raggedright, singlelinecheck=false, width=\textwidth}
    \caption{\textbf{The steady-state abundance of name $A$ under the union rule.} The vertical axis $p$ corresponds to the top $pN$ of nodes identified by different measures, which is abbreviated at the top of the column. Specifically, Rand: Random, 2-Bet: 2-betweenness, 2-Clos: 2-closeness, 2-KCore: 2-k-coreness, 2-Eig: 2-eigenvector, 2-Deg: 2-degree, HC-R: hyper-coreness-$R$, HC-$R_w$: hyper-coreness-$R_w$, Eig-L: eigenvector-linear, Eig-M: eigenvector-max, H-Deg: hyper-degree, $ IPS^{HNG}_{1}$: $ IPS^{HNG}_{1}$.}
    \label{union}
\end{figure}

We explore the cross-paradigm transferability of the IPS framework by applying it to two types of higher-order naming game dynamics (HNG), i.e., the unanimity rule and the union rule.
Figure~\ref{unanimity} presents the cases of the unanimity rule.
Specifically, each heatmap shows the steady-state abundance of the name $A$ (denoted as $\rho_A$) under different $(\beta,p)$.
In each grid cell, the vertical axis ($p$) indicates that the top $pN$ nodes are selected as committed individuals adopting the name $A$, through centrality measures or through the random selection.
For a more realistic comparison, the committed individuals are re-chosen randomly for each value of $p$, instead of truncating the top-$pN$ ranked nodes in a single randomly generated ranking.
To integrate those $p_c$ corresponding to different values of $\beta$, we use the area of $\rho_A=0$ (black regions) to indicate the effectiveness of the measure.
A larger black region means a smaller critical mass $p_c$, which indicates that fewer committed individuals are needed to lead to the adoption of the name $A$ in the whole population, thereby demonstrating greater effectiveness of the measure. 
We find that IPS consistently triggers the smallest $p_c$ in almost all hypergraphs. This finding validates the transferability of IPS in the higher-order naming game model. In addition, IPS also performs the best when considering the union rule in the higher-order naming game (see Fig.~\ref{union}).

\newpage
\section{The performance of IPS in epidemic containment}\label{S6}
The immunization strategy is a classical problem in both complex network science research and human society. 
Its core objective is to find the optimal sequence for node (or edge) removal, a process that is fundamentally based on centrality measures~\cite{imm1}. 

Here, we apply the IPS framework to design immunization strategies for higher-order epidemic containment.
Following the general experimental procedures, we take the SIS type HCP model (denoted as HCP-SIS), where $I$ state nodes convert to the $S$ state with probability $\mu$, rather than to the $R$ state.
We monitor the disease prevalence $\rho$ to evaluate different methods~\cite{imm2}.
In particular, we use different centrality measures to identify important nodes to immunize. The smaller the ratio of immunized nodes required to control the disease, the more effective the centrality measure is.
Generally, the immunization threshold $p_c$ is the minimum node immunization fraction $p$ that satisfies $\rho<1/N$ (a commonly used disease control threshold). 

\begin{figure}[H]
    \centering
    \includegraphics[]{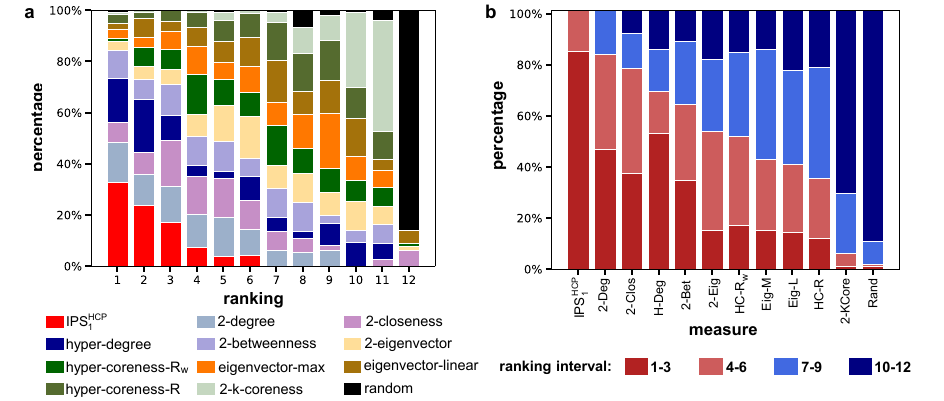}
    \captionsetup{justification=raggedright, singlelinecheck=false, width=\textwidth}
    \caption{\textbf{The overall performance of different measures in epidemic containment. } \textbf{a}: The ranking of control thresholds is obtained under different dynamical settings and different hypergraphs, and we present the proportion of each method in each ranking. \textbf{b}: For all situations, the proportion that the method is in top 6 is as follows: $IPS_{1}^{HCP}$: $100\%$, 2-degree: $83.13\%$, 2-closeness: $77.50\%$, hyper-degree: $68.75\%$, 2-betweenness: $63.75\%$, 2-eigenvector: $53.13\%$, hyper-coreness-$R_w$: $51.25\%$, eigenvector-max: $42.50\%$, eigenvector-linear: $40.63\%$, hyper-coreness-$R$: $35.00\%$, 2-k-coreness:$6.25\%$, random: $1.88\%$.}
    \label{imm-bar}
\end{figure}

\begin{figure}[H]
    \centering
    \includegraphics[]{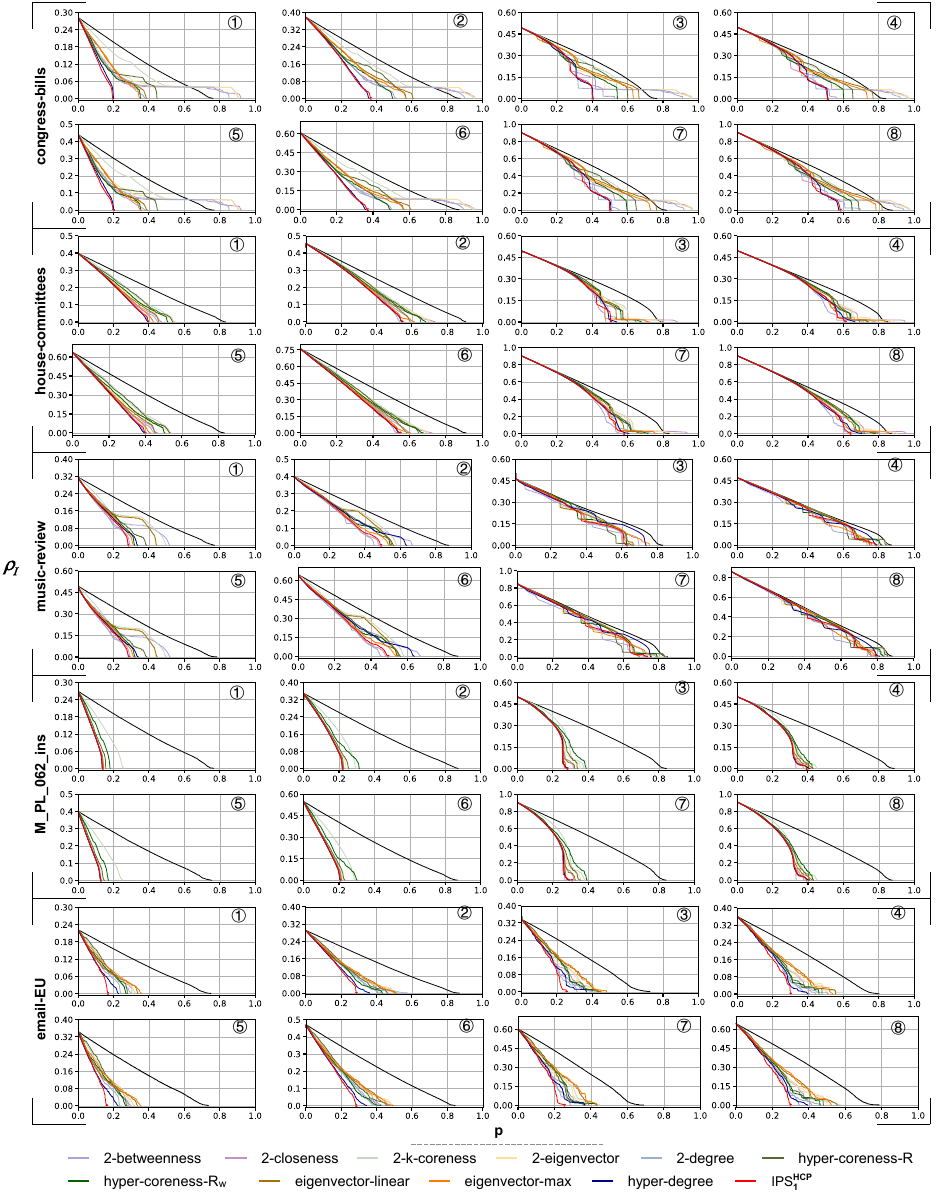}
    \captionsetup{justification=raggedright, singlelinecheck=false, width=\textwidth}
    \caption{\textbf{Prevalence under different immunization strategies ($\mu=1$, Part 1).} The prevalence is shown against the immunized fraction. Each subplot corresponds to the case under a real hypergraph and a dynamical setting (indicated by the circled number). 
    }
    \label{imm-1}
\end{figure}

\begin{figure}[H]
    \centering
    \includegraphics[]{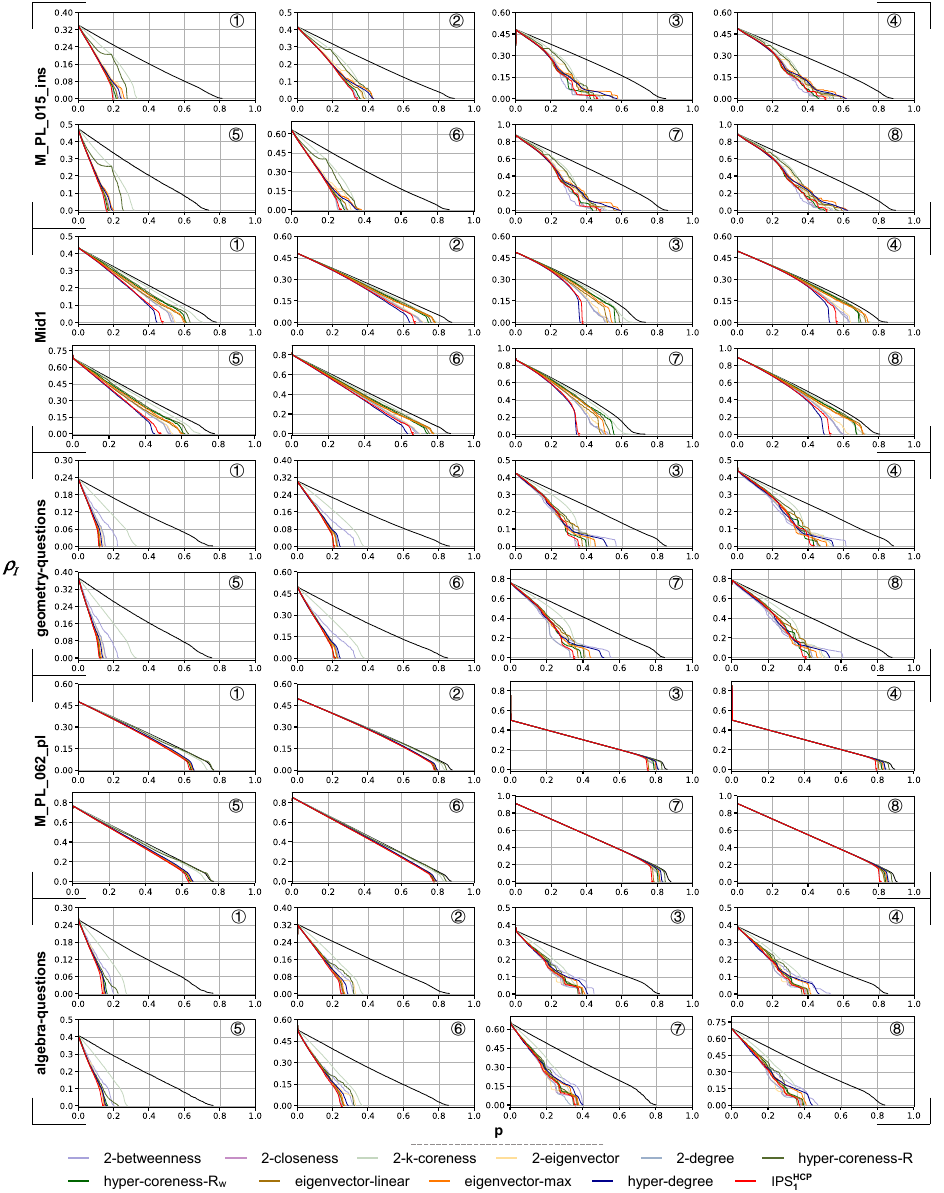}
    \captionsetup{justification=raggedright, singlelinecheck=false, width=\textwidth}
    \caption{\textbf{Prevalence under different immunization strategies ($\mu=1$, Part 2).} The prevalence is shown against the immunized fraction. Each subplot corresponds to the case under a real hypergraph and a dynamical setting (indicated by the circled number). 
    }
    \label{imm-2}
\end{figure}

\begin{figure}[H]
    \centering
    \includegraphics[]{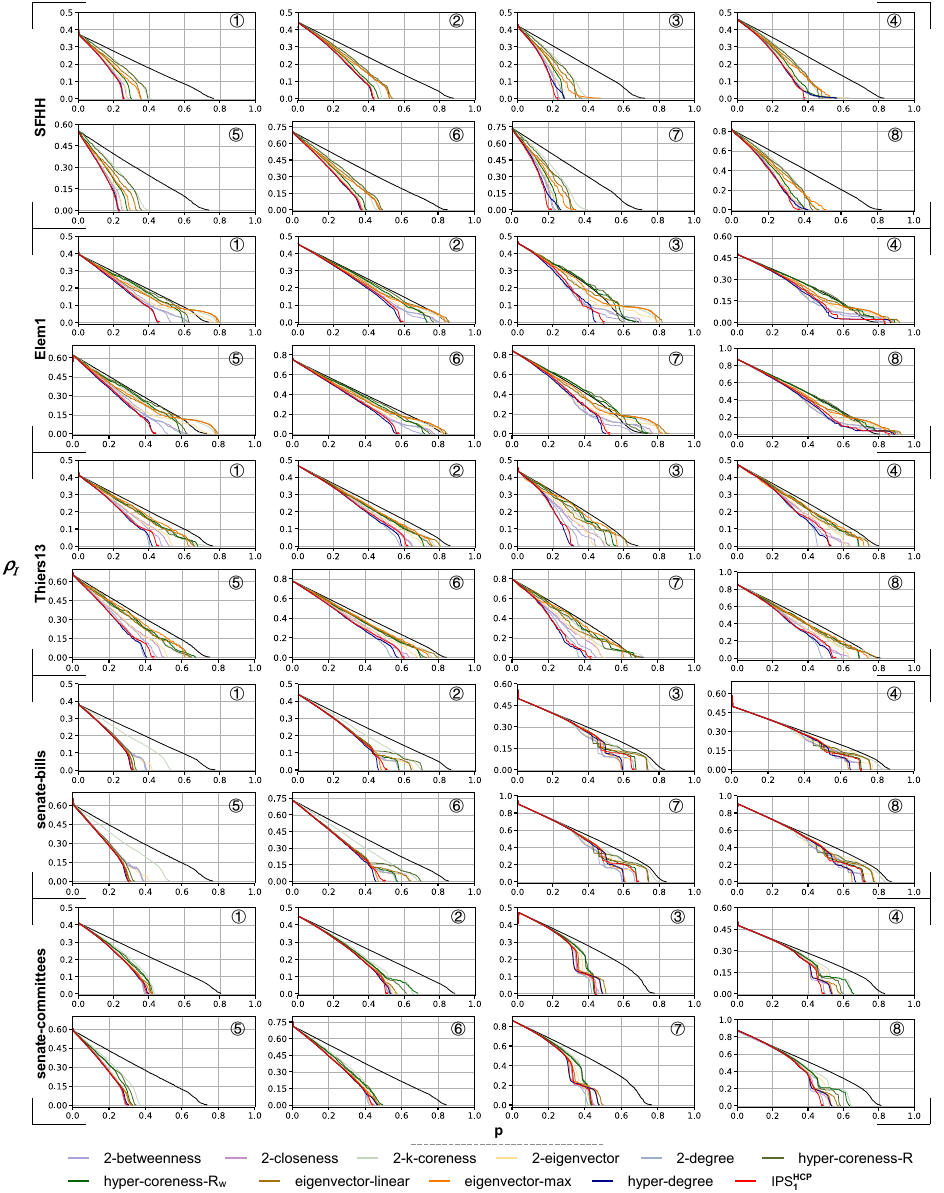}
    \captionsetup{justification=raggedright, singlelinecheck=false, width=\textwidth}
    \caption{\textbf{Prevalence under different immunization strategies ($\mu=0.1$, Part 1).} The prevalence is shown against the immunized fraction. Each subplot corresponds to the case under a real hypergraph and a dynamical setting (indicated by the circled number). 
    }
    \label{imm-3}
\end{figure}

\begin{figure}[H]
    \centering
    \includegraphics[]{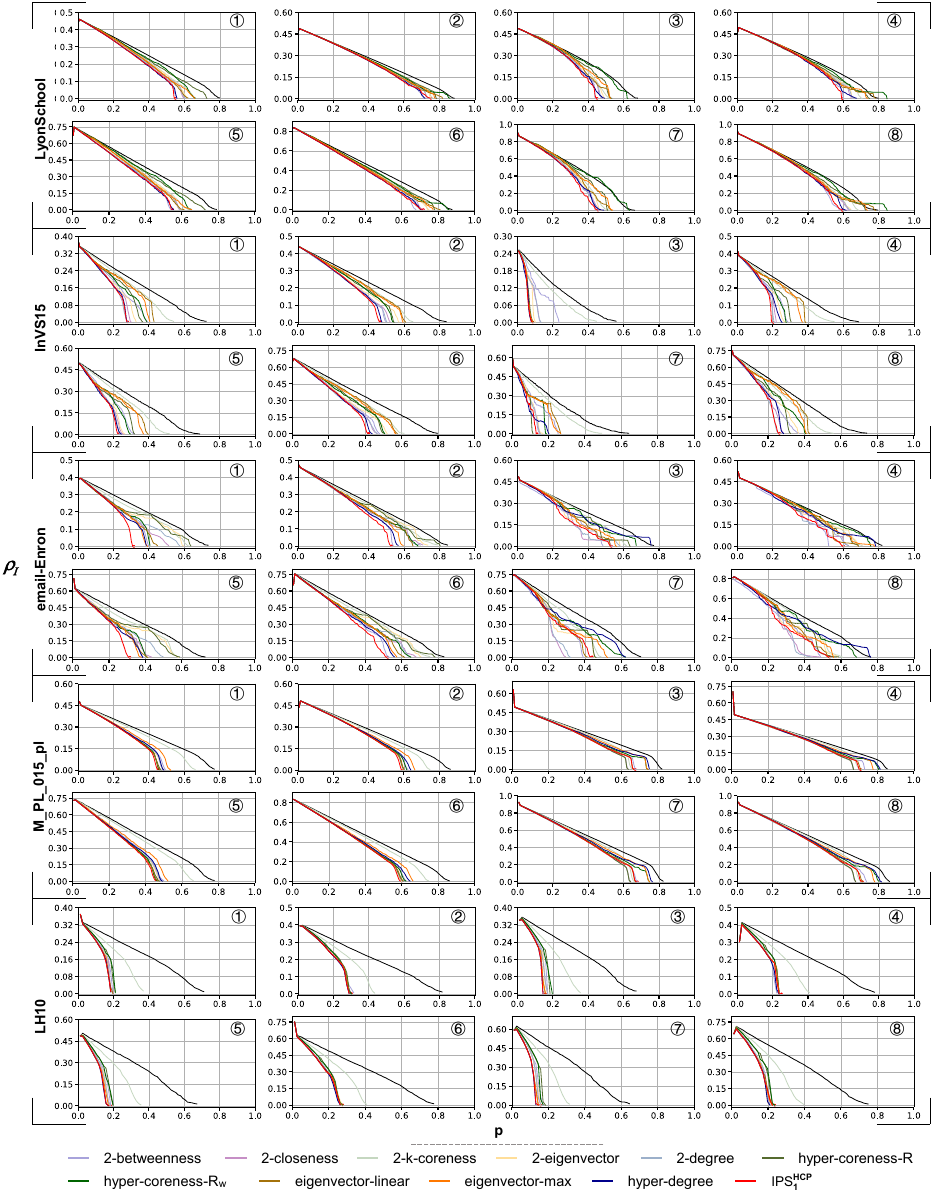}
    \captionsetup{justification=raggedright, singlelinecheck=false, width=\textwidth}
    \caption{\textbf{Prevalence under different immunization strategies ($\mu=0.1$, Part 2).} The prevalence is shown against the immunized fraction. Each subplot corresponds to the case under a real hypergraph and a dynamical setting (indicated by the circled number). 
    }
    \label{imm-4}
\end{figure}

We conduct extensive experiments on 20 real-world hypergraphs. For each hypergraph, we test eight groups of parameters (shown in Tab.~\ref{immpara}). 
In each scenario, we rank these methods by their control threshold in ascending order (i.e., descending order of performance). Figure~\ref{imm-bar} summarizes the performance of each method in all situations.
Results show that the IPS measure is always in the top 6, thereby performing overall best, while others drop out of the top 6 in some cases.
In addition, unlike the previous cases of spreading promotion (i.e., identifying influential nodes in SIR-type propagation), 2-degree, 2-closeness, and 2-betweenness, perform well here in epidemic containment. That is, there is a certain mismatch between spreading influence and the ability to block spreads. Notably, IPS achieves good performance in both objectives, indicating its outstanding ability in locating critical nodes.
The detailed results are provided in Fig.~\ref{imm-1}--\ref{imm-4}, where the circled number in the upper right corner of each subplot indicates the corresponding propagation parameters shown in Tab.~\ref{immpara}. 

\begin{table}[h]
\centering
\setlength{\tabcolsep}{3pt}
\renewcommand{\arraystretch}{1.6} 
\begin{tabular}{@{}c@{\hspace{0.8em}}c@{}} 
\begin{tabular}[t]{|l|cc|cc|cc|cc|}   
\hline
\multirow{5}{*}{Dataset}  & \multicolumn{8}{c|}{Parameters}\\
\hhline{~|--------|}
 & \multicolumn{4}{c|}{$\mu=1$} & \multicolumn{4}{c|}{$\mu=0.1$}\\
\hhline{~|--------|}
 & \multicolumn{2}{c|}{$\nu=1$} & \multicolumn{2}{c|}{$\nu=3$} & \multicolumn{2}{c|}{$\nu=1$} & \multicolumn{2}{c|}{$\nu=3$} \\
\hhline{~|--------|}
 & $\lambda_1$ & $\lambda_2$ & $\lambda_1$ & $\lambda_2$ & $\lambda_1$ & $\lambda_2$ & $\lambda_1$ & $\lambda_2$\\
\hhline{~|--------|}
 &\textcircled{1} &\textcircled{2} &\textcircled{3} &\textcircled{4} &\textcircled{5} &\textcircled{6} &\textcircled{7} &\textcircled{8} \\
\hline
congress-bills      & $ 4.0\!\times\!10^{-4}$ & $ 8.0\!\times\!10^{-4}$ & $ 1.0\!\times\!10^{-4}$ & $ 2.0\!\times\!10^{-4}$ & $ 4.0\!\times\!10^{-5}$ & $ 8.0\!\times\!10^{-5}$ & $ 1.0\!\times\!10^{-5}$ & $ 2.0\!\times\!10^{-5}$ \\ 
\hline
house-committees    & $ 1.0\!\times\!10^{-2}$ & $ 2.0\!\times\!10^{-2}$ & $ 5.0\!\times\!10^{-4}$ & $ 1.0\!\times\!10^{-3}$ & $ 1.0\!\times\!10^{-3}$ & $ 2.0\!\times\!10^{-3}$ & $ 3.5\!\times\!10^{-5}$ & $ 7.0\!\times\!10^{-5}$ \\ 
\hline
music-review        & $ 1.0\!\times\!10^{-2}$ & $ 2.0\!\times\!10^{-2}$ & $ 1.5\!\times\!10^{-3}$ & $ 3.0\!\times\!10^{-3}$ & $ 1.0\!\times\!10^{-3}$ & $ 2.0\!\times\!10^{-3}$ & $ 1.0\!\times\!10^{-4}$ & $ 2.0\!\times\!10^{-4}$ \\ 
\hline
M\_PL\_062\_ins     & $ 4.5\!\times\!10^{-3}$ & $ 9.0\!\times\!10^{-3}$ & $ 1.0\!\times\!10^{-3}$ & $ 2.0\!\times\!10^{-3}$ & $ 4.0\!\times\!10^{-4}$ & $ 8.0\!\times\!10^{-4}$ & $ 5.0\!\times\!10^{-5}$ & $ 1.0\!\times\!10^{-4}$ \\ 
\hline
email-EU            & $ 4.5\!\times\!10^{-3}$ & $ 9.0\!\times\!10^{-3}$ & $ 1.0\!\times\!10^{-3}$ & $ 2.0\!\times\!10^{-3}$ & $ 4.5\!\times\!10^{-4}$ & $ 9.0\!\times\!10^{-4}$ & $ 5.0\!\times\!10^{-5}$ & $ 1.0\!\times\!10^{-4}$ \\ 
\hline
M\_PL\_015\_ins     & $ 2.0\!\times\!10^{-2}$ & $ 4.0\!\times\!10^{-2}$ & $ 1.5\!\times\!10^{-3}$ & $ 3.0\!\times\!10^{-3}$ & $ 1.5\!\times\!10^{-3}$ & $ 3.0\!\times\!10^{-3}$ & $ 1.0\!\times\!10^{-4}$ & $ 2.0\!\times\!10^{-4}$ \\ 
\hline
Mid1                & $ 3.0\!\times\!10^{-3}$ & $ 6.0\!\times\!10^{-3}$ & $ 1.5\!\times\!10^{-3}$ & $ 3.0\!\times\!10^{-3}$ & $ 3.0\!\times\!10^{-4}$ & $ 6.0\!\times\!10^{-4}$ & $ 1.0\!\times\!10^{-4}$ & $ 2.0\!\times\!10^{-4}$ \\ 
\hline
geometry-questions  & $ 3.0\!\times\!10^{-3}$ & $ 6.0\!\times\!10^{-3}$ & $ 2.5\!\times\!10^{-4}$ & $ 5.0\!\times\!10^{-4}$ & $ 3.0\!\times\!10^{-4}$ & $ 6.0\!\times\!10^{-4}$ & $ 1.0\!\times\!10^{-5}$ & $ 2.0\!\times\!10^{-5}$ \\ 
\hline
M\_PL\_062\_pl      & $ 4.0\!\times\!10^{-3}$ & $ 8.0\!\times\!10^{-3}$ & $ 4.5\!\times\!10^{-4}$ & $ 9.0\!\times\!10^{-4}$ & $ 4.0\!\times\!10^{-4}$ & $ 8.0\!\times\!10^{-4}$ & $ 3.0\!\times\!10^{-5}$ & $ 6.0\!\times\!10^{-5}$ \\ 
\hline
algebra-questions   & $ 1.0\!\times\!10^{-2}$ & $ 2.0\!\times\!10^{-2}$ & $ 1.0\!\times\!10^{-3}$ & $ 2.0\!\times\!10^{-3}$ & $ 1.0\!\times\!10^{-3}$ & $ 2.0\!\times\!10^{-3}$ & $ 4.0\!\times\!10^{-5}$ & $ 8.0\!\times\!10^{-5}$ \\ 
\hline
SFHH                & $ 4.0\!\times\!10^{-2}$ & $ 8.0\!\times\!10^{-2}$ & $ 2.5\!\times\!10^{-2}$ & $ 5.0\!\times\!10^{-2}$ & $ 3.5\!\times\!10^{-3}$ & $ 7.0\!\times\!10^{-3}$ & $ 2.0\!\times\!10^{-3}$ & $ 4.0\!\times\!10^{-3}$ \\ 
\hline
Elem1              & $ 3.0\!\times\!10^{-3}$ & $ 6.0\!\times\!10^{-3}$ & $ 1.0\!\times\!10^{-3}$ & $ 2.0\!\times\!10^{-3}$ & $ 3.0\!\times\!10^{-4}$ & $ 6.0\!\times\!10^{-4}$ & $ 1.0\!\times\!10^{-4}$ & $ 2.0\!\times\!10^{-4}$ \\ 
\hline
Thiers13           & $ 4.0\!\times\!10^{-2}$ & $ 8.0\!\times\!10^{-2}$ & $ 2.0\!\times\!10^{-2}$ & $ 4.0\!\times\!10^{-2}$ & $ 4.0\!\times\!10^{-3}$ & $ 8.0\!\times\!10^{-3}$ & $ 2.0\!\times\!10^{-3}$ & $ 4.0\!\times\!10^{-3}$ \\ 
\hline
senate-bills       & $ 2.5\!\times\!10^{-4}$ & $ 5.0\!\times\!10^{-4}$ & $ 3.5\!\times\!10^{-5}$ & $ 7.0\!\times\!10^{-5}$ & $ 2.5\!\times\!10^{-5}$ & $ 5.0\!\times\!10^{-5}$ & $ 2.0\!\times\!10^{-6}$ & $ 4.0\!\times\!10^{-6}$ \\ 
\hline
senate-committees  & $ 1.5\!\times\!10^{-2}$ & $ 3.0\!\times\!10^{-2}$ & $ 2.0\!\times\!10^{-3}$ & $ 4.0\!\times\!10^{-3}$ & $ 1.0\!\times\!10^{-3}$ & $ 2.0\!\times\!10^{-3}$ & $ 1.0\!\times\!10^{-4}$ & $ 2.0\!\times\!10^{-4}$ \\ 
\hline
LyonSchool         & $ 1.0\!\times\!10^{-2}$ & $ 2.0\!\times\!10^{-2}$ & $ 3.5\!\times\!10^{-3}$ & $ 7.0\!\times\!10^{-3}$ & $ 1.0\!\times\!10^{-3}$ & $ 2.0\!\times\!10^{-3}$ & $ 2.5\!\times\!10^{-4}$ & $ 5.0\!\times\!10^{-4}$ \\ 
\hline
InVS15             & $ 3.5\!\times\!10^{-2}$ & $ 7.0\!\times\!10^{-2}$ & $ 1.0\!\times\!10^{-2}$ & $ 2.0\!\times\!10^{-2}$ & $ 3.0\!\times\!10^{-3}$ & $ 6.0\!\times\!10^{-3}$ & $ 1.0\!\times\!10^{-3}$ & $ 2.0\!\times\!10^{-3}$ \\ 
\hline
email-Enron        & $ 3.5\!\times\!10^{-2}$ & $ 7.0\!\times\!10^{-2}$ & $ 1.0\!\times\!10^{-2}$ & $ 2.0\!\times\!10^{-2}$ & $ 3.5\!\times\!10^{-3}$ & $ 7.0\!\times\!10^{-3}$ & $ 3.0\!\times\!10^{-4}$ & $ 6.0\!\times\!10^{-4}$ \\ 
\hline
M\_PL\_015\_pl     & $ 2.0\!\times\!10^{-2}$ & $ 4.0\!\times\!10^{-2}$ & $ 2.0\!\times\!10^{-3}$ & $ 4.0\!\times\!10^{-3}$ & $ 2.0\!\times\!10^{-3}$ & $ 4.0\!\times\!10^{-3}$ & $ 1.0\!\times\!10^{-4}$ & $ 2.0\!\times\!10^{-4}$ \\ 
\hline
LH10               & $ 3.0\!\times\!10^{-2}$ & $ 6.0\!\times\!10^{-2}$ & $ 1.5\!\times\!10^{-2}$ & $ 3.0\!\times\!10^{-2}$ & $ 2.5\!\times\!10^{-3}$ & $ 5.0\!\times\!10^{-3}$ & $ 1.0\!\times\!10^{-3}$ & $ 2.0\!\times\!10^{-3}$ \\ 
\hline
\end{tabular}
\end{tabular}
\captionsetup{justification=raggedright, singlelinecheck=false, width=\textwidth}
\caption{\textbf{Parameter settings in immunization experiments.} We set $\mu=0.1,1$ and $\nu=1,3$, respectively. For each pair of $\mu$ and $\nu$, we take two values of $\lambda$, satisfying $\lambda_2=2\lambda_1$.}
\label{immpara}
\end{table}

\newpage
\bibliography{sn-bib}